\documentclass[12pt]{article}
\usepackage[utf8]{inputenc}
\usepackage[margin=0.8in]{geometry}
\usepackage{amsmath, amssymb, amsfonts, amsthm, csquotes, graphicx, booktabs, enumitem}
\usepackage{libertine}
\usepackage{longtable}
\usepackage{pdflscape}

\usepackage{setspace}
\usepackage{natbib}
\usepackage{xurl}
\usepackage[citecolor=blue]{hyperref}
\usepackage{bbm}
\usepackage{tabulary}
\usepackage{float}
\usepackage{tikz}

\usepackage{caption}
\usepackage{makecell}
\usepackage{xr}
\usepackage{animate}
\usepackage{subcaption} 

\usepackage[affil-it]{authblk}
\usepackage[para,online,flushleft]{threeparttable}
 
\usepackage{tikz}
\newcommand{\tikzxmark}{%
\tikz[scale=0.23] {
    \draw[line width=0.7,line cap=round] (0,0) to [bend left=6] (1,1);
    \draw[line width=0.7,line cap=round] (0.2,0.95) to [bend right=3] (0.8,0.05);
}}

\usetikzlibrary{shapes.geometric, arrows}

\renewenvironment{abstract}{%
    \hfill\begin{minipage}{1\textwidth} 
    \textbf{\abstractname}
    \vspace{0.2cm}\par\noindent
}{%
    \end{minipage}%
}

\title{  \large{How Gender and Birth Order Affect Educational attainment Inequality within-Families: Evidence from Benin\footnote{I express my gratitude to my advisors Marc Henry and Andres Aradillas-Lopez, Keisuke Hirano, Michael Gechter, and Ewout Verriest for their invaluable feedback and insightful suggestions during the review process of this paper. Their expertise and thoughtful comments have greatly contributed to the refinement and enhancement of the research.}} }
\author{Christelle Zozoungbo\footnote{Pennsylvania State University: Department of Economics (cfz5115@psu.edu)}}
\date{August, 2024}

\begin{document}

\maketitle

The latest version of this paper is available \href{https://drive.google.com/file/d/1cmHgjTxVfq2MpVbhfV3nJDp_cLwW0sM5/view?usp=sharing}{here}.

\begin{center}

\begin{abstract}
{\small 
 \setstretch{0.9}
This paper examines how gender, birth order, and innate ability shape within-household disparities in children's educational attainment in developing countries. Using data from Benin, I find that in households with non-educated parents, gender and birth order drive over two-thirds of the average educational attainment disparities among adult children, while their influence decreases to one-third in households with college-educated parents. Furthermore, average inequality, measured by the range of children's educational attainment is twice as high among non-educated parents compared to college educated parents. I propose and estimate a structural model of educational attainment choices within-family. Using the model, I show that the absence of gender and birth order effects does not lead to a significant reduction in the average within-family disparities in children's educational attainment. Additionally, in theory, ensuring that every child has at least one year of education lowers average within-family educational inequality. Yet, even in this scenario, daughters tend to receive less education than sons, and practical efforts to achieve universal entry are less effective than the theoretical model.

}
\end{abstract}
    
\end{center}

\textbf{JEL classification:} I220, I240, H520

\textbf{Keywords:} Education policy, Inequality, Equality of education opportunities, gender gap.




\newpage
\section{Introduction}
\paragraph{}
The trade-off between the number of children one chooses to have and the average education of those children has been a subject of extensive research. The quantity-quality trade-off model predicts a negative relationship between the quantity and the average quality of children. Numerous empirical studies have provided support for this model in various contexts, consistently revealing a negative relationship between the number of children, and the average educational attainment of those children (\cite{becker1973interaction},  \cite{montgomery1995tradeoff} and \cite{li2008quantity}). The current state of the literature on fertility choices (\cite{conley2006parental}, \cite{maralani2008changing}, \cite{li2008quantity}, \cite{weng2019family}) states that rich and educated families tend to have fewer children while allocating greater investments in the education of those children, in comparison to less affluent and less educated families. However, in contexts where fertility can not be controlled by parents--- low access to contraceptive methods--- or when the quality of and access to public education system is poor or there is no clear understanding of the economic benefits of schooling, this trade-off seems to vanish (\cite{montgomery1995tradeoff}, \cite{black2005more}, \cite{maralani2008changing}).

When we extend the analysis beyond the choice of the aggregate education of children to how it is distributed among children within a family, we evidence that, not only does the average change across households, but so does the variance of education. In addition, the variance is most likely non-zero for a majority of households when there are budget constraints and no or poorly enforced compulsory education laws. 
In such circumstances, the educational outcomes of children are strongly influenced by their observed and unobserved individual characteristics, leading to disparities in the amount of education they receive. Several studies have shed light on these disparities, with findings indicating that girls tend to receive less education due to factors such as gender bias or gender preference (as documented by \cite{biswas2000gender} and \cite{ota2007within}). Additionally, birth order can play a role in the educational opportunities afforded to children, with elder siblings benefiting or facing disadvantages, as observed in studies like \cite{ota2007within}, \cite{weng2019family}, \cite{fergusson2006birth}, \cite{de2010birth}, \cite{moshoeshoe2016birth}, and \cite{esposito2020importance}. Furthermore, children with higher abilities are more likely to receive increased educational opportunities and educational attainment, as suggested by the research of \cite{becker1976child}, \cite{dizon2019parents}, and \cite{giannola2023parental}.
However, there exists a notable gap in the research landscape, as there are few studies that comprehensively analyze all these various sources of disparities within the same analytical framework. Such a framework, capable of simultaneously examining gender-based differences, birth order effects, and the impact of individual unobserved abilities on educational inequality within-households, holds the potential for the analysis of the effectiveness of educational counterfactual scenarios in reducing inequality. Specifically, it enables us to explore how these factors interact with one another to shape inequality. Furthermore, this comprehensive approach is crucial for estimating the proportion of inequality attributable to gender and birth order and the portion caused by other unobserved factors such as the cognitive ability of a child compared to their siblings. The former allows us to design educational policies which target inequality due to gender or birth order effects.

This micro-level analysis of educational disparities is particularly relevant to the broader context of educational challenges in Sub-Saharan Africa, where lack of education remains a critical obstacle to development. Despite numerous reforms, the region continues to struggle with high illiteracy rates, significant educational inequality, and have a substantial proportion of the world’s out-of-school children (\cite{unesco2021}). While primary school completion rates are approaching or exceeding 90\% in most regions globally, Sub-Saharan Africa lags behind, with only two out of three children completing primary school (\cite{unesco2021}). Educational disparities are the results of various barriers, which can be categorized into three types: situational (life circumstances), dispositional (personal attitudes), and institutional (structural conditions) (\cite{unesco2021}). A crucial but often overlooked factor contributing to these disparities is the within-family inequality in children's educational attainment, which accounts for approximately 40\% of the variation in educational attainment in the developing world (\cite{giannola2023parental}).
 
Using this regional evidence of educational inequality as a foundation, this paper focuses on the specific context of Benin. Benin is an ideal location for this study for three main reasons: the non enforcement of compulsory education laws, the significant variability in educational attainment among individuals within the same household, and the observed disparities in educational opportunities based on gender and birth order. For this exercise, I focused on households where there are two adult children residing in the same household as their parents\footnote{The extension to households with more than 2 children is straightforward. All the estimates and analysis in the paper are also done after including households with 3 children and the results are presented in the appendix of the paper.}. Among those households there are differences in how much educational resources parents have to distribute and the education attainment of the head of household. Taking that into account, I perform my analysis on households with different observed characteristics separately.

In the first part of the paper, I establish two key stylized facts about intra-household educational inequality. First, there is a non-linear, hump-shaped relationship between household-level mean and variance of educational attainment. Second, I decompose the average within-household educational attainment disparities into gender, birth order, and unobserved residual effects, finding that among households with non-educated heads and one child of each gender, over two-thirds of the average disparities in children's educational attainment is due to gender and birth order, while among college-educated parents, only one-third is due to these factors. Furthermore, average inequality, measured by the range of children's educational attainment is twice as high among non-educated parents compared to college educated parents.
I then propose a structural model of household educational resources allocation to rationalize the observed inequality with budget constraints and parents' preferences. The model accounts for gender effects by introducing a difference in the marginal utility of education between sons and daughters, with the marginal utility being higher for sons when birth order and other unobserved factors are held constant. Additionally, birth order effects are integrated through variations in the marginal cost of education by birth order, holding gender and unobserved factors fixed. Each child's unique unobserved characteristics that influence their educational outcomes are modeled through an unobserved weight on the benefit derived from that child’s education. I estimate the model using the Simulated Method of Moments, and analyze diverse counterfactual scenarios in how they affect average inequality and the share of gender and birth order effects in the average inequality among non-educated parents. The first counterfactual (1) which theoretically set gender and birth order effects to zero; did not lead to a significant reduction in the average inequality in the sample. The second counterfactual (2) which theoretically removed barriers to school entry for all children reduces the average inequality in the sample by about half, but does not reduce the share of gender effect. The third counterfactual (3) which is more practical, relaxes the financial burden of non-educated parents to the same level as their college-educated counterparts. This only leads to a 10\% reduction in the average inequality in the sample, with more than 50\% of the remaining inequality due to gender effects.

The subsequent sections of this paper are structured as follows. In Section 2, I present an overview of the data used for this study, while Section 3 presents key empirical evidence and stylized facts derived from the data description. Section 4 is dedicated for the model's setup, outlining the estimation strategy for key parameters, and describing the inference and estimation procedures employed in this study. Lastly, in Section 5, I present counterfactual analysis to further explore the implications of my findings, and Section 7 concludes the paper.




\subsection*{Other relevant literature}

This paper contributes to the literature on within-household schooling decision, particularly factors influencing parents' distribution decision of education resources among siblings. A key determinant of these distribution decisions is the gender of the child and the gender composition of the household. Previous research has shown that daughters are less likely to receive some education; or have lower educational attainment on average. Studies have shown that, while the presence of elder sisters tends to increase the likelihood of schooling, the presence of younger brothers may decrease it (\cite{biswas2000gender}, \cite{ota2007within}, \cite{ombati2012gender}, \cite{osadan2014gender}, \cite{psaki2018measuring}). Another influential factor is the birth order of children, with mixed findings in previous studies. Some papers suggest a positive effect of birth order on children's education (\cite{ota2007within}, \cite{weng2019family}), while others have shown that later-born children have lower educational attainment (\cite{fergusson2006birth}, \cite{de2010birth}, \cite{moshoeshoe2016birth}, \cite{esposito2020importance}). Finally, a child's innate ability or talent plays a role in parental distribution decisions. Studies have demonstrated that parents invest more in the human capital of high-ability children and allocate more nonhuman capital to low-ability children (\cite{becker1976child}, \cite{dizon2019parents},  \cite{giannola2023parental}). When parents are compelled to invest in the nonhuman capital--- for example inheritance in form of land or financial assets--- of low-ability children, this leads to an inefficient equilibrium, where the investment in the human capital of high-ability children is not optimized (\cite{nerlove1984investment}). This paper adds to this existing literature in two significant ways. First, it examines a context where parents are not constrained to compensate lower ability children by investing in their nonhuman capital but, instead, rely on family taxes (\cite{wantchekon2015education}). Second, this paper proposes a household educational resources distribution model which allows for a more flexible analysis of the distribution of education resources within the household. In this model, the assumption of equal distribution is relaxed, enabling a detailed exploration of the interactions between gender and, birth order effects, and the innate abilities of children in influencing household distribution decisions.

This paper also contributes to the literature on educational Kuznets curve theory ( \cite{londono1990kuznetsian} and \cite{ram1990educational}, \cite{thomas2003measuring}, \cite{morrisson2013kuznets}). Previous studies have analyzed the relationship between the mean and variance of education using cross country data or within-country time series data. This paper contributes to that literature by analyzing the relationship between the mean and variance of education using within-country cross household data. Specifically, it shows that in Benin at the household level; the relationship mean-variance of education is inverted U shaped with the peak occurring at approximately 7 years of education.. 

Finally, this paper contributes to the literature on within-household inequality in children's human capital (\cite{giannola2023parental}). \cite{giannola2023parental} has shown in the context of India that observed inequality within-households is partly explained by parents investing more in the human capital of high-achieving children, especially when they are financially constrained. This behavior stems from the fact that parents are not particularly averse to inequality and tend to reinforce the gap in learning created by innate ability rather than correcting it.
This paper contributes to that literature by first building upon the result that parents unequally invest in the human capital of high-achieving children in contexts where the education system is better tailored to serve high-achieving students. Second, this paper interacts with that result and examines how it relates to other sources of inequality, such as gender and birth order.




\section{Data Description and Definition of Key Variables}

\subsection{Sample and Data}

In this section, I present the data used. I use data from the 2013 Population and Habitation Census of Benin. This census data provides information on all households and their members. It is conducted by the National Bureau of Statistics of Benin, and has information at both the household and individual levels. For the purpose of this paper, the focus is directed towards individuals who identify themselves as the children of household heads, enabling to get information on parental \footnote{Parents here refers to one of the parents, either the mother or the father. This because it is not possible to have both for household with single parents and to identify the biological mother for polygamous households.} and sibling characteristics for a sub-sample of siblings. 
The variable ``Number of children" represents the observed number of children within each household \footnote{It does not include children who moved out of the family house before the census.}. For the primary analysis, only households with children aged between 25 and 40 years are included. This age range is chosen to ensure that the children have either completed their education or nearly achieved maximum educational attainment. The summary statistics of all individuals between 25 and 40 years living in the same household as their parents is presented in Table \ref{table:desc_stats} in Appendix A.

The inequality analysis focuses on households with at least 2 such children falling within the specified age range and at least one child with some educational attainment. This specific condition on the sample is motivated by the goal to examine the reasons for providing equal education to all children, as opposed to the alternative of not educating any children. In particular households with only non-educated children do not offer any information about the distribution of education resources which this paper aims to analyze. The resulting sample comprises approximately $90,000$ individuals and $ \approx 33,000$ households, serving as the basis for further investigation. The sample description is as follows:

\begin{enumerate}   
    \item \textbf{Sample 0:} All households with children between 25-40 yrs old ($\approx 160,000$ households)
    \item \textbf{Sample 1:} Households in sample 0 with at least 2 children between 25 and 40 years old. ($\approx$ 51,600 households) 
     \item \textbf{Sample 2:} Households in sample 1 with at least one educated child between 25 and 40 years old. ($\approx$ 32,800 households)
\end{enumerate}

I use sample 1 for stylized facts, estimation of the model, counterfactual analysis, and  comparative statics, and sample 2 only for counterfactual analysis, and  comparative statics.

\subsection{Key variables and measurement}
The data set contains several key variables used in this paper, including gender, age, religion, area of residence, family size, household wealth index, and educational attainment of individuals, as well as their parents' and a subset of their siblings' variables. Apart from these variables, I also created measures for within-household inequality, within-household average years of educational attainment of children, and gender composition of children within a household. A description of each variable and their measurement is as follows:

\textbf{Within household inequality:} Is the disparity in the educational attainment of children within a given household. It is measured by the within-household range of children's education attainment for households with 2 children. For households with more than 2 children, it is measured by the standard deviation of education.


\textbf{Number of children:} It is the total number of people who identify as children of a the head of household. This variable is denoted by $N_c$.

\textbf{Within-household average years of educational attainment of children:} Is the average education of children between 25 and 40 years for a given household. It serves as a metric for accessing the average quality of children within the household, and is used as proxy for educational resources available. A related variable is the \textbf{Within-household total years of educational attainment of children}, which is the simple sum of children's years of education. It is used as a proxy for the household's total investment in education. The within-household total and average years of educational attainment  of children are denoted by $q_T$ and $\Bar{q}$ respectively.

\subsection{Descriptive statistics}

Table \ref{table:decriptive} displays the descriptive statistics for the key variables. First, among the observed offspring\footnote{Observed offspring refers to adult children living in the same household as their parents.} in sample 2, 38\% are female, their average education level is 8 years, with 80\% having completed at least one year of education. Second, 40\% of heads of household have at least one year of education. 

\begin{table}[H]
    \begin{center}
    \caption{Descriptive Statistics}
    
{
\begin{tabular}{@{\extracolsep{5pt}}lcccc} 
 \noalign{\global\arrayrulewidth=1.2pt}
\\[-1.8ex]\hline 
 \noalign{\global\arrayrulewidth=1.2pt}
\textbf{Statistic} & \multicolumn{1}{c}{\textbf{Mean}} & \multicolumn{1}{c}{\textbf{St. Dev.}} & \multicolumn{1}{c}{\textbf{Min}} & \multicolumn{1}{c}{\textbf{Max}} \\ 
  \noalign{\global\arrayrulewidth=1.2pt}
\hline \\[-1.8ex] 
Age & 29.452 & 3.993 & 25 & 40 \\ 
Female &  0.380 & 0.485 & 0 & 1 \\ 
Years of education & 7.760 & 5.802 & 0 & 21 \\ 
At least one years of education  & 0.776 & 0.417 & 0 & 1 \\ 
Range of children's education  & 6.821 & 4.873 & 0 & 21 \\ 
Standard deviation of children's education  & 3.874 & 2.744 & 0.000 & 14.142 \\ 
Educated head of household  & 0.390 & 0.488 & 0 & 1 \\ 
Number of children between 25 and 40  & 3.055 & 1.557 & 2 & 16 \\ 
Number of children &6.340 & 4.354 & 2 & 79 \\ 
$\Bar{q}$ & 7.760 & 4.444 & 0.143 & 20.250 \\ 
$q_T$ & 22.354 & 15.660 & 1 & 148 \\

\textbf{Number of observation} & \multicolumn{4}{c}{89,594}\\
\hline \\[-1.8ex] 
 \noalign{\global\arrayrulewidth=1.2pt}

\end{tabular} 
}
    \label{table:decriptive}
    \end{center}
\end{table}

\begin{figure}[H]
\begin{center}
 \begin{minipage}[b]{0.45\linewidth}
   \includegraphics [width=8.5cm]{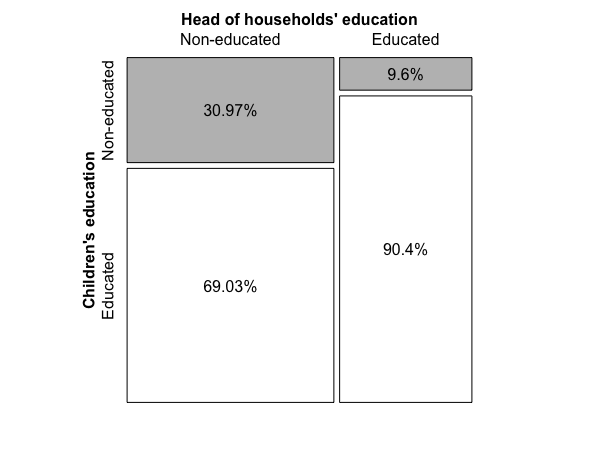}
   (a) Children's education as function of parents' education
 \end{minipage}
 \quad
 \begin{minipage}[b]{0.45\linewidth}
    \includegraphics [width=8.5cm]{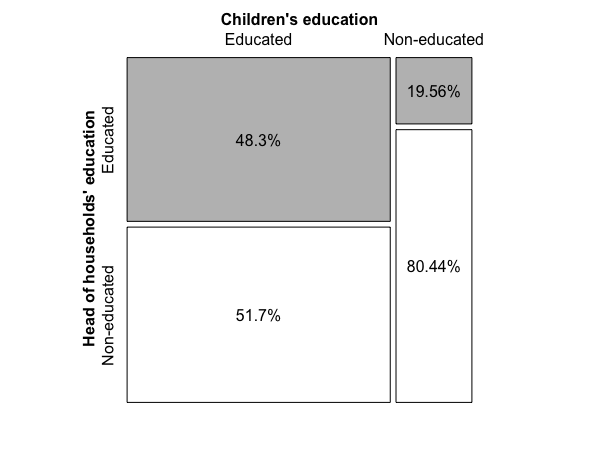}
     (b) Parents' education as function of children's education
\end{minipage}
  \caption{Parents and children's education.}
 \label{parents_children_educ} 
\end{center}
\end{figure}

About 80\% of children without any schooling have parents who also lack formal education, whereas this percentage decreases to 50\% for children with schooling. Conversely, approximately 31\% of parents without schooling have children who likewise lack schooling, compared to only $\approx$ 10\% for parents with schooling (See Figure \ref{parents_children_educ}). These statistics provide suggestive evidence of both inter-generational educational mobility \footnote{Children are more educated than their parents} and inter-generational educational persistence\footnote{Children's education is correlated with their parents' education.}. Third, regarding within-family inequality, the average within-household range in children's education is about 7 years with a maximum of 21 years of education. In addition the standard deviation of the within-family range of children's education is $\approx 4$ years of education, signaling high variability in the with-household inequality across households.

\section{Empirical Evidence}

\textit{\textbf{Empirical Evidence 1:} $2/3$ of the variation in educational attainment in the sample arises from disparities within households. }

I compute the average within-household variation in educational attainment and compare that to the overall variation in educational attainment in the sample. Furthermore, I examine how within and between-household variances in education are related. Let $q_{h} = (q_{h,1}, q_{h,2}, \dots, q_{h,N_{c_h}})$ be the vector of adult children's educational attainment in household $h$ with $N_{c_h}$ adult children, and let $q = (q_1, \dots, q_n)$ be the educational attainment of adult children in the sample. 
\[ Var(q) = Var \Big[\mathbb{E}[q|h]\Big] + \mathbb{E} \Big[Var[q|h]\Big]. \] 
The variance of $q$ is the sum of the average within-household variance \big($\mathbb{E} \big[Var[q|h]\big]$\big) and between household variation $\big(Var \big[\mathbb{E}[q|h]\big]\big)$  in $q$. The estimates of these quantities in my sample are the following: \[ \widehat{\mathbb{E}} \Big[s(q|h)\Big]  = 22.63 \text{ and } s(q) = 33.66,\] where $s(q|h)$ is the sample variance of adult children's educational attainment $q_h$ in household $h$, $s(q)$ is the sample variance of adult children's educational attainment $q$ in the whole sample, and $\widehat{\mathbb{E}}$ is the sample average. This indicates that $2/3$ of the variation in $q$ arises from variation within-households. Furthermore, in the absence of within-household inequality \footnote{When we consider households with disparities in their children's educational attainment.}, the estimate of the between-household variance of children's educational attainment is 20.9. However, in the presence of within-household inequality, the estimate of the between-household variance of children's educational attainment is 11.2. These statistics suggest that, on average, households with some degree of within-household inequality exhibit lower between-household inequality compared to households with no within-household inequality. In conclusion, the analysis highlights on one hand the substantial contribution of within-household inequality to the overall inequality in educational attainment. On the other hand, no within-household variation in educational attainment of children is associated with higher between households variance. 

\textit{\textbf{Empirical Evidence 2:} Within-household disparities in children's educational attainment is heterogeneous across households.}

\begin{figure}[H]
\begin{center}
 \begin{minipage}[b]{0.45\linewidth}
   \includegraphics [width=8cm]{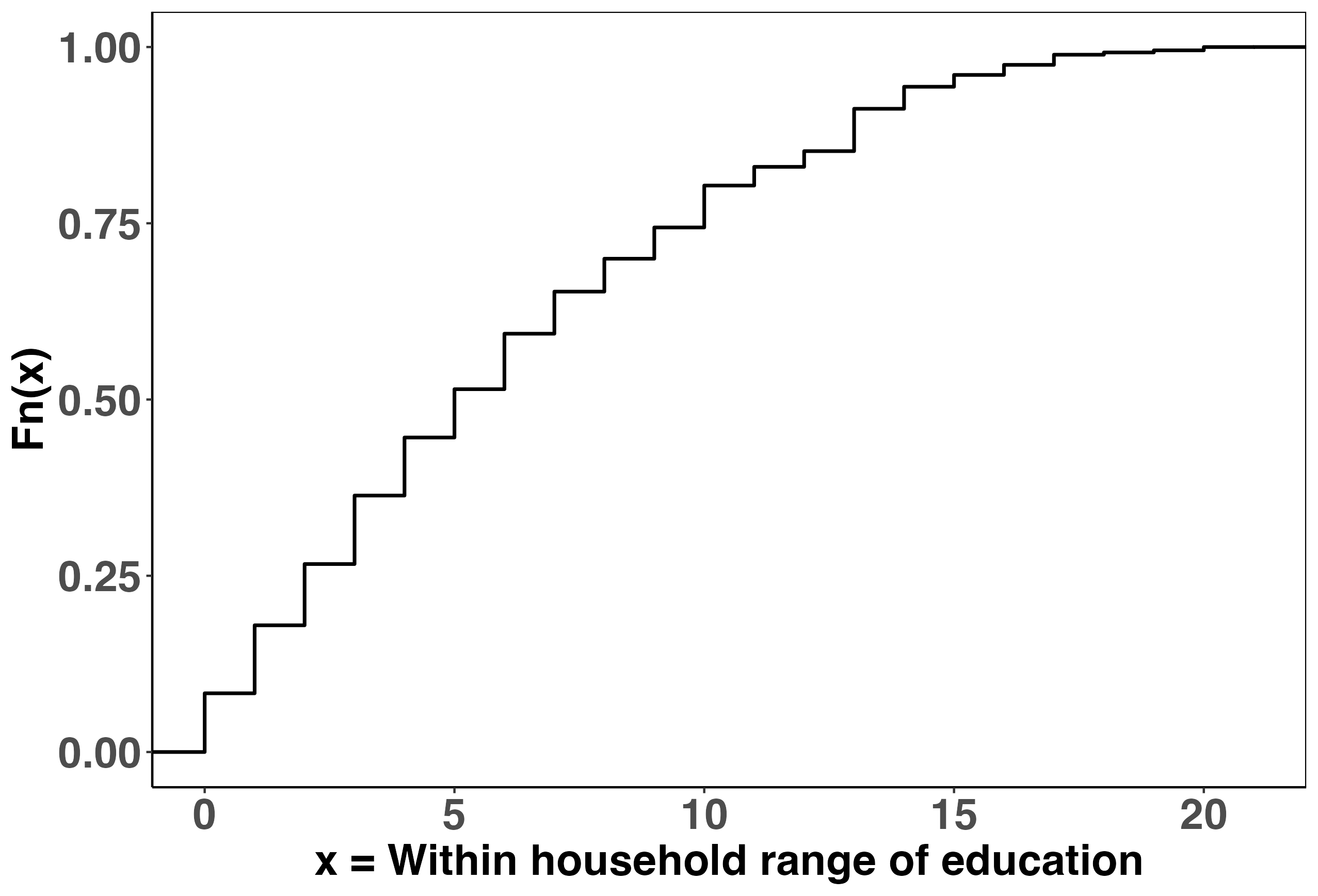}
    (a) Range
 \end{minipage}
 \quad
 \begin{minipage}[b]{0.45\linewidth}
    \includegraphics [width=8cm]{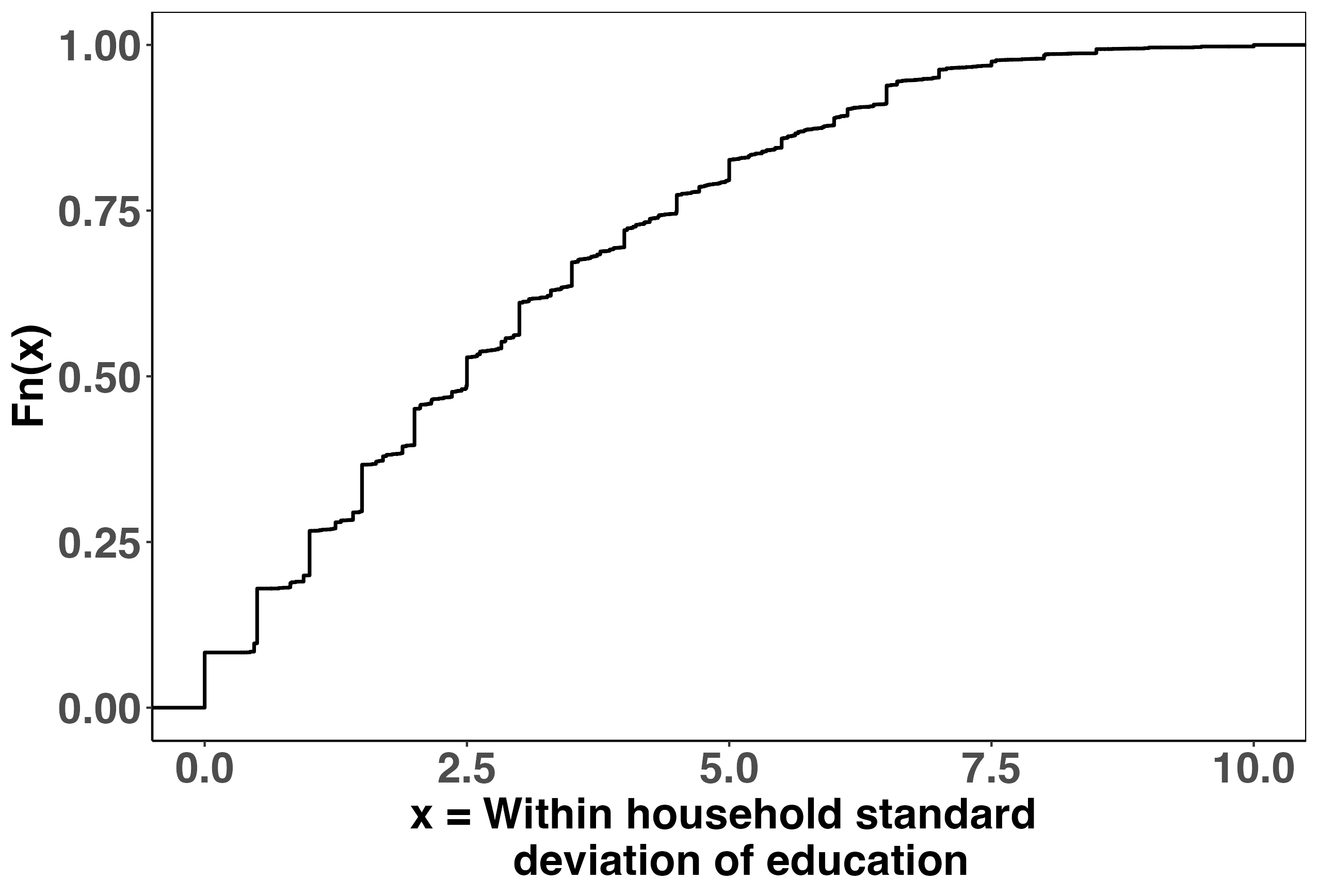}
     (b) Standard Deviation 
\end{minipage}
  \caption{Empirical cdf of within-household range and standard deviation of education attainment of children.}
 \label{range}
 \end{center}
\end{figure}

This empirical evidence focuses on the extent of variation in within-household inequality across households. Understanding these differences can provide valuable insights into the factors that contribute to within-household inequality and the potential mechanisms that can be employed to reduce it. 
Figure \ref{range} depicts the empirical distribution of the within-household range and standard deviation of the educational attainment of adult children. This figure reveals that the magnitude of inequality varies across households, with some household having all of their children with the same education attainment while some have at least a child with some college education and at least a child with no formal education.

 These findings highlight the importance of considering household-level dynamics when addressing educational attainment inequality and suggest that interventions aiming at reducing disparities in education must be tailored to the unique circumstances of each household. 
The inequality is present even within gender, although in lower magnitude. The average within household variance of daughters' educational attainment  (resp. sons) is 4.97 (resp. 8.77).

\begin{figure}[H]
\begin{center}
 \begin{minipage}[b]{0.45\linewidth}
   \includegraphics [width=8cm]{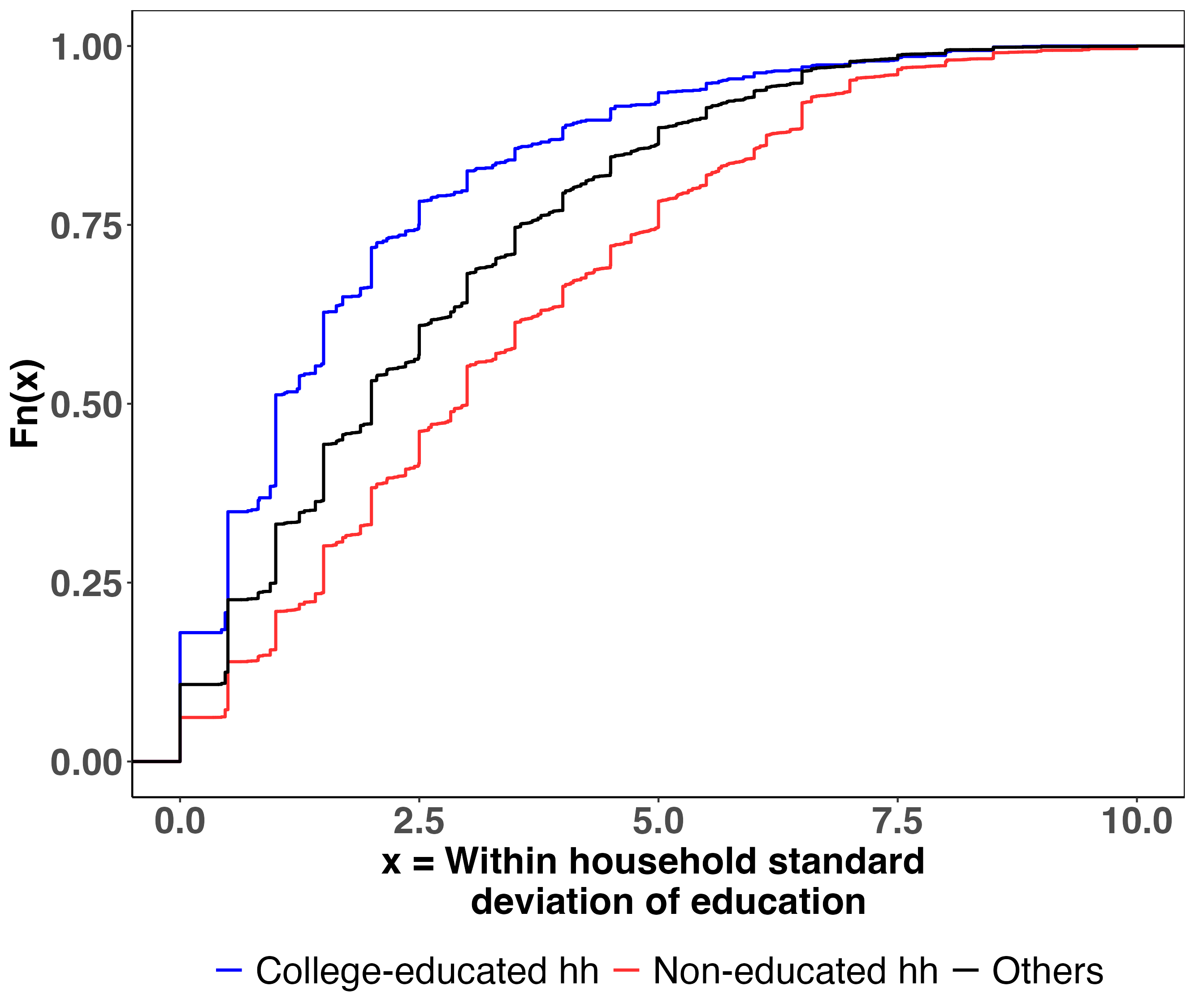}
    (a) Parents Education
 \end{minipage}
 \quad
 \begin{minipage}[b]{0.45\linewidth}
    \includegraphics [width=8cm]{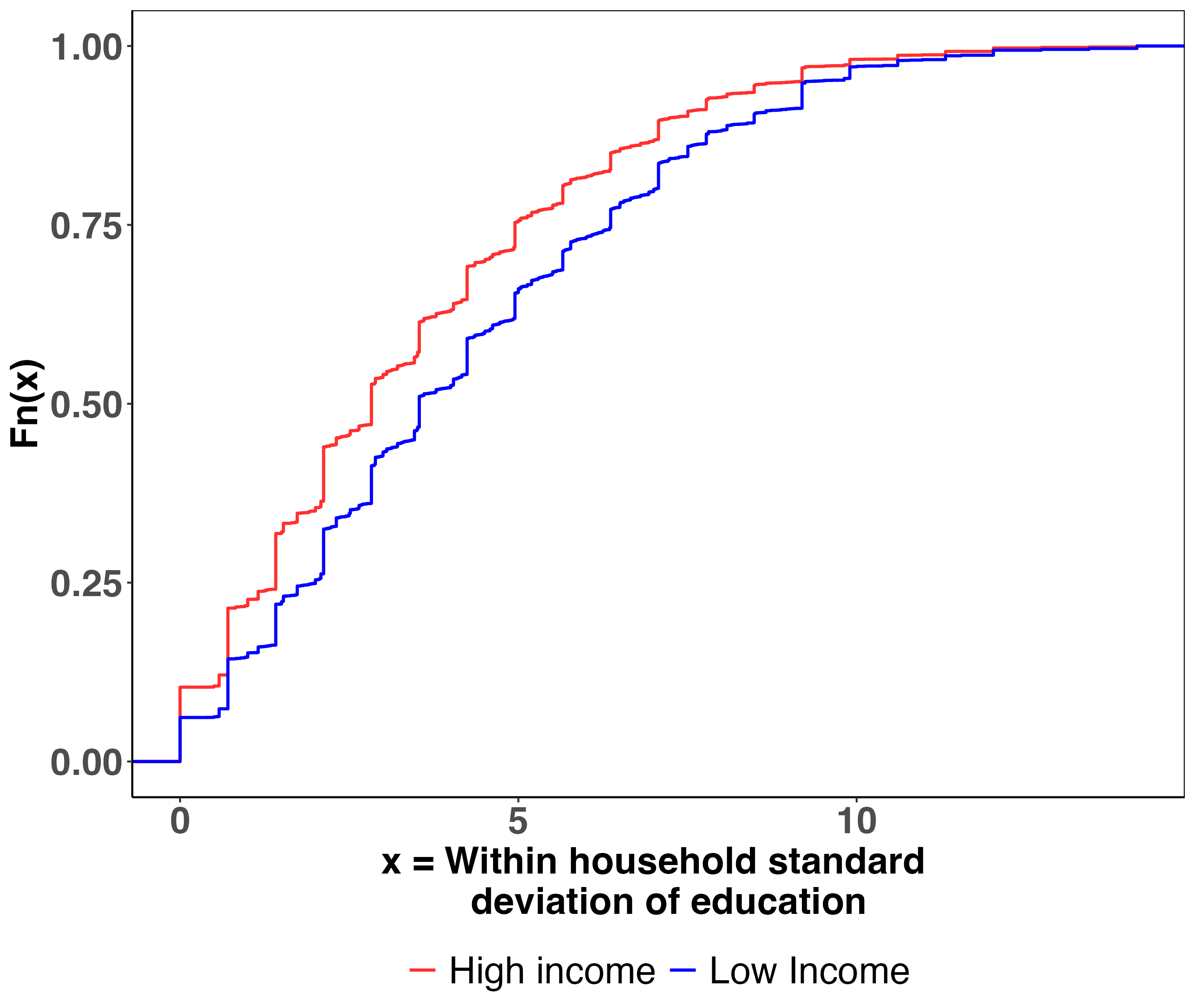}
     (b) Household Wealth Index (HWI) 
\end{minipage}
  \caption{Distribution of inequality by socio-economic groups.}
 \label{het_range}
 \end{center}
\end{figure}

The within-household variance of children's education is non-zero on average for all level of parents' education and household wealth (see Figure \ref{het_range}). However, it appears that within-household variance of adult children's educational attainment decreases with parents' education level and household wealth. Specifically, we observe a first order stochastic dominance between the empirical cdf of within-household inequality in children's educational attainment of college educated (resp. high wealth index) and non-college educated (resp. low wealth index) parents.

\textit{\textbf{Empirical Evidence 3:} At the household level, a negative association emerges between the maximum educational attainment among adult children within the household and the proportion of adult children within that household who have achieved that maximum education level.}

Consider the OLS regression of the within-household maximum years of educational attainment of adult children on the proportion of children with educational attainment equal to that maximum. 
\begin{equation}
    q^{\max}_h = \beta_0 + \beta_1 \frac{1}{N_c} \sum_{i = 1}^{N_c} 1\{ q_i = q^{\max}_h\} + \gamma'X_h + \varepsilon_h,
\end{equation}
where $X_h$ include number of children, HWI, area of residence, religion, gender composition of children, and head of household' s education. $q_h^{\max}$ is the maximum educational attainment of children in household $h$, $\frac{1}{N_c}\sum_{i = 1}^{N_c} 1\{ q_i = q^{\max}_h\}$, is the proportion of children with that maximum educational attainment within the household.

 The estimation results in column (3)- (4) of Table \ref{table:table1} indicate that, on average, households with a 0.5 higher proportion of children attaining the maximum years of education within the household tend to have around 1.3 years lower maximum educational attainment for the children within the household.

 \begin{table}[H]
    \begin{center}
    \caption{Regression of within- household maximum years of education on within-household inequality and of within-household standard deviation of children's education on households' characteristics}
    {
\begin{tabular}{l c c c c c c c}
\\[-1.8ex]\hline 
\hline \\[-1.8ex] 
& \multicolumn{4}{l}{\textbf{Maximum years of education}}  & \multicolumn{3}{l}{\textbf{Standard deviation}} \\
 & (1) & (2) & (3) & (4) & (5) & (6) & (7)\\
\hline
(Intercept)                          & $9.40^{*}$      & $6.16^{*}$      & $11.94^{*}$       & $10.24^{*}$       & $4.24^{*}$        & $1.46^{*}$        & $0.87^{*}$        \\
Standard deviation                                  & $0.50^{*}$      & $0.66^{*}$      &                   &                   &                   &                   &                   \\
  $\frac{1}{N_c} \sum_{i = 1}^{N_c} 1\{ q_i = q_{\max}\}$                                &                 &                 & $\textcolor{red}{-1.35^{*}}$       & $\textcolor{red}{-2.64^{*}}$       &                   &                   &                   \\
hh Educ = Primary                 &                 & $1.05^{*}$      &                   & $0.71^{*}$        & $\textcolor{blue}{-0.77^{*}}$       & $-0.62^{*}$       & $\textcolor{blue}{0.02}$            \\
hh Educ = Junior HS                 &                 & $2.62^{*}$      &                   & $2.14^{*}$        & $\textcolor{blue}{-1.17^{*}}$       & $-0.66^{*}$       & $\textcolor{blue}{0.86^{*}}$        \\
hh Educ = Senior HS                &                 & $4.24^{*}$      &                   & $3.56^{*}$        & $\textcolor{blue}{-1.59^{*}}$       & $-0.47^{*}$       & $\textcolor{blue}{2.98^{*}}$        \\
hh Educ = College                &                 & $5.74^{*}$      &                   & $4.93^{*}$        & $\textcolor{blue}{-1.94^{*}}$       & $0.24^{*}$        & $\textcolor{blue}{4.75^{*}}$        \\

Average years of education  ($\Bar{q}$)                           &                 &                 &                   &                   &                   & $0.91^{*}$        & $1.16^{*}$        \\
$\Bar{q}^2$                      &                 &                 &                   &                   &                   & $-0.06^{*}$       & $-0.07^{*}$       \\
 hh Educ = Primary:$\Bar{q}$      &                 &                 &                   &                   &                   &                   & $-0.29^{*}$       \\
hh Educ = Junior HS:$\Bar{q}$      &                 &                 &                   &                   &                   &                   & $-0.56^{*}$       \\
hh Educ = Senior HS:$\Bar{q}$      &                 &                 &                   &                   &                   &                   & $-0.94^{*}$       \\
hh Educ = College:$\Bar{q}$      &                 &                 &                   &                   &                   &                   & $-1.14^{*}$       \\
 hh Educ = Primary:$\Bar{q}^2$ &                 &                 &                   &                   &                   &                   & $0.02^{*}$        \\
hh Educ = Junior HS:$\Bar{q}^2$ &                 &                 &                   &                   &                   &                   & $0.04^{*}$        \\
hh Educ = Senior HS:$\Bar{q}^2$ &                 &                 &                   &                   &                   &                   & $0.05^{*}$        \\
hh Educ = College:$\Bar{q}^2$ &                 &                 &                   &                   &                   &                   & $0.06^{*}$        \\
Number of children                                   &                 & $0.04^{*}$      &                   & $0.05^{*}$        &                   & $0.04^{*}$        & $0.05^{*}$        \\
HWI                                  &                 & $0.45^{*}$      &                   & $0.38^{*}$        &                   & $-0.10^{*}$       & $-0.10^{*}$       \\
Urban                                &                 & $0.89^{*}$      &                   & $0.71^{*}$        &                   & $-0.32^{*}$       & $-0.32^{*}$       \\
Christian                            &                 & $0.90^{*}$      &                   & $0.74^{*}$        &                   & $-0.28^{*}$       & $-0.27^{*}$       \\
Both gender                       &                 & $0.31^{*}$      &                   & $0.42^{*}$        &                   & $0.20^{*}$        & $0.22^{*}$        \\

\hline
R$^2$                                & $0.10$          & $0.31$          & $0.00$            & $0.16$            & $0.04$            & $0.22$            & $0.24$            \\
Num. obs.                            & $32729$         & $32729$         & $32729$           & $32729$           & $32729$           & $32729$           & $32729$           \\
\hline
\multicolumn{8}{l}{\scriptsize{$^*$ Null hypothesis value outside the confidence interval.}}
\end{tabular}
}
    \label{table:table1}
    \end{center}
\end{table}

 Additionally, an OLS regression of the within-household maximum years of children's educational attainment on the within-household standard deviation of children's educational attainment indicates that households characterized by higher levels of educational inequality demonstrate, on average, higher within-household maximum educational attainment (see column (1)- (2) of Table \ref{table:table1}). These findings suggest a trade-off involved in households' educational decision. 
The same argument as \cite{becker1973interaction} applies here, i.e. on one hand an increase in quality \footnote{Here quality refer to the within-household maximum years of education of children} is more expensive if there are more children with that quality. On the other hand, an increase in quantity\footnote{Quantity refers to the number of children with the within-household maximum years of education of children} is more expensive if children are of high quality. This trade-off is a direct effect of the limited education resources available to households. In conclusion, due to financial constraints within the household, parents are facing a trade-off between reducing inequality within the household or reducing inequality between them and other households.


\textit{\textbf{Empirical Evidence 4:} The relationship between household-level mean and standard deviation of children's education is inverted U-shaped.}

The level of education attained by the head of a household has been found to be a significant factor associated with the level of inequality in children's educational attainment within that household. In particular, an increase in the head of household's education level is associated with a decrease in inequality. However, it remains unclear whether this is a direct result of high educated parents' aversion for inequality or an indirect result of their preference for education. To shed light on this issue, this section will investigate the factors that contribute to the observed negative correlation between parents' educational attainment and within-household inequality. 

\begin{figure}[H]
\begin{center}
 \begin{minipage}[b]{0.45\linewidth}
   \includegraphics [width=7.5cm]{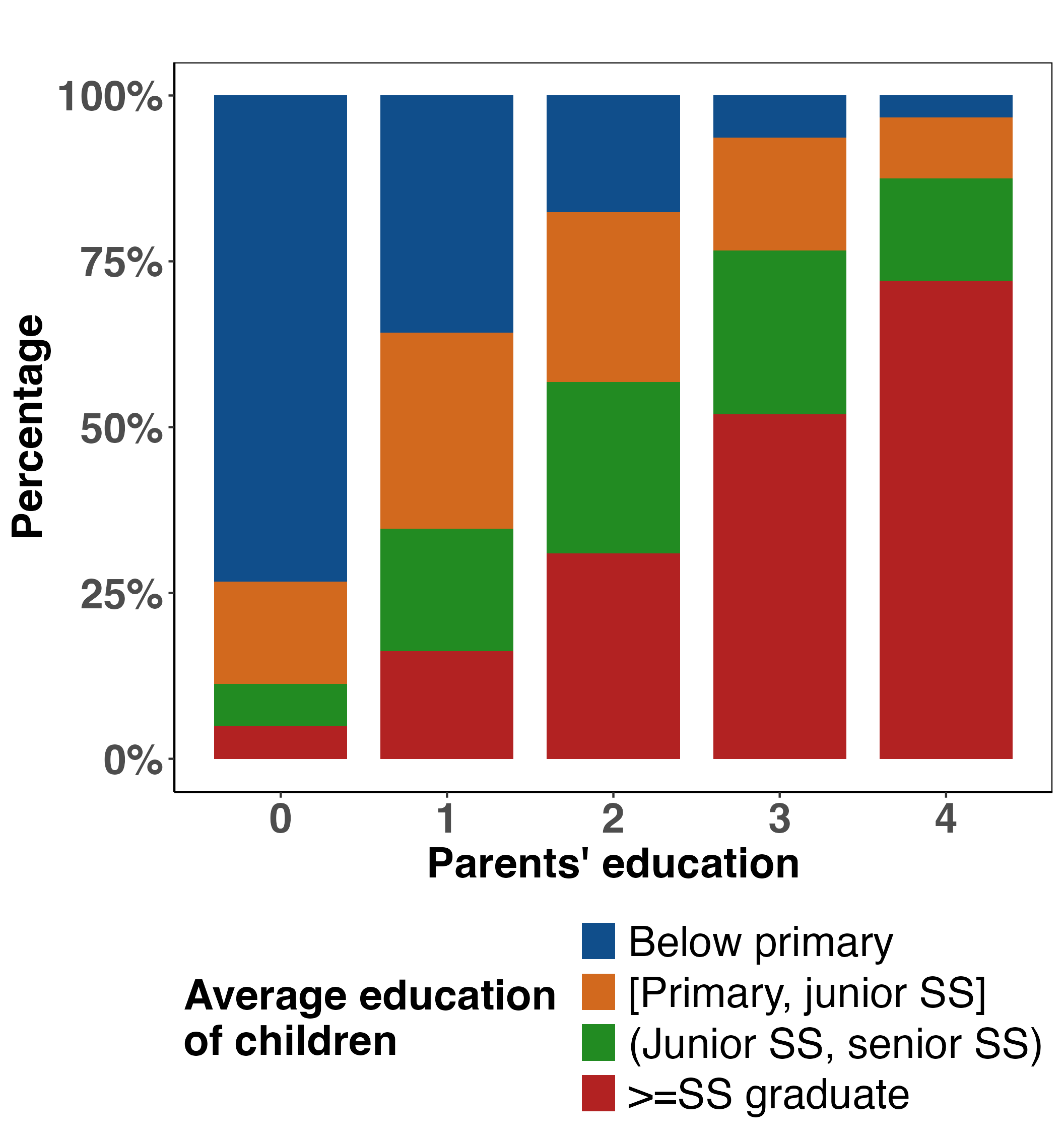}
 \end{minipage}
  \begin{minipage}[b]{0.45\linewidth}
   \includegraphics [width=7.5cm]{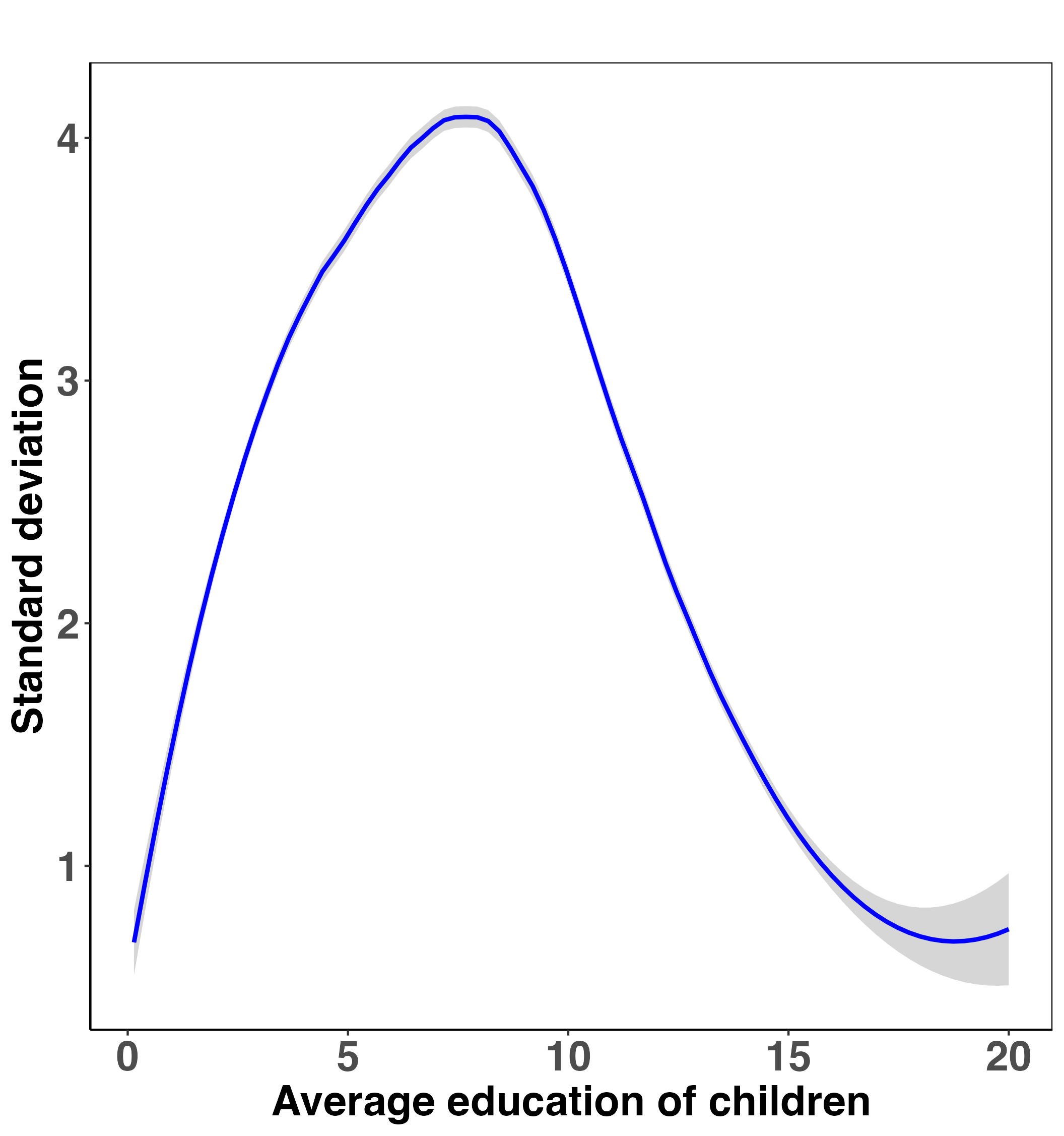}
 \end{minipage}
  \caption{Distribution of average education attainment of children.}
 \label{het_q_T}
  \end{center}
\end{figure}

In addition to having lower level of inequality, households with more educated head of household also tend to have higher average years of education for their children (See Panel (a) of Figure \ref{het_q_T}). This observation is particularly interesting given the hump-shaped relationship between inequality and average educational attainment of children (See Panel (b) of Figure \ref{het_q_T}). 
This inverted U-shaped relationship between average and standard deviation of children's education is consistent with the educational Kuznets curve theory (\cite{thomas2003measuring}). According to the Kuznets curve theory with education distribution, as we move from zero to maximum level of education, the variance first increases and then decreases. This is empirically shown for a set of developing countries in \cite{londono1990kuznetsian} and \cite{ram1990educational}. To investigate this relationship further, I estimate an OLS regression model of within-household inequality on average education of children, and parents' level of education, with a quadratic interaction between between this two variables. 
\begin{equation}
    \text{Inequality}_h = \alpha + \beta_1 \Bar{q}_h + \beta_2 \Bar{q}^2_h + \beta_3 \text{hh\_Educ}_h + \beta_4 \Bar{q}_h \text{hh\_Educ}_h + \beta_5 \Bar{q}^2_h \text{hh\_Educ}_h + \gamma' X_h + \varepsilon_h,
\end{equation}
where $X_h$ include number of children, HWI, area of residence, religion and gender composition. $\text{Inequality}_h$ is the standard deviation of children's educational attainment in household $h$, $\Bar{q}_h$ is the average education of children in household $h$, and $\text{hh\_Educ}_h$ is the education of the head of household $h$. The estimation results in column (5)- (7) of Table \ref{table:table1} suggest that the negative dependence between parents' education and within-household inequality is a result of both variable being correlated with the within-household average educational attainment of children. In particular, the positive correlation between parents' education and the within-household average educational attainment of children combined with the hump shaped relation between within-household inequality and the within-household average education of children is translated into the observed spurious negative relationship between parents' education and within-household inequality.

\textit{\textbf{Empirical Evidence 5:} Daughters with brothers and firstborn children receive on average less education compared to other children.}

In the preceding sections, I have presented evidence at the household level, revealing that various factors contribute to the heterogeneity observed in the level of educational inequality across households. Notably, factors such as budget constraints, total investment in education, and parents' education play significant roles.
In this section, the focus is on exploring the observed characteristics of children who received less education compared to their siblings. The examination of these characteristics is essential for developing effective strategies to address inequality and promote equality of opportunity for all children.

\begin{figure}[H]
\begin{center}
   \includegraphics [width=13cm]{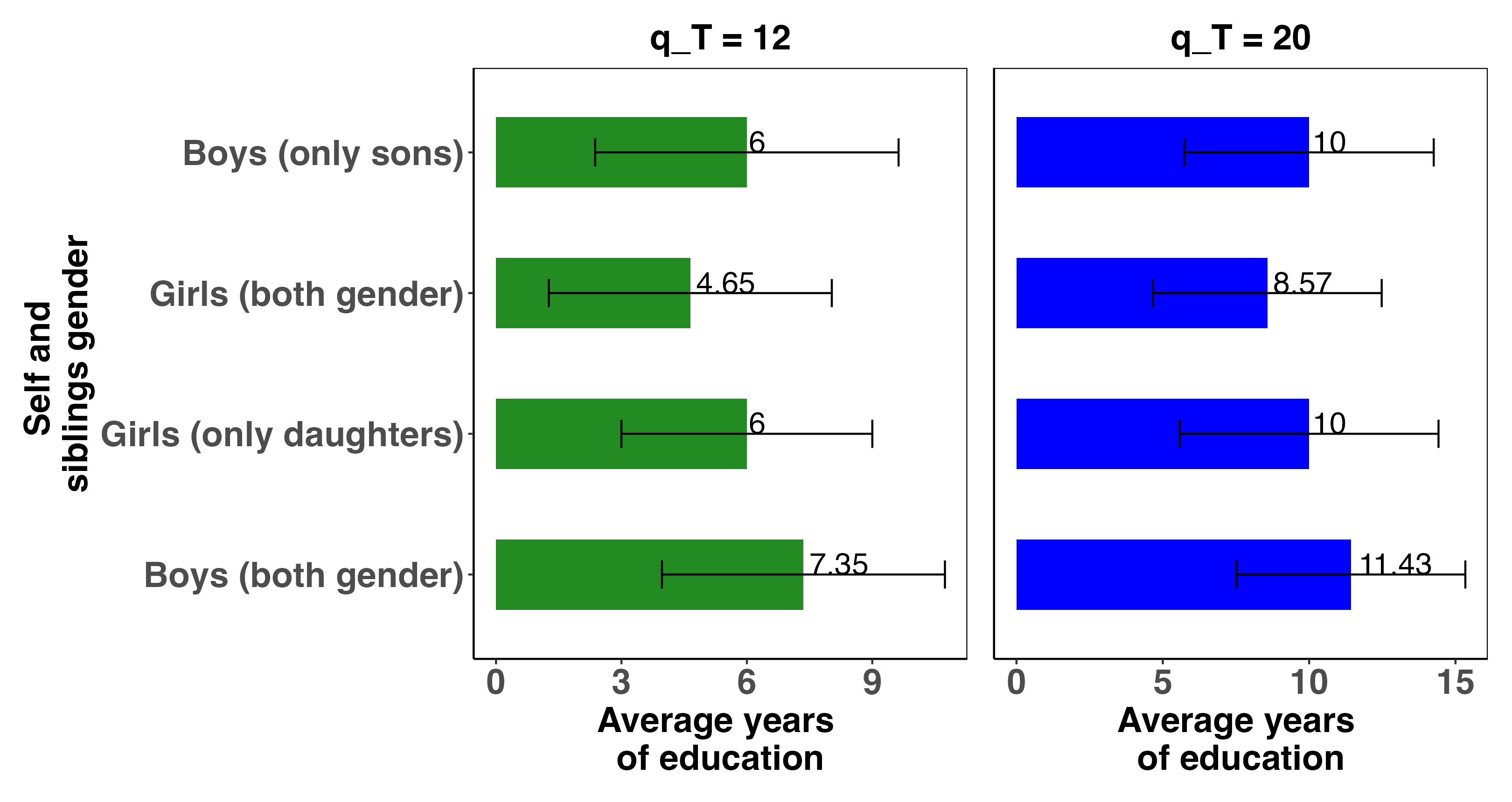}
\caption{Average years of education by gender and households gender composition (for $N_c = 2$ and total within household educational attainment equal 12 and 20)}
 \label{ineq_gender}
  \end{center}
\end{figure}

Figure \ref{ineq_gender} graphs the average years of education based on the gender of children and the gender composition of households. To ensure accurate comparisons, the graph holds the within-household average educational attainment of children constant. In the first panel, the analysis centers around households that are only able to finance primary school education for all their children. In the second panel, households that can only afford to provide education up to junior high school level are considered. The figure reveals that, girls from only-daughter households, on average, have the same level of education as the household average, while boys from only-son households have similar education levels as well. However, in both-gender households, girls' average education is lower than the household average, whereas boys' average education is higher. These findings suggest that there is discrimination against daughters when it comes to the allocation of education resources, when the alternative of giving more to a son is available.

\begin{figure}[H]
\begin{center}
   \includegraphics [width=13cm]{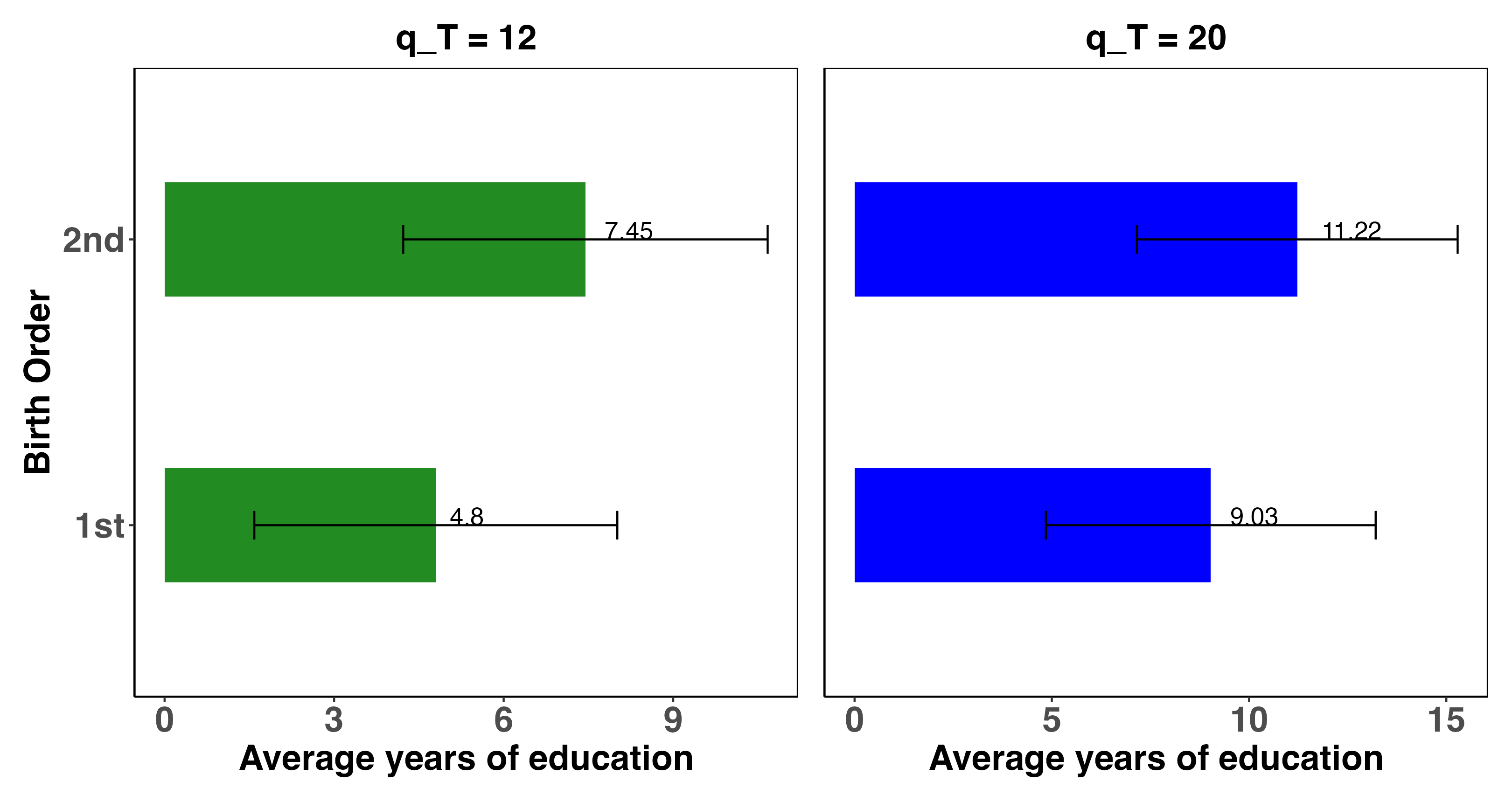}
\caption{Average years of education by birth order (for $N_c = 2$ and total within household educational attainment equal 12 and 20)}
 \label{ineq_birth_order}
  \end{center}
\end{figure}

Figure \ref{ineq_birth_order} allows similar analysis in terms of children's birth order after holding fix the number of children, and the within-household average years of educational attainment of children.

 In panel (a) of Figure \ref{ineq_birth_order}, the plot is for households that can afford to educate all their children up to primary education, and for households that can afford to educate all their children up to junior HS education is in panel (b). The figure demonstrates that the average years of education for the firstborn children is below the household average for both type of households, whereas the average years of education of the second-born children is above the household average. This monotonic increase in educational attainment by birth order applies to any family size (See Appendix B). The findings of Figure \ref{ineq_birth_order} suggest that there is disadvantage in birth order regarding the allocation of educational resources.

\begin{figure}[H]
\begin{center}
 \begin{minipage}[b]{0.45\linewidth}
   \includegraphics [width=8cm]{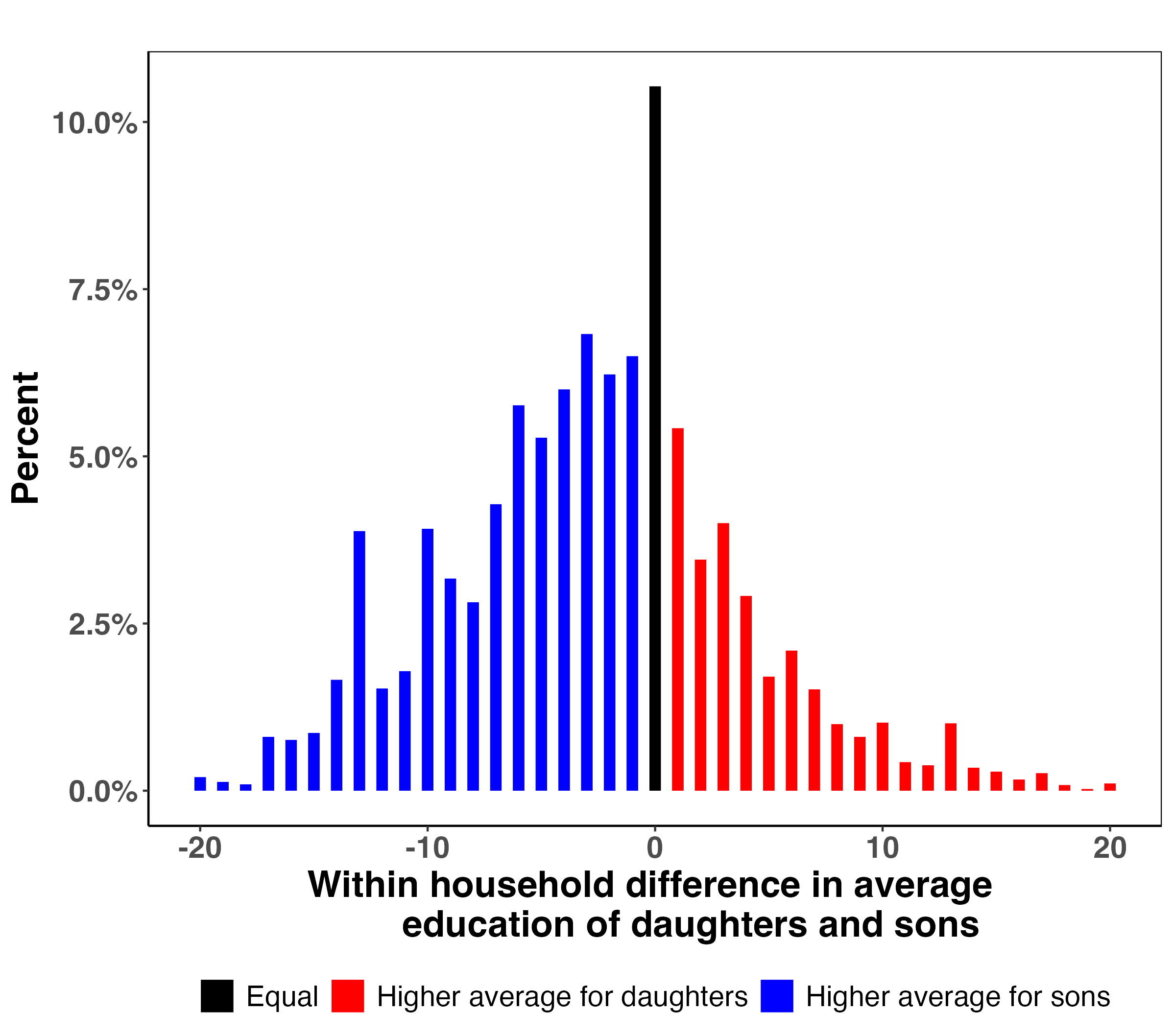}
   
   (a) Between daughters and sons ($N_c = 2$)
 \end{minipage}
 \quad
 \begin{minipage}[b]{0.45\linewidth}
     \includegraphics [width=8cm]{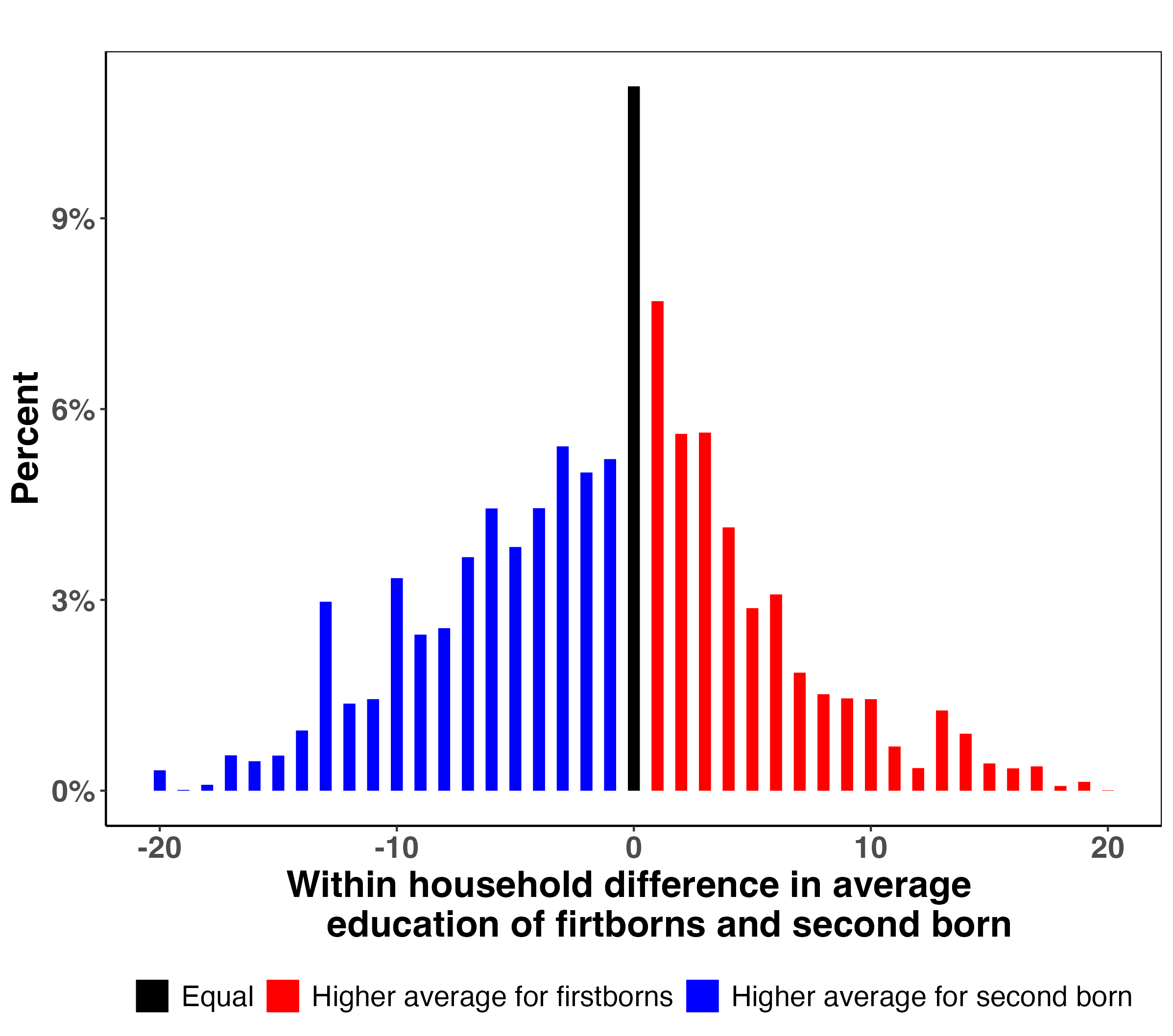}
     
     (b) Between 1st and 2nd born ($N_c = 2$)
\end{minipage}
\caption{Histogram of within-household difference in average education (Benin, 2013)}
 \label{hist_diff_educ}
  \end{center}
\end{figure}

In summary, a child's gender, the gender of their siblings, and their birth order are key determinants of the years of education they receive. Despite taking into account observed household and children characteristics, a significant amount of variation in inequality across households remains unexplained, as evidenced by the $R^2$ value obtained from the regression of within-household standard deviation of children's education on those observed household and children characteristics (See columns (5)-(7) of Table \ref{table:table1}). 
In addition, despite the presence of gender disadvantage against daughters in average educational attainment, it appears that in some households, daughters receive higher education than their brothers (see Figure \ref{hist_diff_educ}). This suggest an heterogeneity in gender discrimination  across households. A modeling assumption which I adopt in this paper stipulates that the unexplained difference in inequality can be attributed to the variance in children's innate abilities, which differs across households. In other words, the fact that some daughters receive higher education compared to their brothers despite gender disadvantages can be attributed to high ability draws by these girls. This is a significant aspect of the household's education distribution model, which I present in the next section.

\textit{\textbf{Empirical Evidence 6-1:} Average within-household inequality in children's education is negatively related to parents' education. Among households with non-educated heads and one child of each gender, over two-thirds of the average inequality is due to gender and birth order, while among college-educated parents, only one-third is due to these factors.}

In the previous section, I have presented some empirical evidence about the observed characteristics of children which explain the within-household inequality in their educational attainment. In this section, I will provide a decomposition of the average within-households inequality, categorizing it into components associated with gender disparity, birth order effects, and variations in children's unobserved abilities (or any unobserved factors affecting educational resources distribution). The decomposition is conducted across various within-household average educational levels on one hand and parents' education level on the other hand. I used a household fixed-effect regression approach to achieve this breakdown.

\paragraph{Regression with Household Fixed Effects\\}

To decompose the average within-household inequality into components categorized as gender and birth order effects and unobserved differences, I consider the following regressions: 
\begin{equation}
\label{fe_reg_b}
    \text{Educ}_{i, h} = \beta_1 \text{Female}_{i, h} + \beta_2 \text{Firstborn}_{i, h} + \beta_3 \text{Female}_{i, h} \times \text{Firstborn}_{i, h} + \nu_h + \varepsilon_{i, h}
\end{equation}
\begin{equation}
\label{fe_reg_s}
    \text{Educ}_{i, h} =  \beta_1 \text{Firstborn}_{i, h} + \nu_h + \varepsilon_{i, h}
\end{equation}

where $\text{Educ}_{i, h}$ is the years of education of child $i$ in household $h$, $\text{Female}_{i,h}$ is a gender indicator variable equal to 1 if child $i$ in household $h$ is a daughter, $\text{Firstborn}_{i, h}$ is a birth order indicator variable equal to 1 if child $i$ in household $h$ is a firstborn, and $\nu_h$ is the household fixed effect. Equation \ref{fe_reg_b} is for households with both sons and daughters, while equation \ref{fe_reg_s} is for households with either only sons or only daughters. 

The estimates from equation \ref{fe_reg_b}  and \ref{fe_reg_s} are presented in Table \ref{table:reg_bias_fe} by average education of children and in Table \ref{table:bias_reg_fe_hh_educ} by parents' education. The results suggest, on one hand, that about $63\%$ of the observed within-household inequality in children's education is due to gender and birth order effects for households with both son and daughter. On the other hand, for households with only daughters or only sons, about 33\% of the observed inequality is due to birth order effects. This change suggests that part of the unobserved sources of inequality is muted by gender disadvantage.

For the primary analysis, which focuses on households with just two adult children living at home, the reliability of the estimates shown in Tables \ref{table:reg_bias_fe} and \ref{table:bias_reg_fe_hh_educ} may be compromised. This unreliability stems from the incidental parameter problem, a consequence of having only two data points per household for the fixed effect regressions. To validate the initial findings, I use the following alternative regression for a more robust examination.
\begin{equation}
\label{diff}
    \Delta_{\text{daughter-son}} \text{Educ}_{h} = \beta_0 + \beta_1 \text{Firstborn\_daughter}_h + \varepsilon_h,
\end{equation}
where $\Delta_{\text{daughter-son}} \text{Educ}_{h}$ is the average difference in the education of sons and daughters in household $h$, $\text{Firstborn\_daughter}_h$ is an indicator variable equals to 1 if the firstborn in household $h$ is a daughter.

\begin{table}[H]
    \begin{center}
    \caption{Regression of children's education on their gender and birth order with household fixed effect by within-household total years of children's education $(N_c = 2)$}
    {
\scalebox{0.75}{
\begin{tabular}{l c c c c c c}
\hline
  & (1) & (2) & (3)  & (4) & (5) & (6) \\
  & \multicolumn{2}{c}{$q_T = 12$} & \multicolumn{2}{c}{$q_T = 20$} & \multicolumn{2}{c}{All $q_T$} \\
\hline
Female     & $-3.03^{*}$       &                   & $-2.75^{*}$       &                   & $-2.46^{*}$       &                   \\
First born         & $-3.24^{*}$       & $-1.90^{*}$       & $-2.59^{*}$       & $-2.61^{*}$       & $-0.95^{*}$       & $-1.24^{*}$       \\
Firstborn female   & $1.26^{*}$        &                   & $0.38$            &                   & $-0.27^{*}$       &                   \\
\hline
R$^2$      & $0.21$            & $0.09$            & $0.11$            & $0.09$            & $0.67$            & $0.70$            \\
Adj. R$^2$ & $-0.59$           & $-0.82$           & $-0.79$           & $-0.83$           & $0.34$            & $0.40$            \\
Num. obs.  & $1632$            & $300$             & $1558$            & $278$             & $43970$           & $7562$            \\
RMSE       & $4.39$            & $3.91$            & $5.71$            & $5.76$            & $4.52$            & $4.23$            \\
Household fixed effects      & \checkmark           & \checkmark            & \checkmark  & \checkmark           & \checkmark            & \checkmark          \\
Average inequality (Both gender: Firstborn female)      & \multicolumn{2}{c}{6.23}           & \multicolumn{2}{c}{7.22}           & \multicolumn{2}{c}{5.84} \\
Average inequality (Only daughters)      & \multicolumn{2}{c}{3.81}           & \multicolumn{2}{c}{6.01}           & \multicolumn{2}{c}{4.59} \\
Average inequality (Only sons)      & \multicolumn{2}{c}{5.07}           & \multicolumn{2}{c}{6.09}           & \multicolumn{2}{c}{5.08} \\
\hline
& \multicolumn{6}{l}{Explained proportion} \\
Gender & 50.1\% & - & 38.1\% & - & 33.7\% & - \\
Birth order  & 30.5\% & 49.9\% & 35.9\% & 43.4\% & 29.3\% & 32.9\% \\
Unexplained  & 19.6\% & 50.1\% & 26\% & 56.6\% & 37\% & 67.1\% \\
\hline
\multicolumn{7}{l}{\scriptsize{$^*$ Null hypothesis value outside the confidence interval.}}
\end{tabular}
}
}
    \label{table:reg_bias_fe}
    \end{center}
\scriptsize{Note: Columns (2), (4), and (6) are for households with only daughters. For households with only sons the decomposition is 19\% birth order + 81\% ability.}
\end{table}

\begin{table}[H]
    \begin{center}
    \caption{Regression of children's education on their gender and birth order with household fixed effect by parents' education $(N_c = 2)$}
    {
\scalebox{0.75}{
\begin{tabular}{l c c c c c c}
\hline
  & (1) & (2) & (3)  & (4) & (5) & (6) \\
  & \multicolumn{2}{c}{Non-educated} & \multicolumn{2}{c}{College educated} & \multicolumn{2}{c}{All} \\
    & \multicolumn{2}{c}{parents} & \multicolumn{2}{c}{parents} & \multicolumn{2}{c}{} \\
\hline
Female     & $-3.16^{*}$       &                   & $-0.90^{*}$       &                  & $-2.47^{*}$       &                   \\
Firstborn         & $-1.19^{*}$       & $-1.55^{*}$       & $-0.41$           & $-0.13$          & $-0.93^{*}$       & $-1.24^{*}$       \\
Firstborn female  & $-0.39^{*}$       &                   & $0.25$            &                  & $-0.34^{*}$       &                   \\
\hline
R$^2$      & $0.58$            & $0.60$            & $0.68$            & $0.69$           & $0.67$            & $0.70$            \\
Num. obs.  & $22540$           & $3528$            & $1884$            & $478$            & $40884$           & $6956$            \\
Household fixed effects      & \checkmark           & \checkmark            & \checkmark  & \checkmark           & \checkmark            & \checkmark          \\
Average inequality (Both gender: Firstborn female)      & \multicolumn{2}{c}{6.76}           & \multicolumn{2}{c}{3.27}           & \multicolumn{2}{c}{5.84} \\
Average inequality (Only daughters)      & \multicolumn{2}{c}{5.29}           & \multicolumn{2}{c}{3.16}           & \multicolumn{2}{c}{4.59} \\
Average inequality (Only sons)      & \multicolumn{2}{c}{5.71}           & \multicolumn{2}{c}{2.78}           & \multicolumn{2}{c}{5.08} \\
\hline
& \multicolumn{6}{l}{Explained proportion} \\
Gender & 47.2\% & - & 36\% & - & 33.7\% & - \\
Birth order  & 23\% & 29.3\% & 4\% & 4.1\% & 29.3\% & 32.9\% \\
Unexplained  & 29.8\% & 70.7\% & 60\% & 95.9\% & 37\% & 67.1\% \\
\hline
\multicolumn{7}{l}{\scriptsize{$^*$ Null hypothesis value outside the confidence interval.}}
\end{tabular}}
}

    \label{table:bias_reg_fe_hh_educ}
    \end{center}
     \scriptsize{Note: Columns (2), (4), and (6) are for households with only daughters. For households with only sons the decomposition is respectively 17\% birth order + 83\% ability for college educated parents and 21\% birth order + 79\% ability for non-educated parents. For the whole sample it is 18\% birth order + 82\% ability.}
\end{table}

\begin{figure}[H]
\begin{center}
 \begin{minipage}[b]{0.45\linewidth}
   \includegraphics [width=8cm]{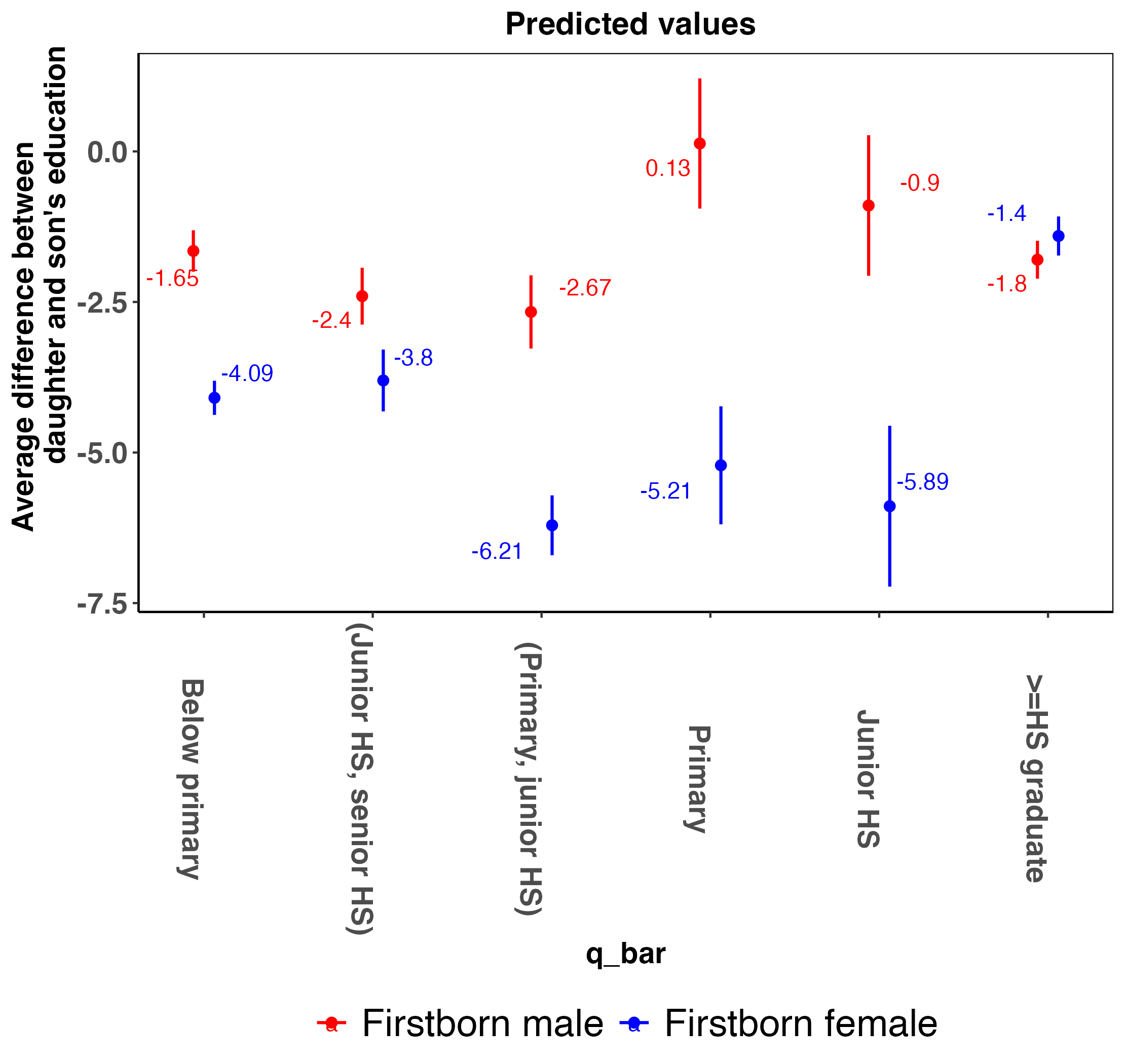}
   
   (a) As function of within-household average education of children
 \end{minipage}
 \quad
 \begin{minipage}[b]{0.45\linewidth}
     \includegraphics [width=8cm]{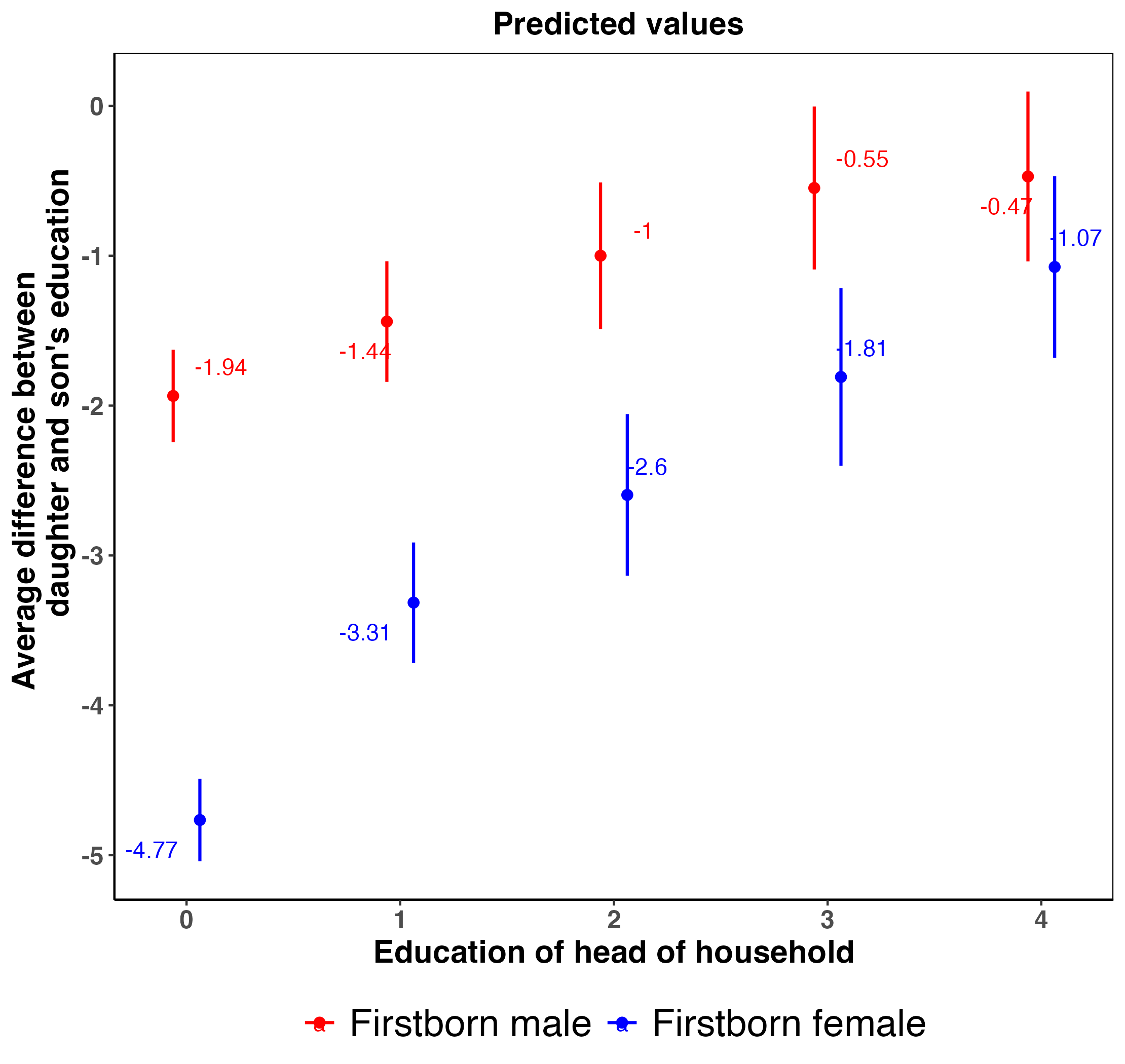}
     
     (b) As function of head of household's education
\end{minipage}
    \caption{Effect of gender and birth order disadvantages on within-household inequality  ($N_c =2$)}
     \label{disadv_effect}
      \end{center}
\end{figure}

The estimates are summarized in Figure \ref{disadv_effect}. Figure \ref{disadv_effect} illustrates the mean disparity in educational attainment between daughters and sons, for households with a firstborn son and a firstborn daughter separately. These measurements are provided across various average educational levels of the children in the panel a) and across education of the head of household in panel b), and are use to decompose the average absolute difference in children's educational attainment \footnote{Let $q_h = (q_{1, h}, q_{2, h})$, and $\text{Range}_h = \max(q_h) - \min(q_h) = |q_{1,h} - q_{2,h}|$. } by household's observable characteristics as follows:
\begin{equation}
\label{blue}
  \beta_0 + \beta_1 =  \text{ average effect of gender } + \text{ average effect of birth order},  \text{ and} 
\end{equation} 
\begin{equation}
\label{red}
   \beta_0 = \text{ average effect of gender } - \text{  average effect of birth order},   
\end{equation}

\begin{figure}[H]
\begin{center}
 \begin{minipage}[b]{0.45\linewidth}
   \includegraphics [width=8cm]{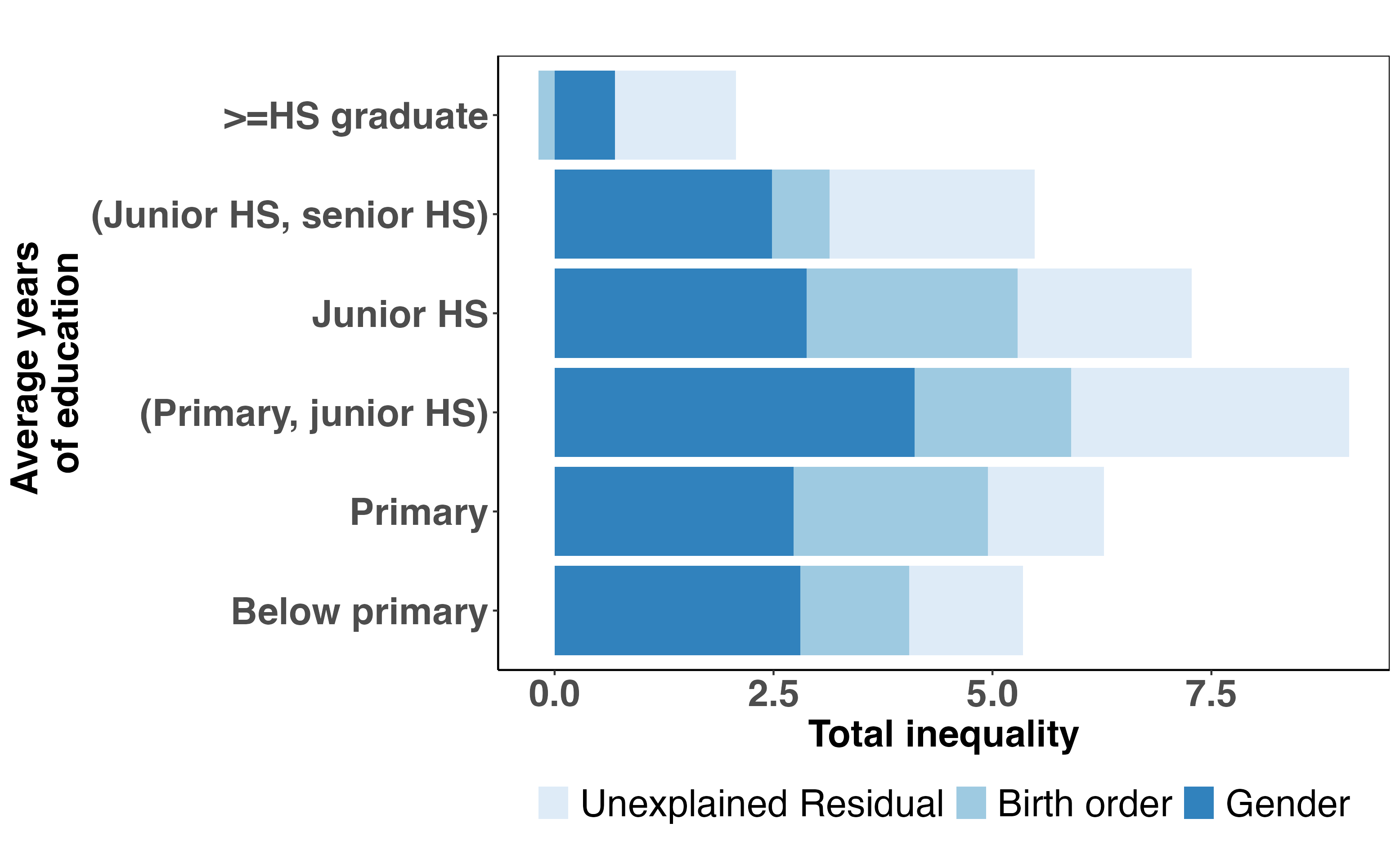}
   
   (a) As function of within-household average education of children
 \end{minipage}
 \quad
 \begin{minipage}[b]{0.45\linewidth}
     \includegraphics [width=8cm]{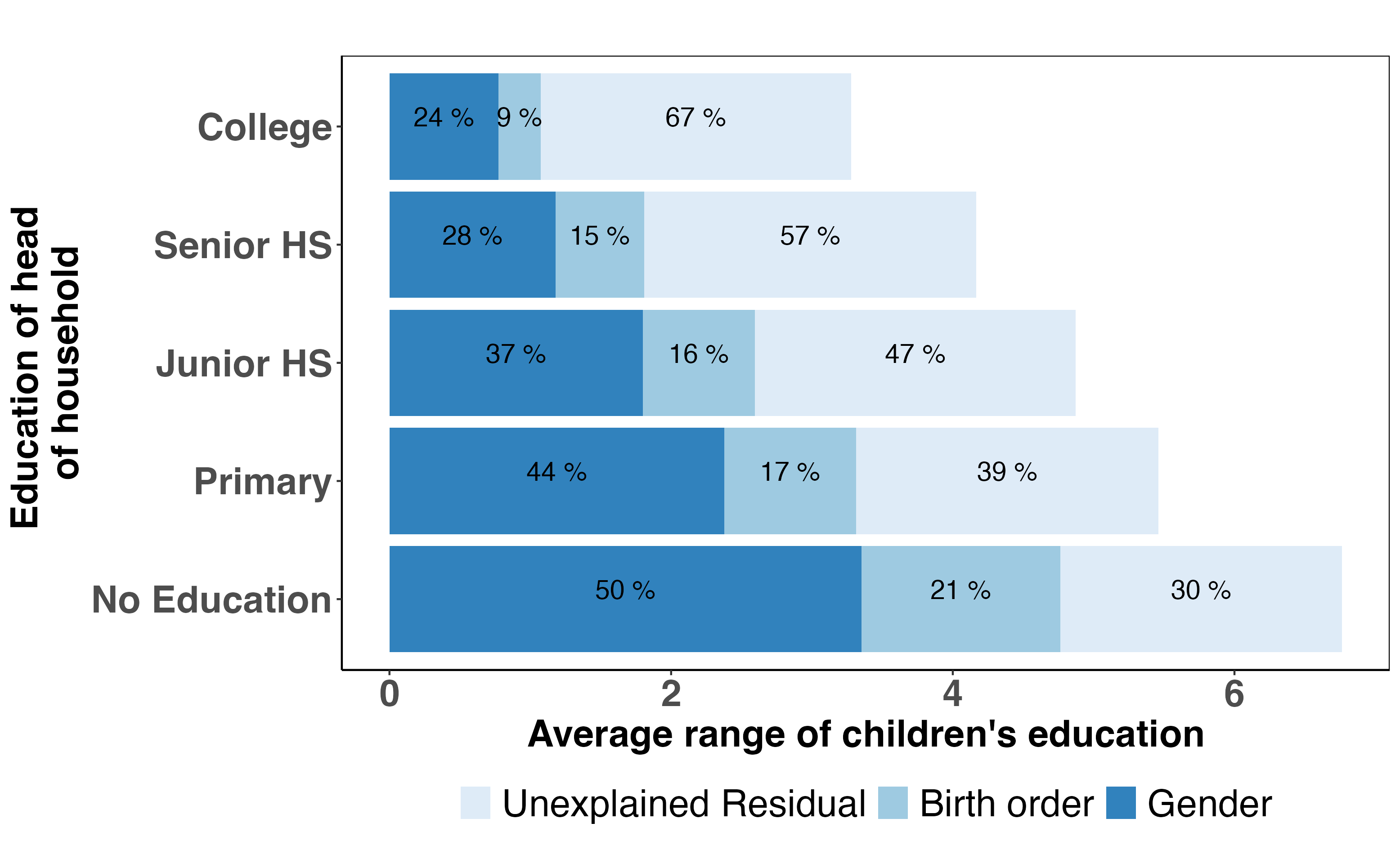}
     
     (b) As function of head of household's education
\end{minipage}
    \caption{Inequality decomposition ($N_c =2$)}
     \label{decomposition}
      \end{center}
\end{figure}

I use  equations \ref{blue} and \ref{red} to get the average effect of gender and the average effect of birth order on within-family disparities in educational attainment. Note that in households with firstborn daughters, the same child is affected by both gender and birth order disadvantages. Given that $|\beta_0 + \beta_1|$ is the average effects of gender and birth order on educational attainment disparities within families, $ E[\text{Range}| \text{household has a firstborn daughter}]   - |\beta_0 + \beta_1|$ is the unexplained residual\footnote{ $E[\text{Range}| \text{household has a firstborn daughter} ] \geq |\beta_0 + \beta_1|$}. I use these calculations to break down the average inequality found within-households into the three factors illustrated in Figure \ref{decomposition}. Figure \ref{decomposition} displays how the average inequality is divided among gender effect, birth order effect, and differences in unobserved factors. It reveals that gender disadvantage is the predominant factor contributing to inequality. As the average educational level of children increases, the influence of unobserved ability differences becomes more significant, while the impact of birth order diminishes. Similarly, as parents' education level increases, total inequality is smaller on average, and the shares of gender and birth order disadvantages reduce.  This indicates not only that there is variability in the degree of average inequality in children's educational attainment across different levels of children's average education and parents' education but also in the way it is broken down.

\textit{\textbf{Empirical Evidence 6-2:} Intra-household educational inequality is present both at extensive and intensive margin. Compared to the extensive margins, the unexplained component has higher share in the average inequality for the intensive margin. The decrease in inequality by parents' education is mostly present in the extensive margin.}

It is relevant to analysis how within-household inequality in education is decomposed for the extensive margin compared to the intensive margin. To analyze that, I run the previous fixed effect regression in equations \ref{fe_reg_b} and \ref{fe_reg_s} for households with only educated children--- for the intensive margin analysis---, and the following regression for households with at least one non-educated child--- for the extensive margin analysis.
\begin{equation}
\label{fe_reg_ext_b}
    1\{Educ_{i, h} > 0\} = \beta_1 Female_{i, h} + \beta_2 Firstborn_{i, h} + \beta_3 Female_{i, h} \times Firstborn_{i, h} + \nu_h + \varepsilon_{i, h}
\end{equation}

\begin{figure}[H]
 \begin{minipage}[b]{0.45\linewidth}
   \includegraphics [width=7cm]{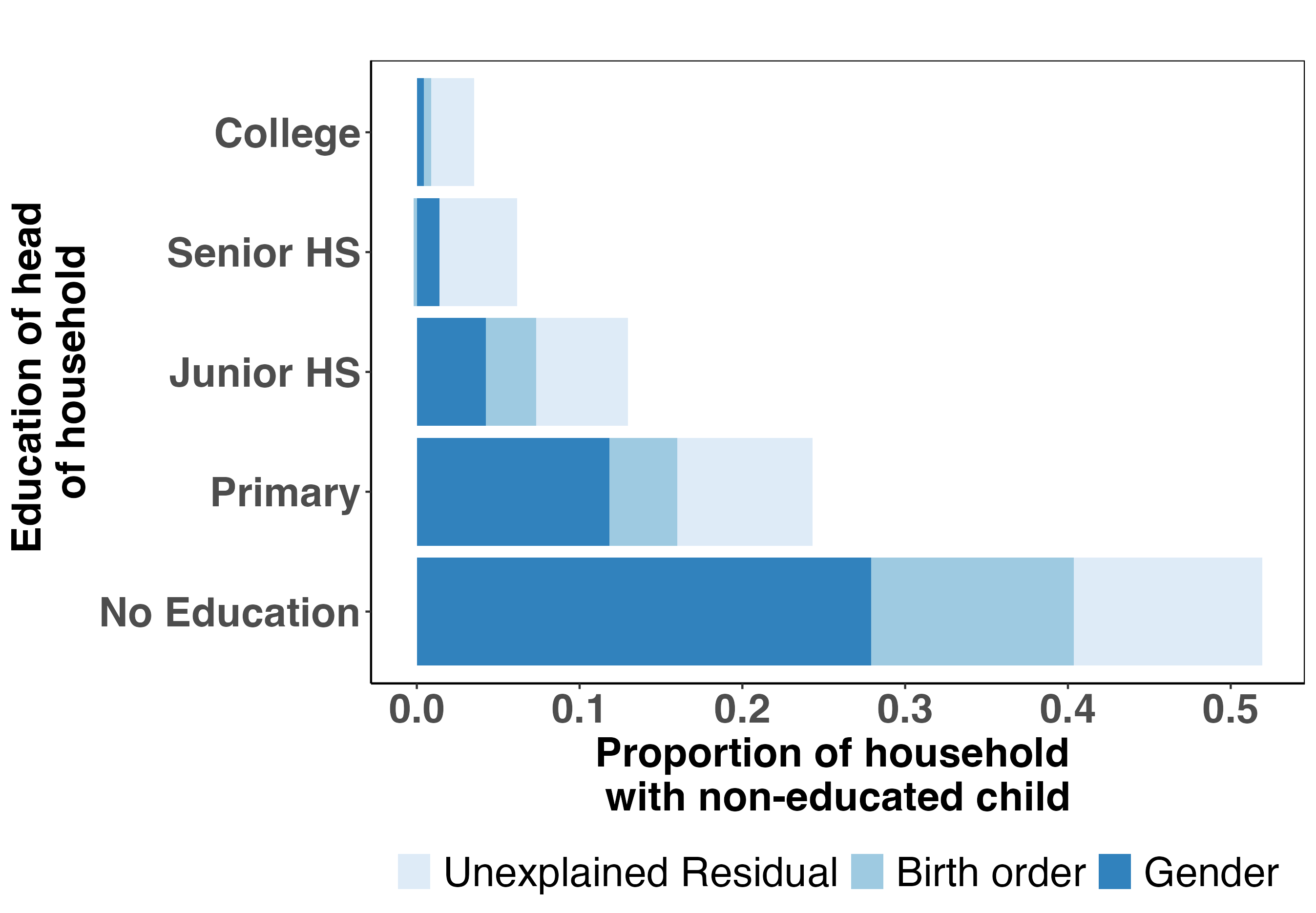}
   
   (a) Extensive Margin
 \end{minipage}
 \quad
 \begin{minipage}[b]{0.45\linewidth}
     \includegraphics [width=7cm]{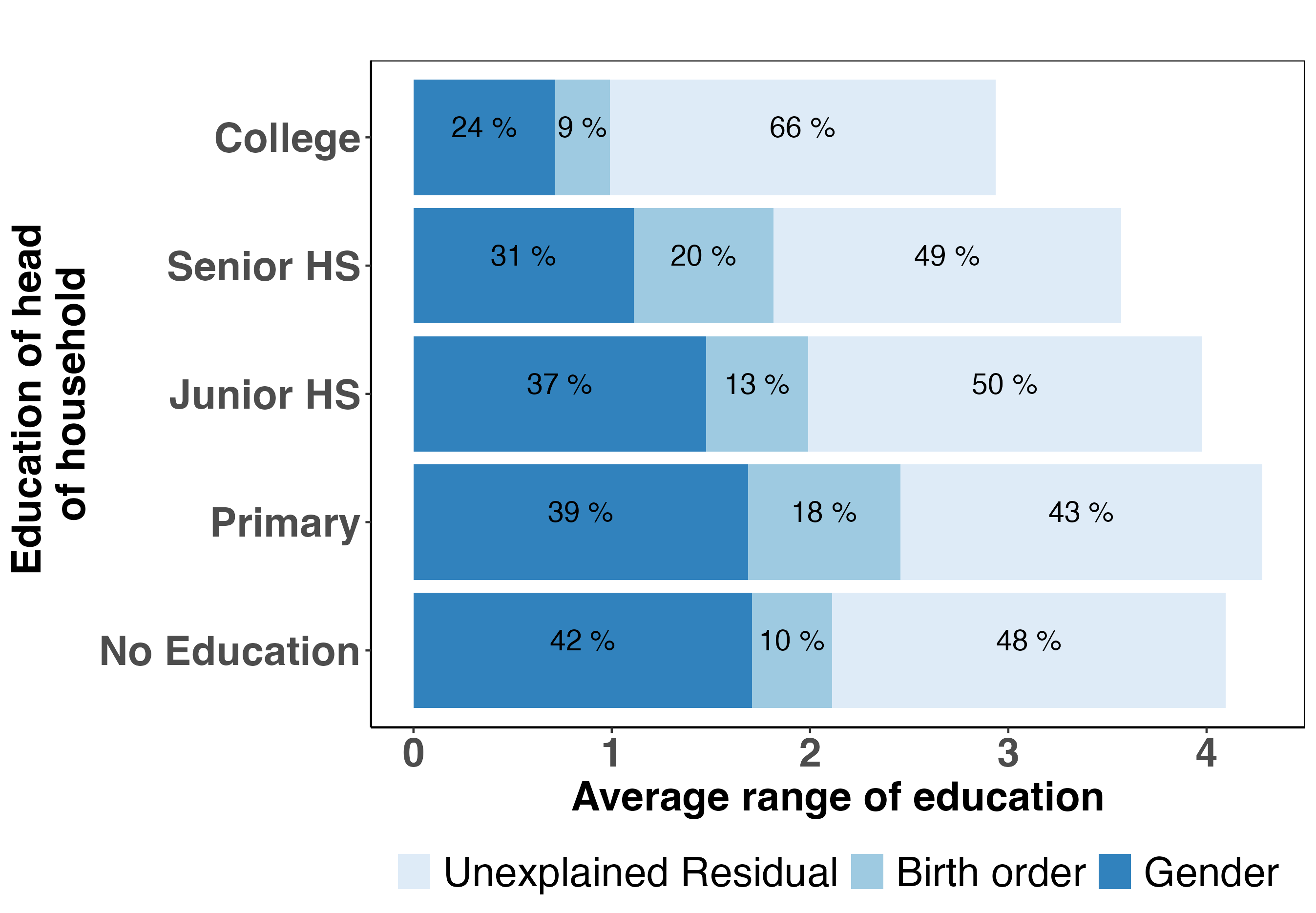}
     
     (b) Intensive margin
\end{minipage}
    \caption{Inequality decomposition as function of head of household's education ($N_c =2$)}
     \label{decomposition_int_ext}
\end{figure}

The estimates are presented in Table \ref{table:fe_reg_int_ext}, and the decomposition of inequality at extensive and intensive margins is presented in Figure \ref{decomposition_int_ext}. The numbers indicate that parents' education is negatively related to inequality in children's education mostly at the extensive margin. In particular, panel a) of Figure \ref{decomposition_int_ext} shows the proportion of households with a non-educated child by parents' education. That number is the highest among non-educated parents ($\approx 50\%$) and close to $0$ ($\approx 3\%$) among college educated parents. There is also a substantial heterogeneity in the decomposition of inequality at the extensive margin. Specifically, for most of households with a non-educated head of household and a non-educated child, the non-educated child is either a daughter or a firstborn. This is not true among college educated parents.

\begin{table}[H]
    \begin{center}
    \caption{Regression of children's education on their gender and birth order with household fixed effect (Extensive vs Intensive margin) $(N_c = 2)$}
    {
\scalebox{0.9}{
\begin{tabular}{l c c c c c c}
\hline
& \multicolumn{2}{c}{Non-educated parents} & \multicolumn{2}{c}{College educated parents} & \multicolumn{2}{c}{All} \\
 & Extensive  &  Intensive   & Extensive  &  Intensive  & Extensive  &  Intensive  \\
 & (1) &  (2) &  (3) &  (4)&  (5) &  (6) \\
\hline
Female     & $-0.54^{*}$       & $-1.73^{*}$       & $0.05$           & $-0.94^{*}$       & $-0.52^{*}$       & $-1.38^{*}$       \\
Firstborn         & $-0.23^{*}$       & $-0.39^{*}$       & $0.11$           & $-0.47^{*}$       & $-0.22^{*}$       & $-0.29^{*}$       \\
Firstborn Female  & $-0.05$           & $0.05$            & $-0.35$          & $0.45$            & $-0.05$           & $-0.25^{*}$       \\
\hline
R$^2$      & $0.21$            & $0.71$            & $0.03$           & $0.71$            & $0.20$            & $0.75$            \\
Num. obs.  & $10166$           & $12374$           & $62$             & $1822$            & $12846$           & $28038$  \\
Household fixed effects      & \checkmark           & \checkmark            & \checkmark  & \checkmark           & \checkmark            & \checkmark          \\
\hline
\multicolumn{7}{l}{\scriptsize{$^*$ Null hypothesis value outside the confidence interval.}}
\end{tabular}}
}
    \label{table:fe_reg_int_ext}
    \end{center}
\end{table}

\section{Structural Model of Educational Attainment Choice}

\subsection{Setup}
The model I propose considers children as investment goods rather than simple consumption goods. In other words, the number of children does not enter parents' utility function directly like in \cite{becker1973interaction}. Parents' choices consist of 2 distinct stages. In the first stage, households make decisions regarding the number of children, denoted as $N_c$, and observe their abilities and other characteristics (which are unobserved to the econometricien), represented by the vector $\omega = \big( \omega_1, \dots, \omega_{N_c} \big )$. They then choose the aggregate total years of educational attainment, denoted $q_T$, for these $N_c$ children. This leads to a within-household average years of education of children, denoted $\Bar{q} = \frac{1}{N_c} q_T$. This initial stage can be viewed as choices derived from solving a fertility choice model, resembling the one described in \cite{becker1976child}, with the distinction that each child is not assumed to receive $\Bar{q}$ years of education. In other words, the decisions made in the first stage are based on the quantity-quality trade-off theory. This leads to different choices on average for parents with different level of education. 

\paragraph{Parents unobserved financial, social and cultural constraints\\}

In addition to allowing an influence of parents' education on the quantity of children and resources devoted to their schooling, I account for an unobserved heterogeneity that reflects parents' financial, social and cultural constraints, leading to limited educational opportunities for some of their children. In particular, there is an unobserved type for parents which creates barriers to school entry for some of their children. These barriers include limited household resources that might not be sufficient to educate all children and labor demands in agricultural families that makes them keep certain children out of school to contribute to farming activities. They also include religious beliefs about formal schooling. The unobserved type dictates the percentage of uneducated children observed within the family. For households with two children, the unobserved type can take 2 possible value \{High (1),  or Low (0)\}. The type ``High" means high financial, social, and cultural constraints and is associated with the presence of an uneducated child, whereas the type ``Low" means low constraints and is associated with the absence of an uneducated child. I model parents' unobserved types using the following binary choice model:

For each household $h$ (Parents' education and number of children hold fixed), let $\nu_h^{\star}$ be the unobserved financial, social, and cultural constraints, and $\nu_h$ be the observed discrete outcome of number of children with no formal education. $\nu_h^{\star} = X_h \beta + \varepsilon_h$, where $X_h$ includes a constant, an indicator for rural area of residence, an indicator for agricultural households, HWI, religion, gender composition, and average education of children. $\nu_h = 1$ if $\nu_h^{\star} > 0$, and $0$ otherwise.
I assume $\varepsilon$ follows a standard normal distribution, and estimate $\beta$ using a probit regression. Given $\hat{\beta}$, for each household $h$, $\nu_h$ is modeled as follows:

\[\nu_h \sim Bernoulli (\Phi[X_h \hat{\beta}]). \]

\begin{figure}[H]
 \centering
 \begin{minipage}[b]{0.45\linewidth}
   \includegraphics [width=8cm]{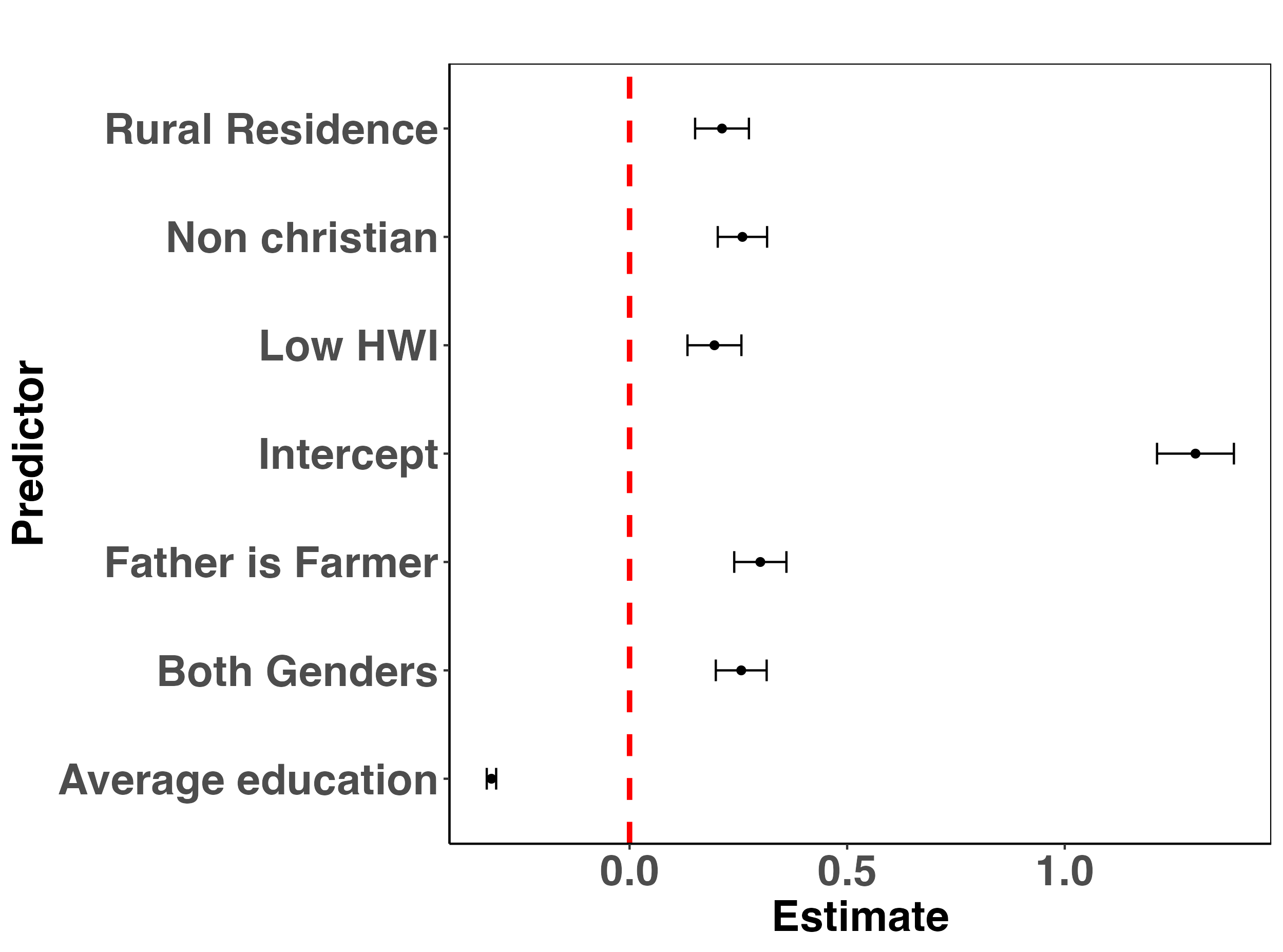}
   
   (a) Estimate
 \end{minipage}
 \quad
 \begin{minipage}[b]{0.45\linewidth}
     \includegraphics [width=8cm]{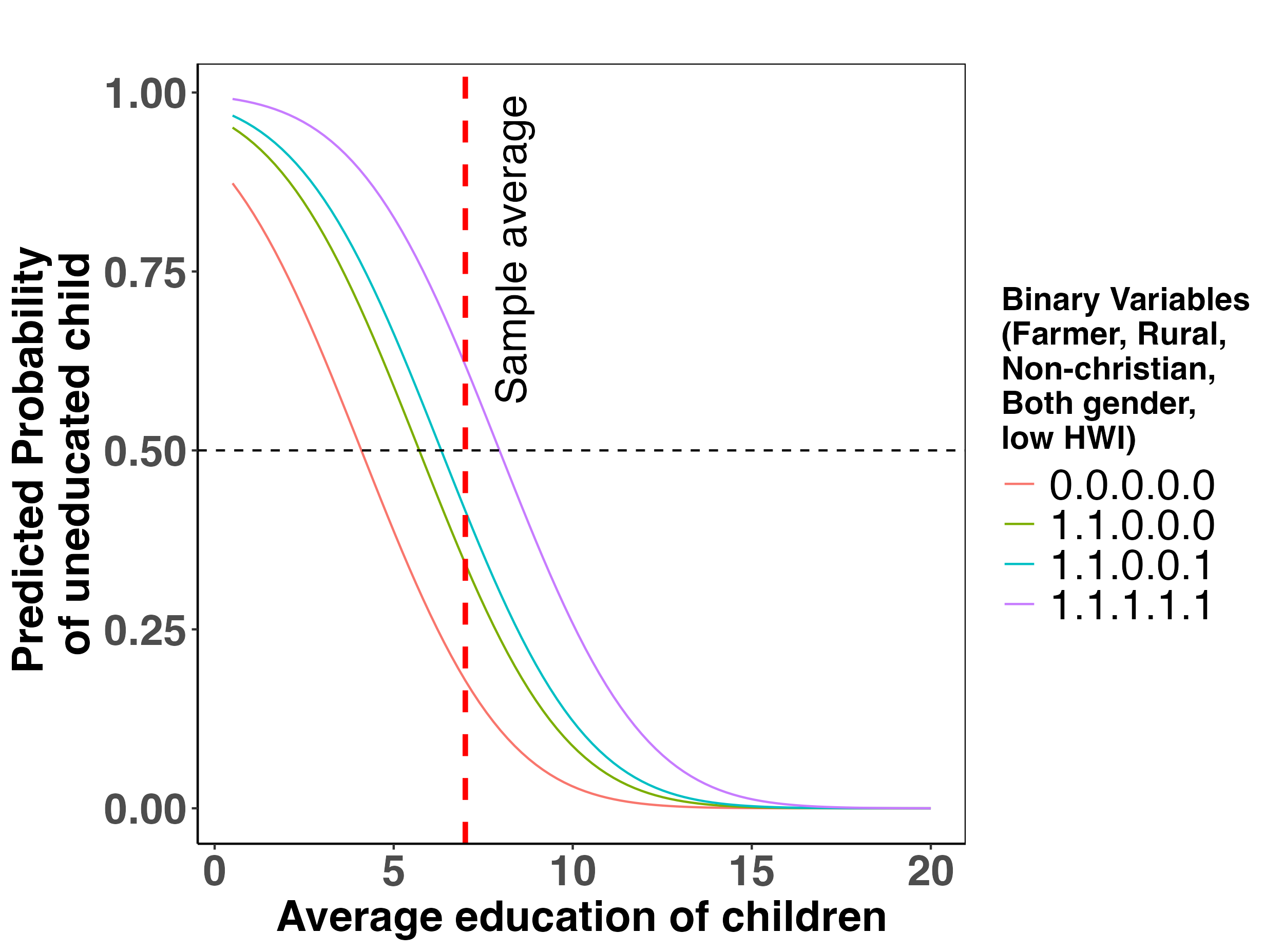}
     
     (b) Predicted probability as function of $\Bar{q}$
\end{minipage}
\caption{Estimates of $\hat{\beta}$}
\label{estimate_beta_hat}
\end{figure}

For number of children $> 2$, I use a multinomial model. \footnote{The estimates of $\hat{\beta}$ are presented in Figure \ref{estimate_beta_hat} }

\paragraph{The maximization problem\\}
In the second stage of their decision, households decide on the distribution of $q_T$. This decision is function of their unobserved type combined with children's observed and unobserved characteristics.  Specifically, each household is characterized by a type $\nu_h$ (their level of financial hardship, social and cultural constraints). Given $\nu_h$, the household distributes $q_T$ among the children taking into consideration their gender, birth order and innate ability/other unobservables. The decision of parents is to choose the distribution $(q_1, \dots q_{N_c})$ of $q_T$, which maximizes the household's utility function. 
\begin{equation}
    \max_{q_i} U ( q, \theta)
    \label{eq1}
\end{equation}
\[\text{subject to } \sum q_i \leq q_T, \text{  } q_i > 0, \text{  } q_i \leq q_{\max} \]

I assume that $U(.)$ is increasing and  concave, and $q_{\max}$ is the maximum years of education a child can receive. $\theta$ is the vector of parameters described in the next section. The model analyzes decisions in the second stage, taking choices in the first stage as given.

\paragraph{Functional form of households' utility [for households with 2 adult children]\\}

\begin{figure}[H]
\begin{center}
 \begin{minipage}[b]{0.45\linewidth}
   \includegraphics [width=8.5cm]{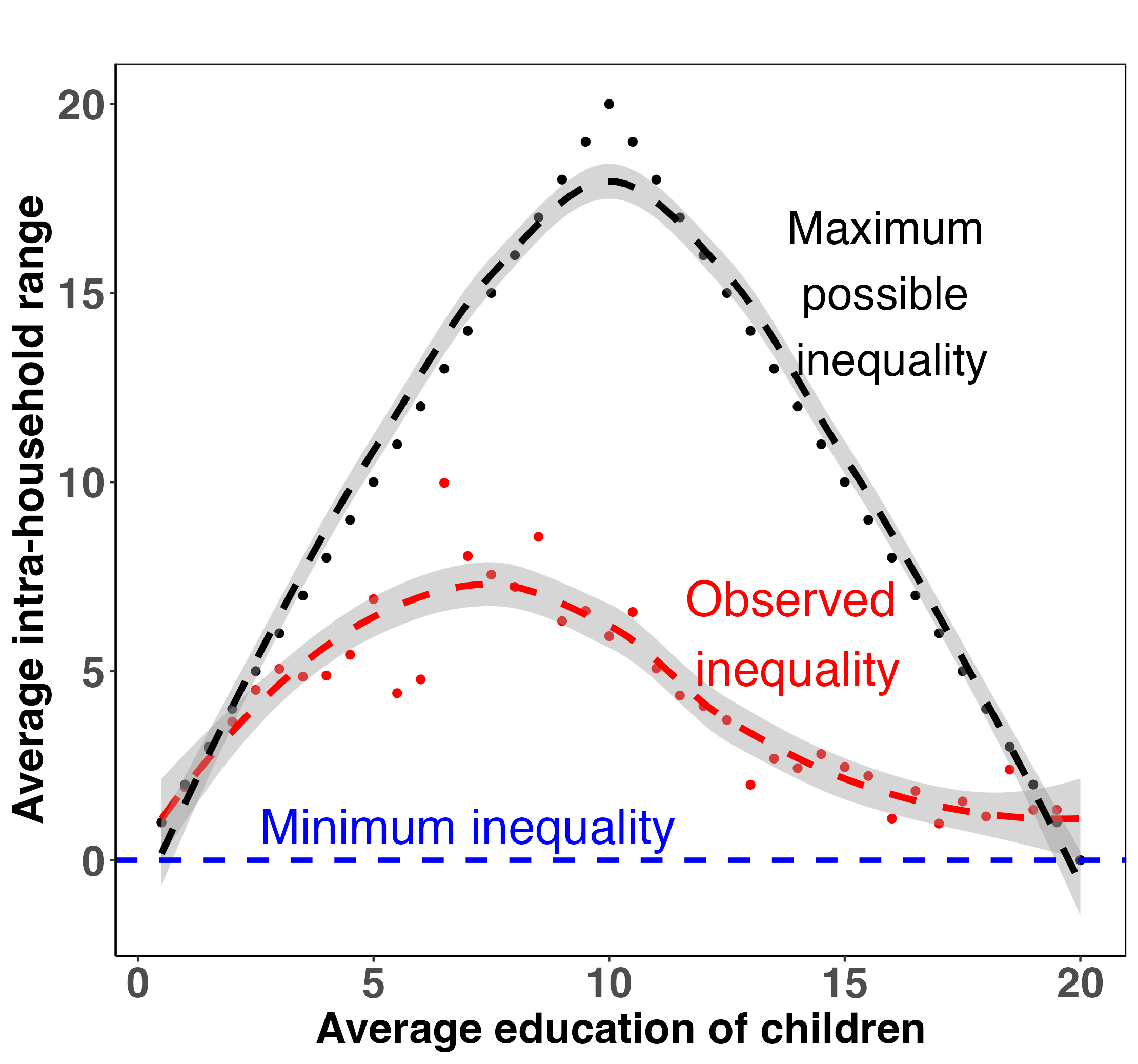}
 \end{minipage}
  \caption{Within-household standard deviation of children's education as function of average education of children (Min vs. Max vs. Observed for $N_c = 2$).}
 \label{ineq_educ_c}
 \end{center}
\end{figure}

I use a generalized utilitarian social welfare function \footnote{The use of a generalized utilitarian social welfare function allows for flexibility in how children's utilities are aggregated within the household. Individual utilities can be weighted or transformed, allowing for a range of preferences about inequality, risk, or the relative importance of individuals' well-being.} to represent parents' utility function. This function incorporates a concave utility function derived from the education levels of each child. I chose this functional form because it better aligns with the observed mean-standard deviation curve, which differs from the one predicted by a linear utility function and more closely resembles a concave utility function from children's educational outcomes (see Figure \ref{ineq_educ_c}). Let $q_h = (q_{1, h}, \dots, q_{N_c, h})$ be the distribution of $q_{T, h}$ in household $h$. The utility function for households with 2 children has the following expression\footnote{The utility function for households with more than 2 children is presented in Appendix B.}:
\begin{equation}
   U(q_h) = (1-\nu_h) \Big[ \sum_{i = 1}^{2} a_{i, h}. (q_{i, h})^{\delta_{i, h}^{\text{low}}} - \alpha_{i}^{\text{low}} q_{i, h} \Big ] +\nu_h \Big \{ \sum_{i = 1}^{2} \Big [ e_{i, h}.  \big ( a_{i, h}. (q_i)^{\delta_{i, h}^{\text{high}}} - \alpha_{i}^{\text{high}} q_{i, h} \big ) \Big]  \Big \}, 
\end{equation}
where,
\begin{itemize}
    \item $a_{i, h} = \frac{\omega_{i, h}}{\sum_{j =1}^{N_{c_h}} \omega_{j, h}} \sim G(.)$ captures the unobserved ability-based weight on child $i$'s education relative to other children in household $h$. By definition $\sum_{i =1}^{2} a_{i, h} = 1$.
    \item $\nu_h$, is parents' level of financial, social, and cultural constraints.
    \item  $e_{i, h} = 1\{a_{i, h}. (q_T)^{\delta_{i, h}^\text{{high}}} - \alpha_{i}^{\text{high}} q_T > a_{j, h}. (q_T)^{\delta_{j, h}^{\text{high}}} - \alpha_{j}^{\text{high}}q_T\}$, $e_{j, h} = 1-e_{i, h}$, $e_{i, h}$ and $e_{j, h}$ are indicator of whether not children $i$ and $j$ in household $h$ have some education.
    \item $\delta_{i, h}^{\text{low}} = \delta(\text{gender}_{i, h}, \text{gender\_comp}_h) = \gamma -  \theta_1^{low} \text{Female}_{i, h}(1 - \text{Female}_{j, h})$.
    \item $\delta_{i, h}^{\text{high}}  = \gamma -  \theta_1^{\text{high}} \text{Female}_{i, h}(1 - \text{Female}_{j, h})$.  [$\gamma$ is normalized to 0.5].
    \item $\delta_{i,h}$ is the marginal benefit from giving an additional year of education to child $i$ in household $h$.
      \item $\alpha_{i}^{\text{low}}$, and $\alpha_{i}^{\text{high}}$ are the costs (financial and opportunity costs) of giving a year of education to $i^{th}$ child at the extensive and intensive margin respectively.  $\alpha_2^{\text{high}}$, and $\alpha_2^{\text{low}}$ are normalized to 0.
      \item $q_{i, h}$ is the years of education of child $i$ in household $h$.
\end{itemize}

The vector of parameters of interest is \[\theta = \Big ( \theta_1^{\text{low}}, \theta_1^{\text{ds, high}}, \theta_1^{\text{sd, high}}, (\alpha_1^{\text{high}} - \alpha_2^{\text{high}}), (\alpha_1^{\text{low}}- \alpha_2^{\text{low}}) \Big ).\]
The utility from providing a $q_{i, h}$ level of education for each child [$u_{i, h} = a_{i, h}. (q_{i, h})^{\delta_{i, h}^{\text{low}}} - \alpha_{i}^{\text{low}} q_{i, h} $ or $u_{i, h} = a_{i, h}. (q_i)^{\delta_{i, h}^{\text{high}}} - \alpha_{i}^{\text{high}} q_{i, h}$] in the parents utility has two parts: the benefit and the cost parts. Note that $a_{i, h}$ and $q_{i, h}$ are complementary in the benefit part of the utility from providing a $q_{i, h}$ level of education to child $i$ in household $h$. In other words, holding everything else fix, parents get higher utility by providing higher $q_{i, h}$ to child $i$ compared to child $j$ if $a_{i, h} > a_{j, h}$.

\paragraph{Assumption 1: $a_{i, h}$ of a child $i$ in household $h$ is drawn from a distribution $G(.)$, with the constraint that $\sum_{i =1}^{N_{c_h}} a_{i, h} = 1$. I assume that $G(.)$ is independent of gender and birth order.}
\paragraph{}

The incorporation of differences in $\delta$ across the children's genders and the household's gender composition within the model allows for the consideration of disadvantage that females face at the extensive and intensive margin in terms of human capital investment when they have a brother. This parameter models the difference in educational attainment by gender, reflecting potential gender disadvantage that may exist within the household--- as evidenced in Figure \ref{ineq_gender}. The assumed functional form is designed to capture the idea that girls with brothers receive a penalty in the distributional decision of the education resources made by parents. Additionally that penalty is an increasing function of the proportion of boys among the siblings.
Similarly, the model allows for differences in $\alpha$ across children's birth order to capture the monotonic increase in educational attainment as birth order advances as observed in Figure \ref{ineq_birth_order}. 
To estimate the vector of parameter $\theta$, I used simulated method of moments. The procedure is outlined in the next section. The parameters are estimated for each level of education of parents holding number of children fixed.



\subsection{Estimation and Inference Strategy}
In this section, I provide an overview of the data moments used to estimate the key parameters in the model. I use two sets of moments for the parameters' estimation. First, the difference in average educational attainment between daughters and sons in households with one child of each gender and no uneducated children, while holding fixed parents' education and number of children; and the average educational attainment by birth order in households with children of the same gender only and no uneducated children, holding fixed the head of household's education and number of children. This moments provide data variations to estimate $\delta^{\text{low}}$ and $\alpha_{(i)}^{\text{low}}$ respectively.

The second set of moments includes the proportion of educated daughters and firstborn children in households with an uneducated child, while holding fixed parents' education and number of children. These moments help estimate parameters in $\delta^{\text{high}}$ and $\alpha_{(i)}^{\text{high}}$. Specifically, the proportion of educated firstborn children in households with only children of the same gender and one uneducated child allow for the estimation of $\alpha_{(i)}^{\text{high}}$. The proportion of educated daughters from households with both genders and one uneducated child  is used to estimate $\theta_{1}^{\text{high}}$.

For the rest of this section, let's define the variables $Y_{h}^d$ as daughters' education in household $h$ and $Y_{h}^s$ as sons' education in the same household $h$. And let $Y_h^1$, $Y_h^2$, be the education of firstborn and second born children respectively. Additionally, let $Z$ be a vector of observables, such as the education of the head of the household, the number of children ($N_c$), the aggregate education of children ($q_T$), and the gender composition of children.
Note that for households with the same observed (Z) and unobserved ($\nu$) types, any differences observed in the variables $Y_{h}^d$ and $Y_{h}^s$, or in $Y_h^1$ and $Y_h^2$ between these households stem from disparities in the unobservable difference in children's characteristics (eg. innate ability). 

Given the defined notations and functional form, the inference procedure proceeds as follows. First, I simulate $H$ households, each with $N_c = 2$ children, possible gender composition (from $\{$only sons, only daughters, firstborn son and second born daughter, firstborn daughter and second born son$\}$), households observed characteristics X \footnote{X includes rural, farmer, HWI, religion, gender composition}, and $q_T$. The simulated  households are drawn from of the empirical joint distribution of these variables. Second, for a fixed $\delta_{i, h}^{\text{high}}$, $\delta_{i, h}^{\text{low}}$, $\alpha_1^{\text{high}}$, and $\alpha_1^{\text{low}}$, I solve the household's maximization problem in equation \ref{eq1} for $s$ draws of $\Big \{(a_{i, h})_{i = 1}^{N_{c_h}}, \text{ with } \sum_{i =1}^{N_{c_h}} a_{i, h} = 1 \Big \}$ for each of the $H$ simulated households.\footnote{Note: H = 1000 and $s = 10$} This procedure yields the following model predictions:
\begin{enumerate}
    \item $S_d^d = s \times H^d$ predictions of the educational attainment of daughters in households with only daughters, where $H^d$ is the number of simulated households with only daughters.
    \item $S_d^b = s \times H^b$ predictions of the educational attainment of daughters in households with both genders, where $H^b$ is the number of simulated households with both genders.
    \item $S_d^i = s \times H^d$ predictions of the educational attainment of the $i^{th}$ born daughter in households with only daughters.
    \item $S_s^i = s \times H^s$ predictions of the education of the $i^{th}$ born son in households with only sons, where $H^s$ is the number of simulated households with only sons.
\end{enumerate}

These predicted educational attainments  represent the educational outcomes based on the given parameter values. I then take the average of the $S_l^m$ predictions for each moment, where $l, m \in$ \{only sons (s), only daughters (d), both gender (b) \}.

To do inference on the parameters in $\delta^{\text{low}}$, the model and data moments are matched across various gender compositions. This process involves normalizing the parameter $\gamma$ to 0.5 and estimating $\theta_1^{\text{low}}$ by matching the model's predictions with the observed data in terms of the difference in educational attainment for daughters and sons from households with both genders.
For the inference on the parameters $(\alpha_{(t)}^{\text{low}})_{t = 1}^{N_c}$ associated with birth order, the model and data moments are matched across different birth orders. This process entails normalizing $\alpha_{(N_c)^{\text{low}}}$ to 0 and estimating $(\alpha_{(t)}^{\text{low}})_{t = 1}^{N_c - 1}$ by comparing the model's predictions to the observed data regarding the difference in educational attainment between $t^{th}$ and $(t + 1)^{th}$ born children.
For the parameters $\delta^{high}$, and $(\alpha_{(t)}^{high})_{t = 1}^{N_c}$, the data moments and the model moments on the proportion of educated firstborn children from one gender households, firstborn daughter, and second born daughters from mix gender households are matched with the model moments. 

Let $\hat{\mu}_l^d( \theta, Z)$ represent the predicted average education attainment of daughters in different household types, where $l \in {d, s, b}$ denotes households with only daughters, only sons, and both genders, respectively. Similarly, let the vector $\hat{\mu}( \theta, Z) = (\hat{\mu}_1( \theta, Z), \dots, \hat{\mu}_{N_c}( \theta, Z))$, be the predicted average education attainment by birth order. Finally, let $\hat{\pi} = (\hat{\pi}_1, \hat{\pi}_{d})$, be the model prediction of the proportions of firstborn children, and daughters for households with an uneducated child. These simulations provide estimates of the model's predictions for various household compositions, gender and birth orders, allowing for the comparison of the model's outcomes with the observed data. The data moments are defined as follows: 
Let $T_{\text{Educ}}$ be the total number of educated children.
\begin{itemize}
    \item $m_1 = E [ Y^s | \text{Gender\_Comp} = b, T_{\text{Educ}} = 2] - E [ Y^d | \text{Gender\_Comp} = b,  T_{\text{Educ}} = 2],$
    \item $m_2 = E [1\{ Y^1 > 0 \} | \text{Gender\_Comp} = s \text{ or } d, T_{\text{Educ}} = 1] ,$
    \item $m_3 = E [1\{ Y^d > 0 \} | \text{Gender\_Comp} = b, T_{\text{Educ}} = 1] ,$
    \item $ m_{t +3} = E [ Y_{t+1} |\text{Gender\_Comp} = s \text{ or } d, T_{\text{Educ}} = 2 \text{ }\& \text{ birth\_order } = t + 1 ] - E [ Y_t | \text{Gender\_Comp} = s \text{ or } d, T_{\text{Educ}} = 2 \text{ } \& \text{ } \text{birth\_order} = t ], \text{ } t \in \{1, \dots, N_c \}.$
\end{itemize}
I matched the following data and model moments to estimate $\theta$.
\[ m_1 = \hat{\mu}^d_d - \hat{\mu}^d_b, \text{ } m_2 = \hat{\pi}_1, \text{ } m_3 = \hat{\pi}_{d}, \text{ and } m_{t+3} = \hat{\mu}_{t +1} - \hat{\mu}_{t}; \text{ } t \in \{1, \dots N_c - 1\}. \]

\begin{figure}[H]
\centering
    \subfloat[$\theta_1^{high}$]{\includegraphics[width=0.45\textwidth]{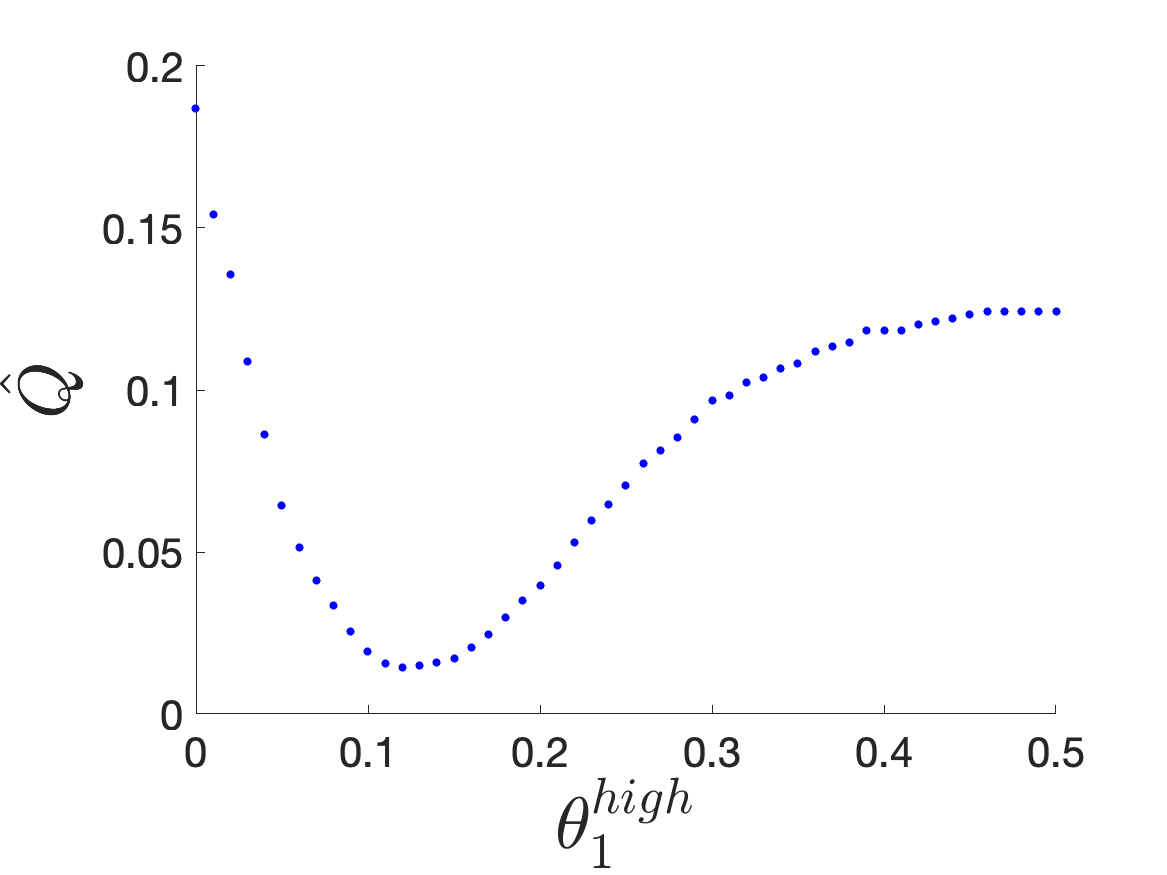}}
    \subfloat[$\alpha_1^{high}$]{\includegraphics[width=0.45\textwidth]{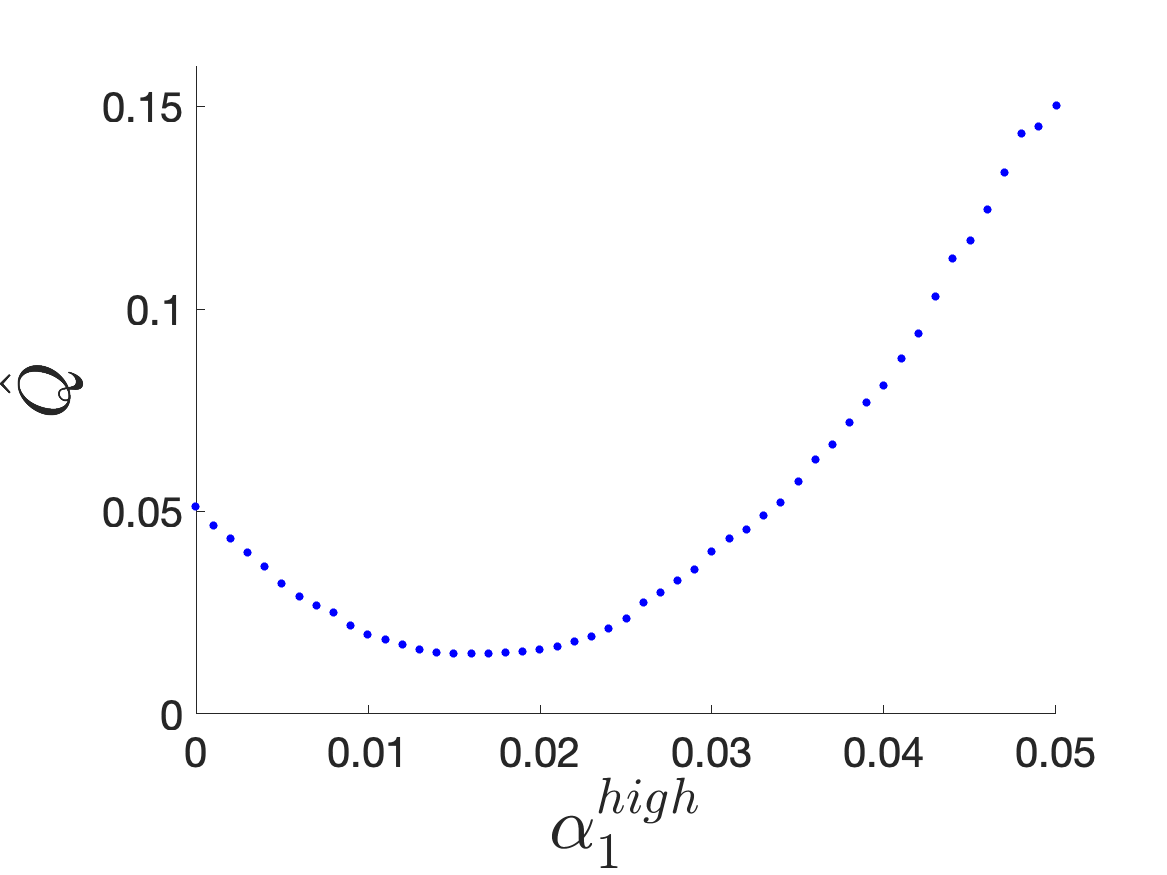}}
    \caption{Plots of the objective function}
    \label{plots}
 \end{figure}

The corresponding sample objective function is the following expression: 
\begin{equation}
    \Hat{Q}(\theta) = ({\Bar{Y}_{d,z}^d - \Bar{Y}_{b, z}^d} - ({\Hat{\mu}_{d, z}^d - \Hat{\mu}_{b, z}^d}))^2 + ({\hat{m}_2} - {\hat{\pi}_1})^2 + ({\hat{m}_3} - {\hat{\pi}_{d}})^2 + 
\sum_{l \in \{d, s\}}  ({\Bar{Y}_{2, z}^l - \Bar{Y}_{1, z}^l} - ({\Hat{\mu}_{2, z}^l - \Hat{\mu}_{1,z}^l}))^2.
\end{equation}
 \[ \hat{\theta} = \text{ argmin}_{\theta \in \Theta} \hat{Q}(\theta).\]
 The sample objective function possess a unique local optimizer (See Figure \ref{plots}). SMM standard errors are computed.

\subsubsection{Estimation of G(.)}
\begin{figure}[H]
 \begin{center}
 \begin{minipage}[b]{0.45\linewidth}
   \includegraphics [width=10cm]{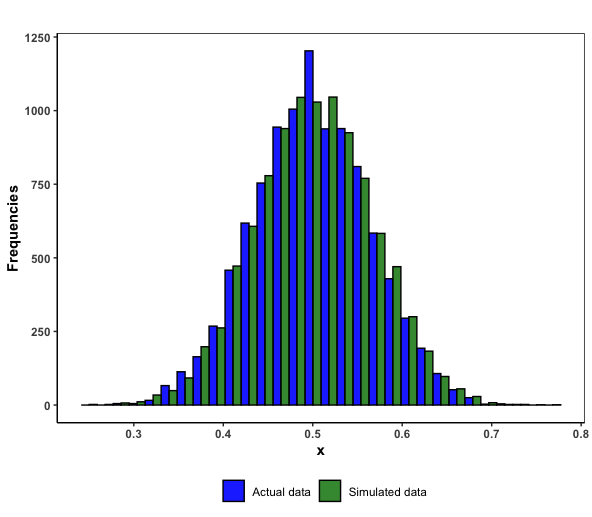}
 \end{minipage}
  \caption{Histogram of relative GPA in junior high school and histogram of random draws from Beta (28.82, 28.78).}
 \label{ability_parameter}
  \end{center}
\end{figure}

I use auxiliary data to estimate the parameters of the distribution $G(.)$ of the ability-based weight on children's education outside of the model. In particular, I assume that the ability-based weights on children's education are i.i.d across household and from a Dirichlet distribution.
        \[a_{h} \sim^{i.i.d}   \text{Dirichlet}(\eta_1, \dots,  \eta_{N_{c_h}}) \text{ where, } a_h = (a_{1, h}, \dots,  a_{N_{c_h}, h}) \]
       $(\eta_1, \dots,  \eta_{N_{c_h}})$ are estimated using auxiliary data. Specifically, I used data on average GPA in junior high school for a sample of student in Benin in 2018 to estimate $(\eta_1, \dots,  \eta_{N_{c_h}})$ using maximum likelihood method. The Dirichlet distribution seems to be a good fit for the distribution of relative ability (See Figure \ref{ability_parameter}).

 \section{Estimation Results and Counterfactual Analysis}

 \subsection{Estimation Results}

 The estimates of $\theta$ are provided in Table \ref{table:estimates} for households with $N_c = 2$ children, for non-educated and college educated parents. 

      \begin{table}[H]
    \begin{center}
   \caption{Estimates of $\Hat{\theta}$,  ($N_c = 2$)}

\scalebox{0.85}{{
\begin{tabular}{lcccccc} 
 \hline
 \noalign{\global\arrayrulewidth=1.2pt}
  & \multicolumn{4}{c}{\textbf{Non-educated parents}}  & \multicolumn{2}{c}{\textbf{College educated parents}} \\ 
   \addlinespace
 & $\hat{\theta}_1^{\text{low}}$ & $\hat{\alpha}_{1}^{\text{low}}$&  $\hat{\theta}_1^{\text{ high}}$ &  $\hat{\alpha}_{1}^{\text{high}}$& $\hat{\theta}_1^{\text{low}}$ & $\hat{\alpha}_{1}^{\text{low}}$ \\
 \addlinespace
\textbf{Estimates} & 0.0238$^{**}$  &0.0014$^{}$  & 0.13$^{**}$ & 0.017$^{**}$ &0.00759$^{**}$ & 0.00045$^{}$ \\ [1ex]
\addlinespace
\textbf{Standard errors} & 0.0017 &0.0004 & 0.0019  & 0.001 &0.0006 & 0.0012 \\ [1ex]
\addlinespace
\textbf{Number of observations} &  \multicolumn{2}{c}{6187} &  \multicolumn{2}{c}{5083} &   \multicolumn{2}{c}{942}  \\ [1ex]

 \hline
 \hline
\end{tabular}}
}

    \label{table:estimates}
    \end{center}
     \footnotesize{$**$ significant at 5\% level of significance.}
\end{table}

 \textit{\textbf{Result 1:} For parents without formal education and with high financial, social and cultural constraints, parents' perceived average utility at high school level of education is $\approx$ 11.5\% higher for a second born child compared to firstborn child of the same gender.}


The estimate of the marginal educational cost difference between firstborn and second born children suggests that, on average, for parents without formal education and with high financial, social and cultural constraints, the likelihood of the firstborn child being educated compared to a second-born child of the same gender is approximately 0.3663, which correspond to an average cost difference of $0.02$. Holding everything else equal, this cost difference translate into a utility gap of $\approx$ 11.5\%  for high school level of education.

 \textit{\textbf{Result 2:} For parents without formal education and with high financial, social and cultural constraints, perceived average utility at high school level of education is $\approx$ 40\% higher for the second born son compared to the firstborn daughter.}

Among parents without formal education and high financial, social and cultural constraints, the marginal utility from an additional year of education is 33\% high if given to a son compared to given to a daughter. After factoring in birth order, these parents' perceived average utility at high school level of education is $\approx$ 40\% higher for the second born son compared to the firstborn daughter. Their perceived average utility of graduating high school is $\approx$ 18\% higher for the firstborn son compared to the second born daughter . Note that estimates for the extensive margin parameters ($\hat{\theta}_{1}^{\text{high}}$ and $\hat{\alpha}_{1}^{high}$) are not provided for college-educated parents, as nearly all of them--- approximately 98\%--- have only educated children.

 \textit{\textbf{Result 3:} Among parents with low financial, social and cultural constraints, the ones without formal education perceive a 6.6\% higher utility on high school of education for sons compared to daughters, while for college-educated parents, the utility gap is approximately 2\%.}

The parameter estimates for parents with low financial constraints indicate those without formal education perceive a 6.6\% higher utility on graduating high school for sons compared to daughters, while for those with college education the difference is approximately 2.2\%. The cost difference between providing a given level of education to firstborn children compared to second born is very small. For both non-educated and college-educated parents the estimates of the marginal costs different are not statistically different from 0 at 5\% significance level. Parents' perceived average utility at high school level of education is $\approx$ .9\% (resp. 0.3\%) higher for a second born child compared to firstborn child of the same gender among non-educated parents (resp. college educated parents).  

\subsection{The Model's Fit}

\begin{figure}[H]

 \begin{minipage}[b]{0.45\linewidth}
   \includegraphics [width=8cm]{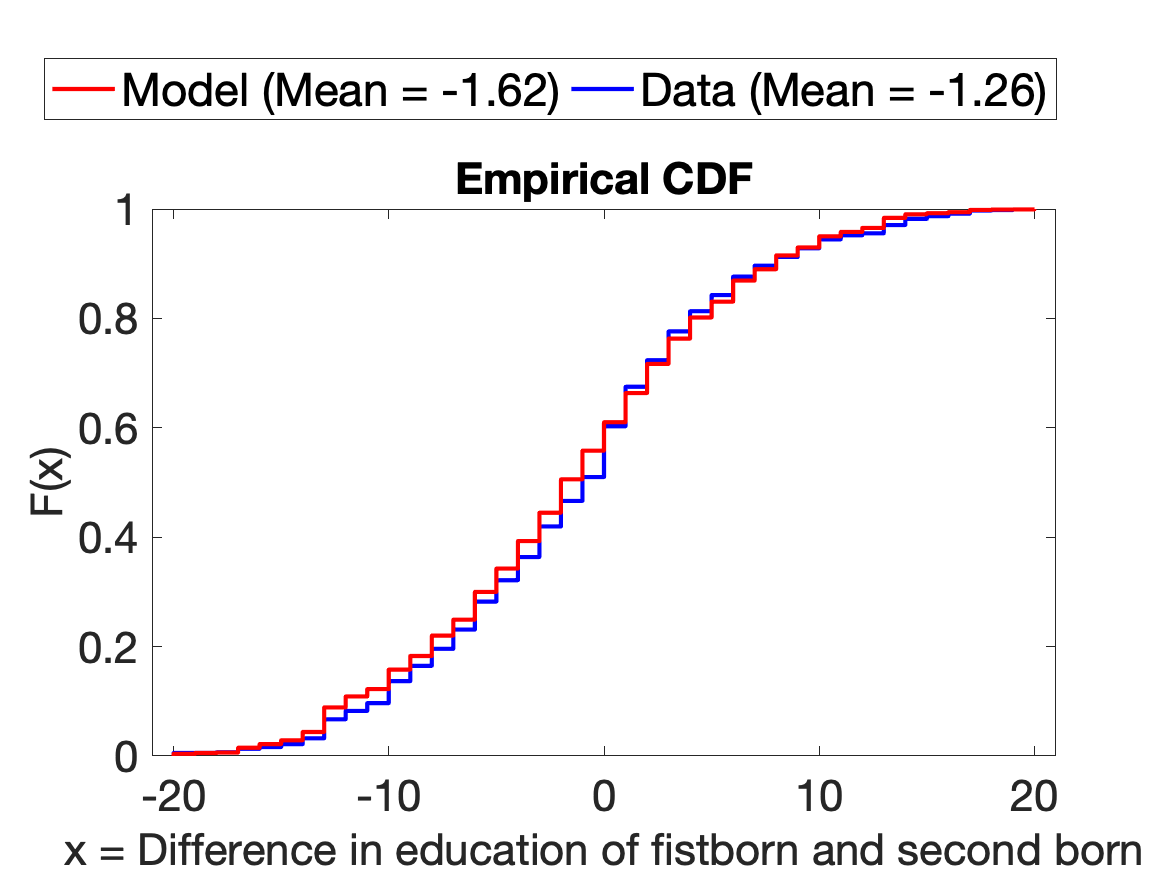}
   
   (a) Firstborn vs. second born
 \end{minipage}
 \quad
 \begin{minipage}[b]{0.45\linewidth}
     \includegraphics [width=8cm]{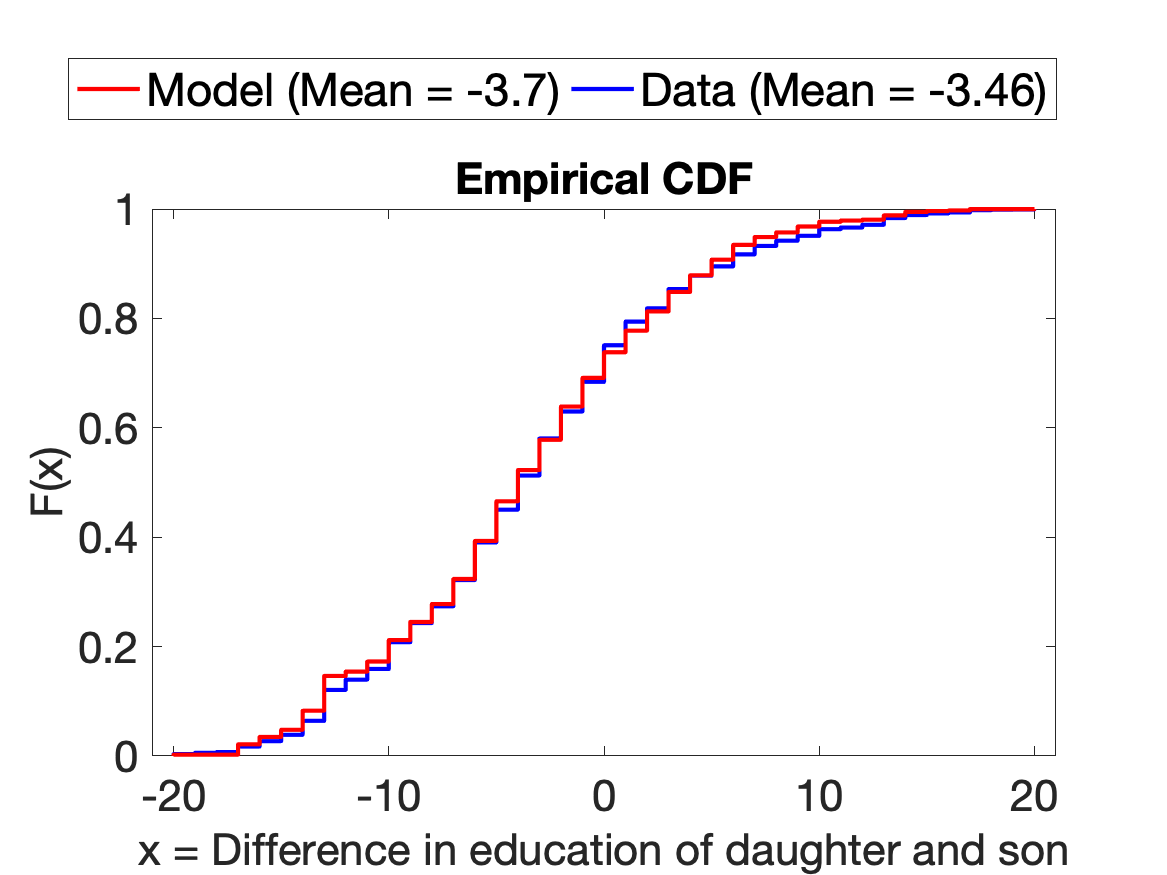}
     
     (b) Daughter vs. son
\end{minipage}
\caption{Empirical distribution of key moments: Data vs. Model ((For non-educated parents))}
 \label{model_data}
\end{figure}

\begin{figure}[H]

 \begin{minipage}[b]{0.45\linewidth}
   \includegraphics [width=8cm]{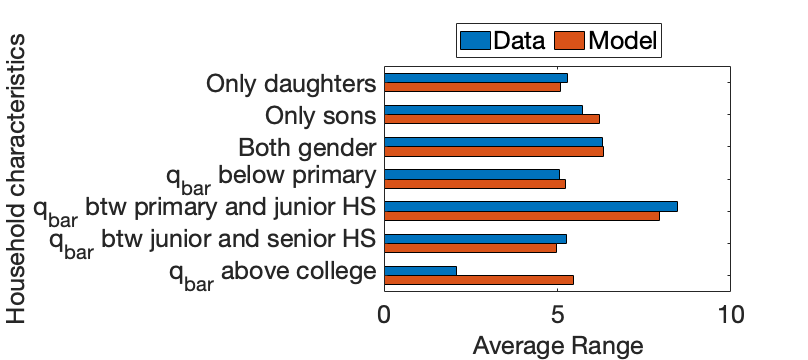}
   
   (a) Average range
 \end{minipage}
 \quad
 \begin{minipage}[b]{0.45\linewidth}
     \includegraphics [width=8cm]{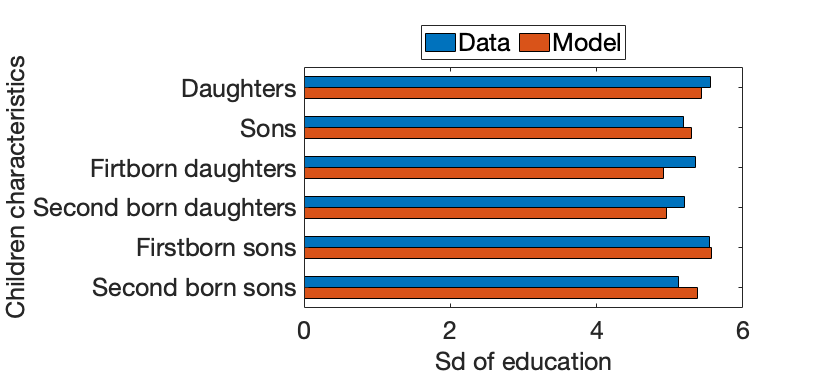}
     
     (b) Standard deviation
\end{minipage}
\caption{Data vs Model moments (For non-educated parents)}
 \label{model_data_moments}
\end{figure}

In this section, I access the fit of the model. Using the estimated parameters $\hat{\theta}$, I solve the household maximization problem, to obtain the optimal distribution $q^{\star}_h$ of $q_{T_h}$ among children within each household $h$ across a simulated sample of $H$ households. I then compare $q^{\star}$ with the observed data to evaluate the model's fit. First, I compute the intra-household educational difference between daughters and sons' educational attainment, as well as between firstborn and second-born children using $q^{\star}$ for the simulated households. The empirical distributions of these within-household differences are compared with the ones from actual educational attainment observed in the data.
Figure \ref{model_data} provides a visual comparison of the model's distributions with the observed data, showing a clear match. Second, I use $q^{\star}$ to derive targeted moments for estimation, alongside selected non-targeted moments. These analytical outcomes are summarized in Figure \ref{model_data_moments}. It shows that there is no significant difference between the model and the data for all the targeted moments and most of the non-targeted moments as well.

\subsection{Counterfactual Analysis}

\textit{\textbf{Counterfactual 1:} Without gender and birth order effects, the distribution of the average educational attainment difference between daughters and sons exhibits first-order stochastic dominance over the distribution in cases with such effects.}

\begin{figure}[H]
 \centering
 \begin{minipage}[b]{0.45\linewidth}
   \includegraphics [width=8.5cm]{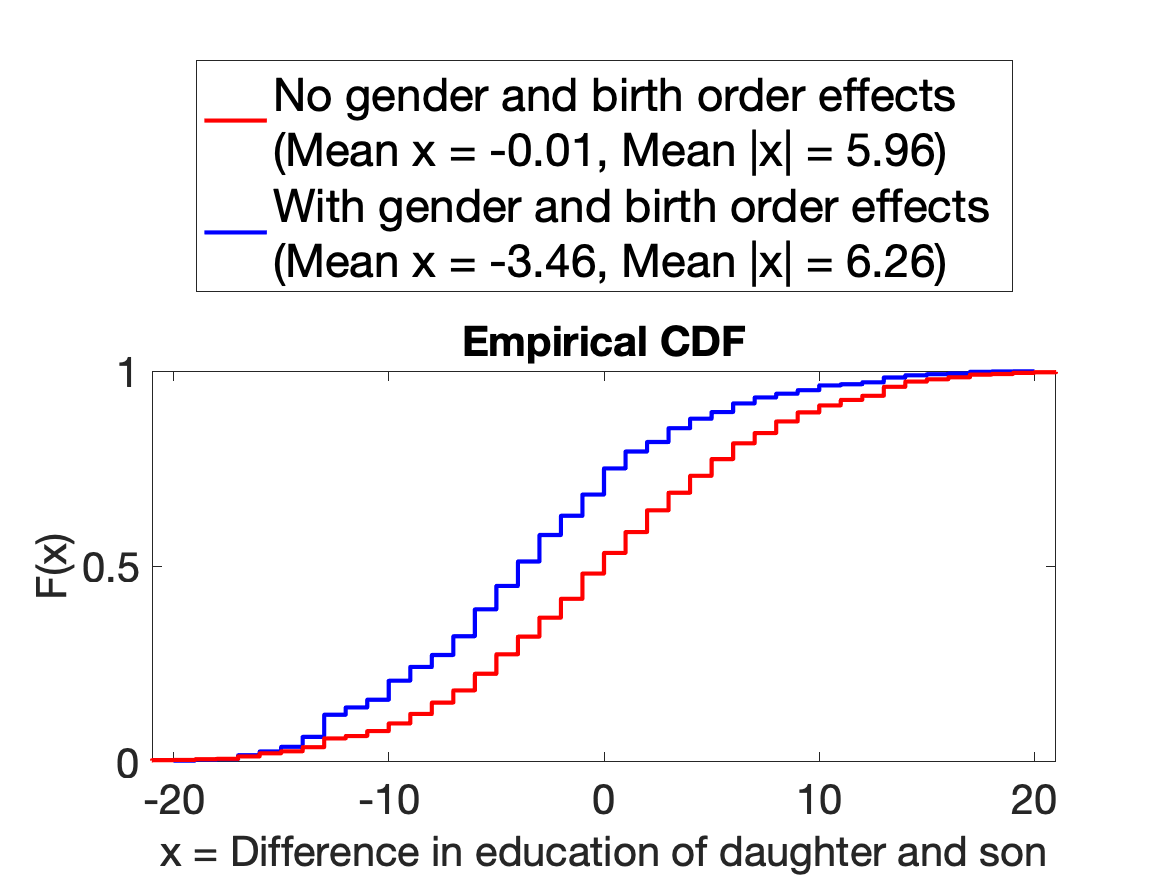}
   
   (a) Daughter v.s son
 \end{minipage}
 \quad
 \begin{minipage}[b]{0.45\linewidth}
     \includegraphics [width=8.5cm]{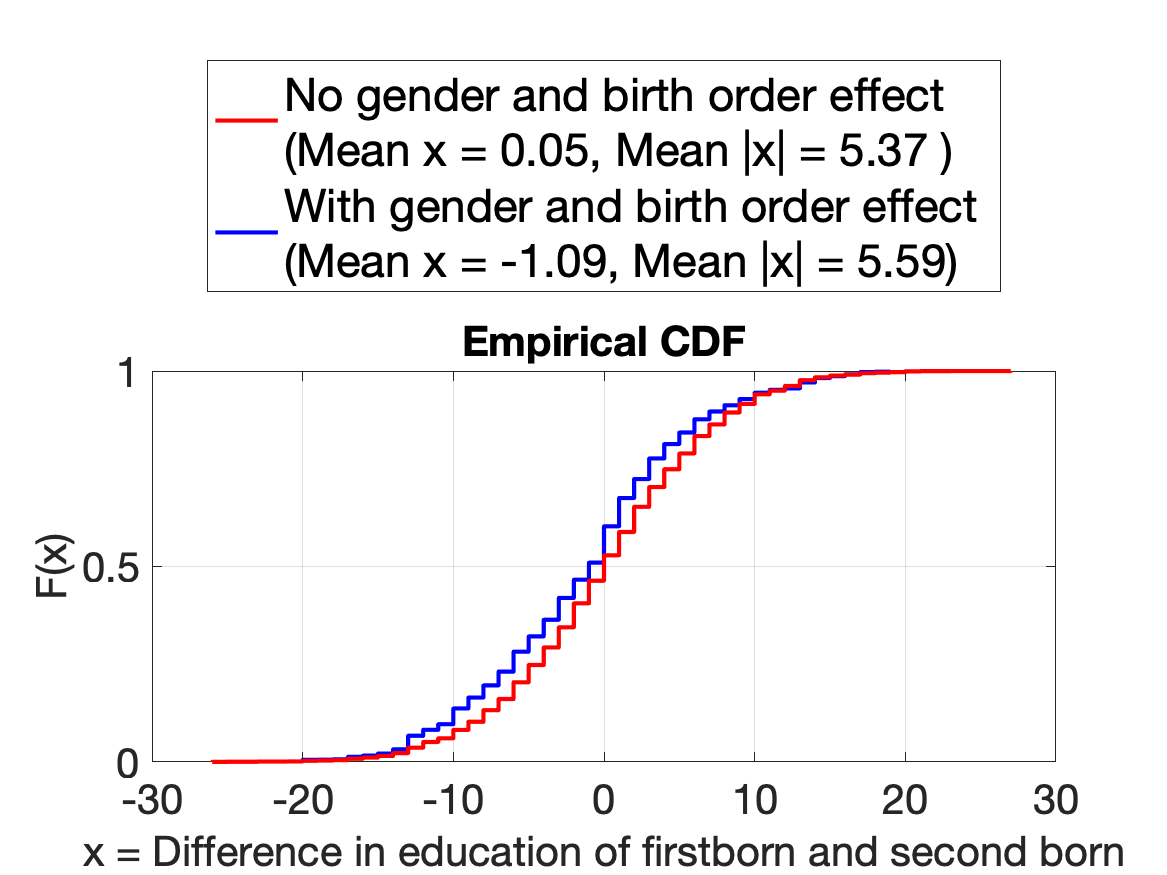}
     
     (b) Firstborn v.s second born
\end{minipage}
\caption{Distribution of the difference in children's education ($N_c = 2$)}
 \label{with_without_bias}
\end{figure}

In the absence of any gender and birth order effects, differences in education within a household between daughters and sons primarily result from variations in their individual unobserved characteristics/abilities. If children's innate abilities are assumed to be distributed independently of gender and birth order, the average educational difference between daughters and sons, without considering gender or birth order effects, follows a symmetric distribution centered around 0. However, this distribution shifts towards the negative side in instances when gender and birth order disadvantages are present. In other words, without gender and birth order effects, the distribution of the average education difference between daughters and sons exhibits first-order stochastic dominance over the distribution in cases with such effects (see Figure \ref{with_without_bias} for non-educated parents). The distribution with no gender and birth order effects can be achieved through the substitution effect of a targeted educational cost reduction policy. Note however that such scenario did not change significantly the overall average within-family disparities in children's educational attainment among non-educated parents.

\textit{\textbf{Counterfactual 2:} At the intensive margin, the average educational attainment difference between firstborn daughters and second born sons is equal to 0 in the presence of gender and birth order effects when the firstborn daughter's ability draw is $\approx 13\%$ (resp. 8\%) higher than the ability draw of the second born son for households with non-educated head of household, (resp. households with college educated head of household).}

In this counterfactual analysis, my primary goal is to assess how the unobserved source of inequality interact with the observed sources in the educational attainment choice within-family. I do this by quantifying the additional ability needed to counterbalance educational inequality due to gender and birth order effects. First, I compute the extra ability needed by daughters and older siblings to offset the effect of gender and birth order on their educational attainment. In order to do that I solve the household maximization problem in equation \ref{eq1} with (using $\Hat{\theta}$) and without (setting $\theta = 0$) gender and birth order effects for a grid of relative ability of children for two-child families with a firstborn daughter and a second-born son, and compute the following quantities:

    \begin{enumerate}
    \item Ability of the firstborn daughter relative to the second born son at which the average difference between daughter's and son's education is equal to zero even in the presence of gender and birth order effects.
    \item The change in inequality due to gender and birth order effects, by level of relative ability of the firstborn daughter.
    \end{enumerate}

\begin{figure}[H]
 \begin{minipage}[b]{0.45\linewidth}
   \includegraphics [width=8.5cm]{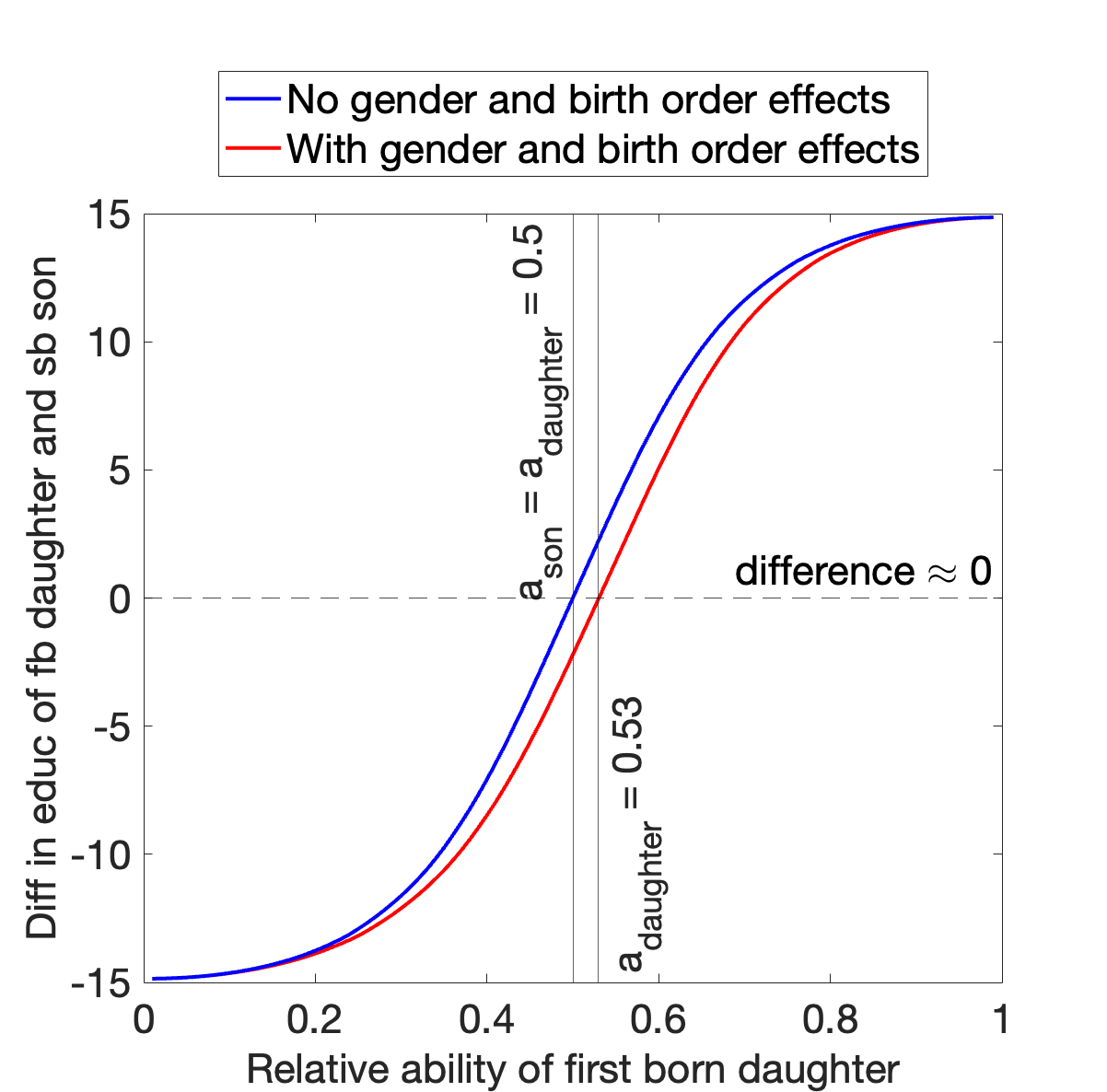}
   
   (a) Non-educated head of household
 \end{minipage}
 \quad
 \begin{minipage}[b]{0.45\linewidth}
     \includegraphics [width=8.5cm]{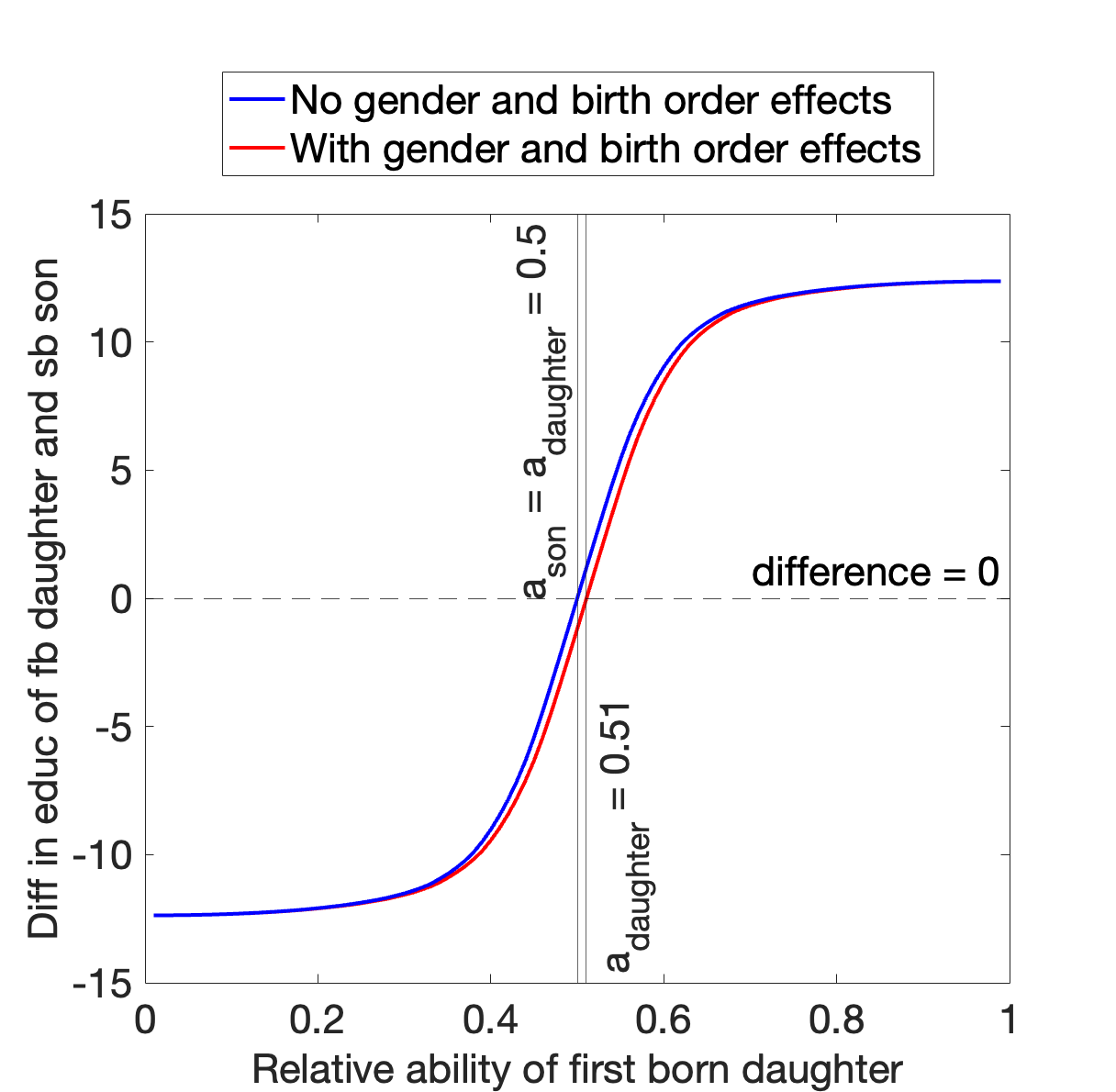}
     
     (b) College educated head of households
\end{minipage}
\caption{Effect of gender and birth order on inequality ($N_c = 2$)}
 \label{gender_bias_ability_hh_educ}
\end{figure}

Figure \ref{gender_bias_ability_hh_educ} presents the results for non-educated and college educated parents with low financial, social and cultural constraints. It suggests two main conclusions. First, for the same ability draws, gender and birth order effects reduce the education attainment of the first born daughter by $\approx 2.2$ years and 1.2 years for non-educated parents and college educated parents respectively. Second, the average education difference between firstborn daughters and second born sons is equal to 0 in the presence of gender and birth order effects when the firstborn daughter's ability draw is $\approx 13\%$ (resp. 8\%) higher than the ability draw of the second born son for households with non-educated head of household, (resp. households with college educated head of household). 

\subsubsection{Counterfactual 3}
This counterfactual is analyzed only among non-educated parents only, and the outcomes are compared to college-educated parents' observed outcomes.

\paragraph{Barriers to School Entry for all Children: In Theory\\}

\begin{figure}[H]
 \centering
 \begin{minipage}[b]{0.45\linewidth}
   \includegraphics [width=8.5cm]{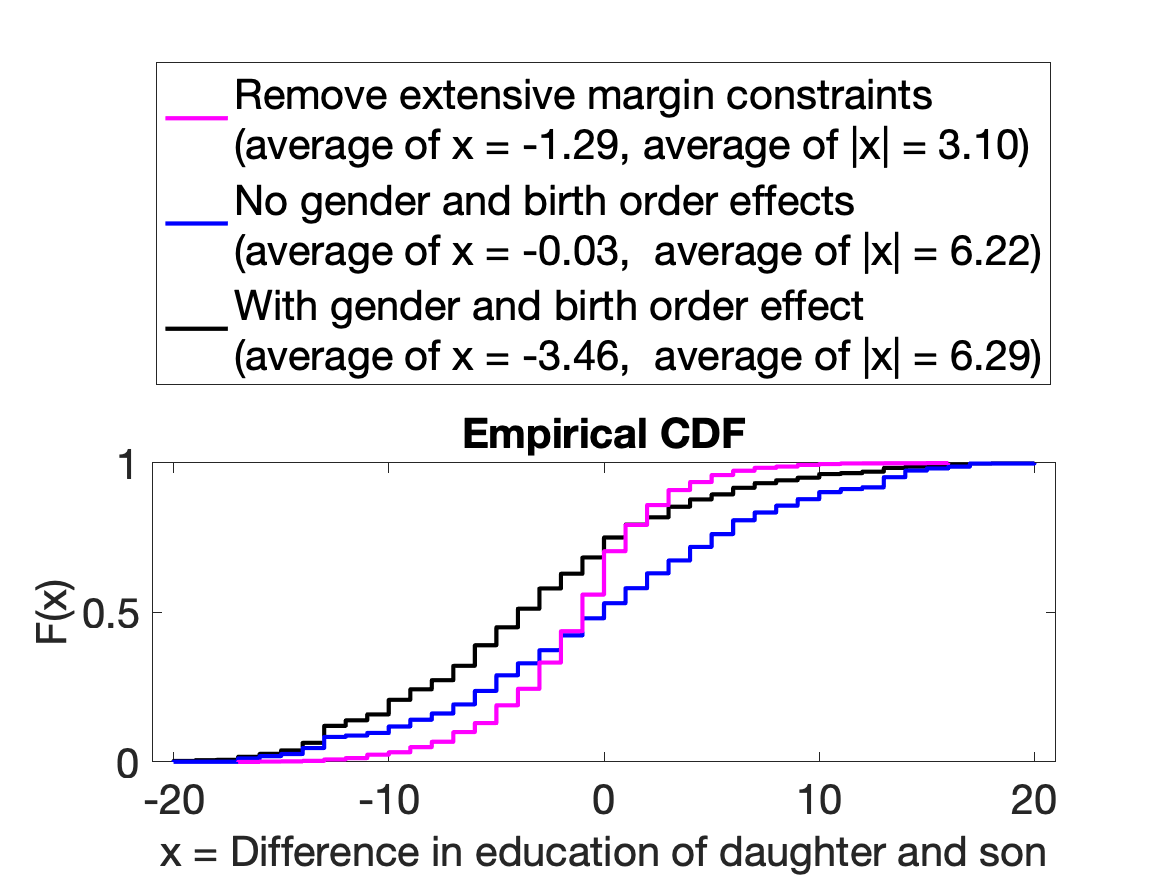}
   
   (a) Daughter v.s son
 \end{minipage}
 \quad
 \begin{minipage}[b]{0.45\linewidth}
     \includegraphics [width=8.5cm]{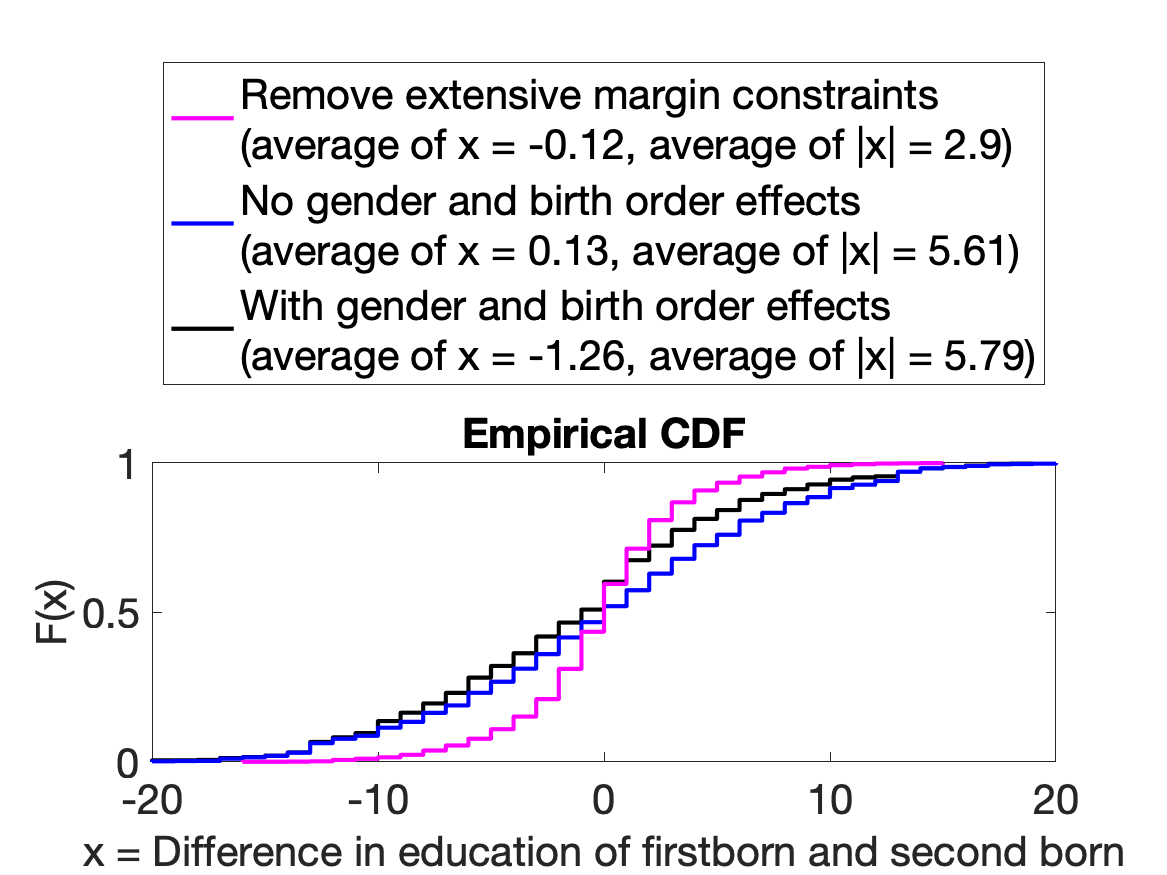}
     
     (b) Firstborn v.s second born
\end{minipage}
\caption{Distribution of the difference in children's education for non-educated parents ($N_c = 2$) [Observed vs. with no barriers to school entry for all children]}
 \label{policy_ce}
\end{figure}

The objective of this section is to examine the effectiveness of a counterfactual focusing on removing barriers to school entry, primarily addressing obstacles arising from parental decisions, to ensure that every child is enrolled in the school system. Theoretically this means setting $\nu_h$ to 0 for all households. Figure \ref{policy_ce} displays the distribution of the difference in education between daughters and sons in panel (a), and between firstborn and second-born children of the same gender in panel (b), across three distinct situations.
The elimination of barriers to school entry reduces gender and birth order effects and overall average within-family disparities in children's educational attainment. In particular, the distribution of the difference between daughters' and sons' (resp. firstborn and second born children's) educational attainments, after removing barriers to school entry for all children, second order stochastically dominates both the distributions with and without gender (and birth order) effects. This means that, compared to the situations there is no gender and birth order effects, removing barriers to school entry leads to more favorable and equitable educational outcomes. The overall distribution shifts in a way that is consistently better, resulting in a notable reduction in average inequality in the sample.

In summary removing barriers to school entry for all children reduces part of the gender and birth order effects on the within-household educational inequality. In addition, overall average within-household inequality is reduced by 50\%.  However, a significant share ($\approx 40\%$) of the average inequality is attributable to gender effect. In particular, it reduces the gender effect by 60\% and birth order effect by 78\%. That is consistent with the fact that gender effect is equally present in both the extensive and intensive margin where as the birth order effect is mostly present in the extensive margin.

\paragraph{Barriers to School Entry for all Children: In Practice\\}

\begin{figure}[H]
 \centering
  \includegraphics [width=15cm]{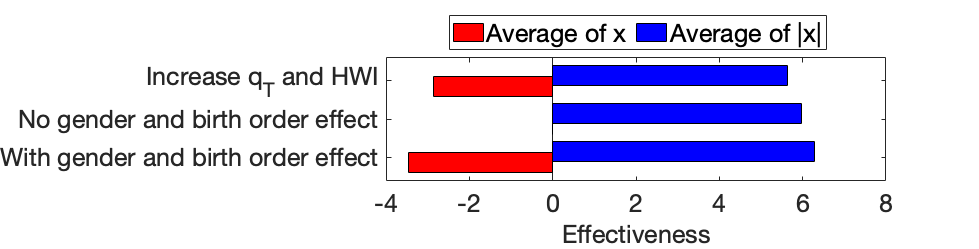}
  \caption{Effects of increase in income and educational resources}
  \label{remover_barriers}
\end{figure}

In the previous section, I set the financial, social and cultural constraint parameter to low for all households and analyze how within-family disparities in children's educational attainment is affected. In this section, I analysis how this theoretical situation can be achieved with practical policy tools. Recall that financial, social and cultural constraints are determined by parents' observable characteristics such as their area of residence (rural vs. urban), their sector of occupation (agricultural vs. non-agricultural), their religion (Christian vs. non-Christian), their level of wealth (low vs. high), and the average educational attainment of their children (a proxy for educational resources available).  Some of these characteristics can be changed through a policy intervention. In particular, the financial constraints---wealth level and education resources. Observed data evidence show that children of college-educated parents with two adult children have an average education level of 14.5 years, compared to 9.2 for children of non-educated parents. In addition, average HWI [Household Wealth Index] is 0.03 among non-educated parents compared to 1.06 among college educated parents.  In this section, I increase these two quantities for non-educated parents to the same level as college educated parents. The results show a decrease by 10\% in the average disparities in children's educational attainment within-family--- not nearly as low as college-educated parents’ level--- with 51\% of the remaining inequality due to gender effects (see Figure \ref{remover_barriers}). What happened is that the increase in income and educational resources was successful at removing barriers to school entry for children with non-educated parents [we observed a significant increase in the proportion of non-educated children], however, it also created more room for gender bias to create inequality in the intensive margin. And that is why it was not as effective as expected. In summary, some combination of awareness campaigns or laws promoting gender equality in education and an increase in households’ educational resources would be effective. 

The following Table summarize the effectiveness of these different education policies counterfactual.

\begin{table}[H]
    \caption{Summary of Education Policies Counterfactual}
    \begin{center}
    {
\begin{center}
\begin{tabular}{l c c} 
 & Eliminate gender \& &  Reduce average    \\
  &   birth order effects &  inequality  \\
    [0.5ex] 
Remove Gender and birth order disadvantages (1) & \checkmark & \tikzxmark \\
 [0.5ex] 
 Remove barriers to school entry (2)&  \tikzxmark  &  \checkmark \\
  [0.5ex] 
 Increase in income and educational resources (3) & \tikzxmark  & \checkmark \\

\end{tabular}
\end{center}
}
    \end{center}
\end{table}




 \section{Conclusion}

In this paper, I examine the interaction between the three empirically known sources of disparities in children's educational attainment within-households. I propose and estimate a structural model of households' distribution of education resources among children, allowing for the influence of factors such as gender, birth order and ability of children. The model not only allows me to decompose, for each relative ability draw, the total observed inequality into parts due to gender, birth order, and ability differences; it also gives a platform for analysing how different education policies affect within-household inequalities.

The construction of the model is motivated by contexts similar to the one of Benin; a setting marked by notable disparities in children's education within-households, coupled with evidence of gender and birth order disadvantages. To ensure tractability, certain aspects of the parental decision-making process regarding education resources distribution are omitted. Notably, the model adopts a static approach, although the education decision of children is inherently dynamic. The primary objective of the paper being to rationalize the observed differences in children's education, attributing them to gender effect, birth order effect, or variations in innate ability draws; despite its static nature, the model proves relevant, as it effectively incorporates and analyzes the interactions among these three factors. Additionally, the paper attributes any unexplained differences in children's education, not accounted for by gender and birth order, to differential draws of innate ability. However, it acknowledges the potential influence of other unobserved factors, such as varying favoritism for wives in polygamous households, which could lead to increased parental investment in the education of specific children. In recognizing this, the interpretation of unexplained inequality within-households is acknowledged as an upper bound of the effect of differential ability.

In light of the findings in this paper, we can expect a reduction in the opportunity cost of girls education such as education support in the form of cash transfers, scholarships, and school kits for girls; to reduce within-household inequality in children's education that is due to gender disadvantage. Additionally, a reduction in the opportunity cost of education for firstborn, such as cash transfers and school kits, to young parents (first-time parents) or scholarships for firstborn children; is expected to reduce within-household inequality in children's education that is due to birth order disadvantage. However, these two policies need to be combined for an effective reduction in disadvantaged-based inequality. This is due to the possibility of displacement of disadvantage from one group to another. In particular, if the policy only targets firstborn children, the disadvantage against daughters might increase, and vice versa. Finally, a compulsory education policy is the most effective in reducing average inequality in the sample. However, as long as there is budget constraint, as we move toward maximum education for everybody, there will always be a positive within-household inequality.
 


\newpage
\section{Tables and Figures}

 \begin{figure}[H]
\begin{center}
 \begin{minipage}[b]{0.45\linewidth}
   \includegraphics [width=8.5cm]{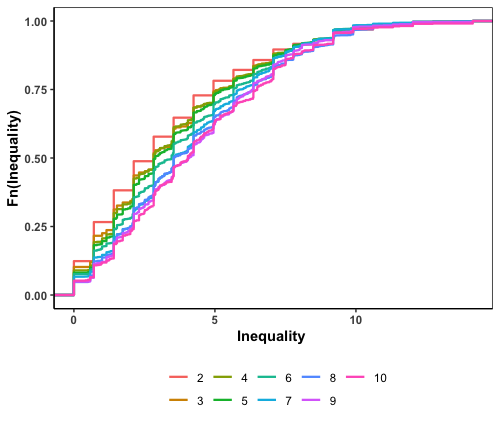}
 \end{minipage}
  \caption{Distribution of inequality by number of children.}
 \label{range_nc}
 \end{center}
\end{figure}

 \begin{figure}[H]
\begin{center}
 \begin{minipage}[b]{0.45\linewidth}
   \includegraphics [width=8cm]{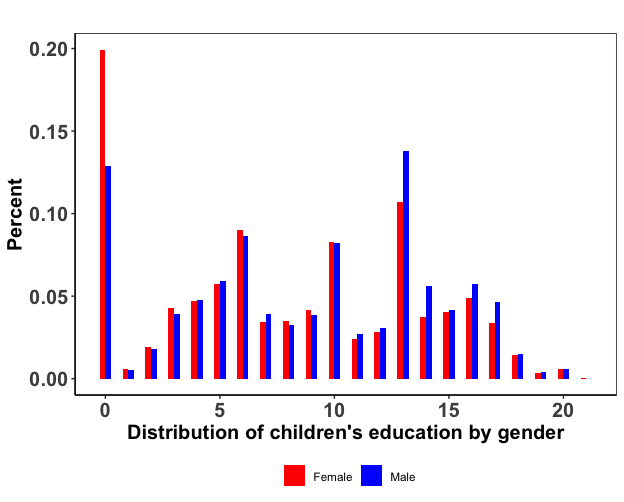}
   
   (a) By gender
 \end{minipage}
 \quad
 \begin{minipage}[b]{0.45\linewidth}
     \includegraphics [width=8cm]{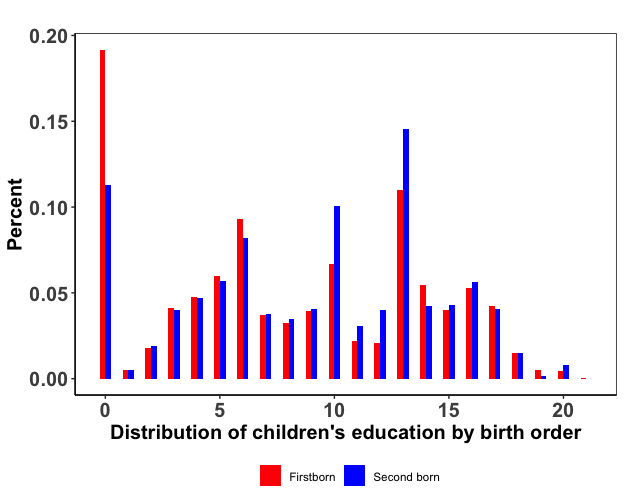}
     
     (b) By birth order
\end{minipage}
\caption{Distribution of children's education for number of $N_c = 2$ (Benin, 2013)}
 \label{dist_educ}
  \end{center}
\end{figure}


\begin{figure}[H]

 \begin{minipage}[b]{0.45\linewidth}
   \includegraphics [width=8cm]{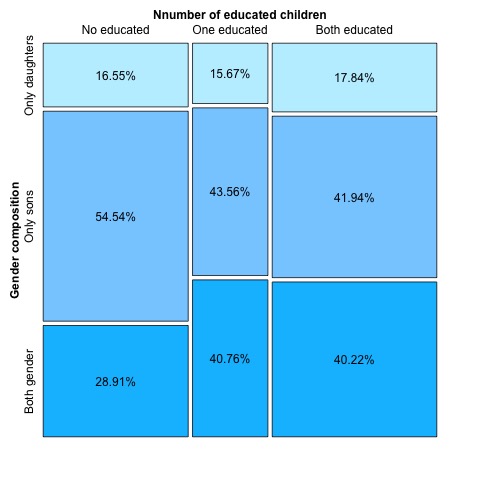}
   
   (a) Gender composition of households as function of number of uneducated children
 \end{minipage}
 \quad
 \begin{minipage}[b]{0.45\linewidth}
     \includegraphics [width=8cm]{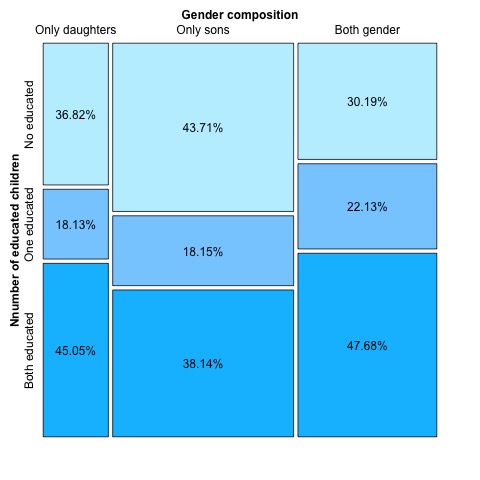}
     
     (b) Number of uneducated children as function of gender composition of households
\end{minipage}
    \caption{Number of uneducated children by gender composition ($N_c =2$)}
     \label{household_type_gender_comp}
\end{figure}


 \begin{figure}[H]
\centering
    \subfloat[$\theta_1^{low}$]{\includegraphics[width=0.45\textwidth]{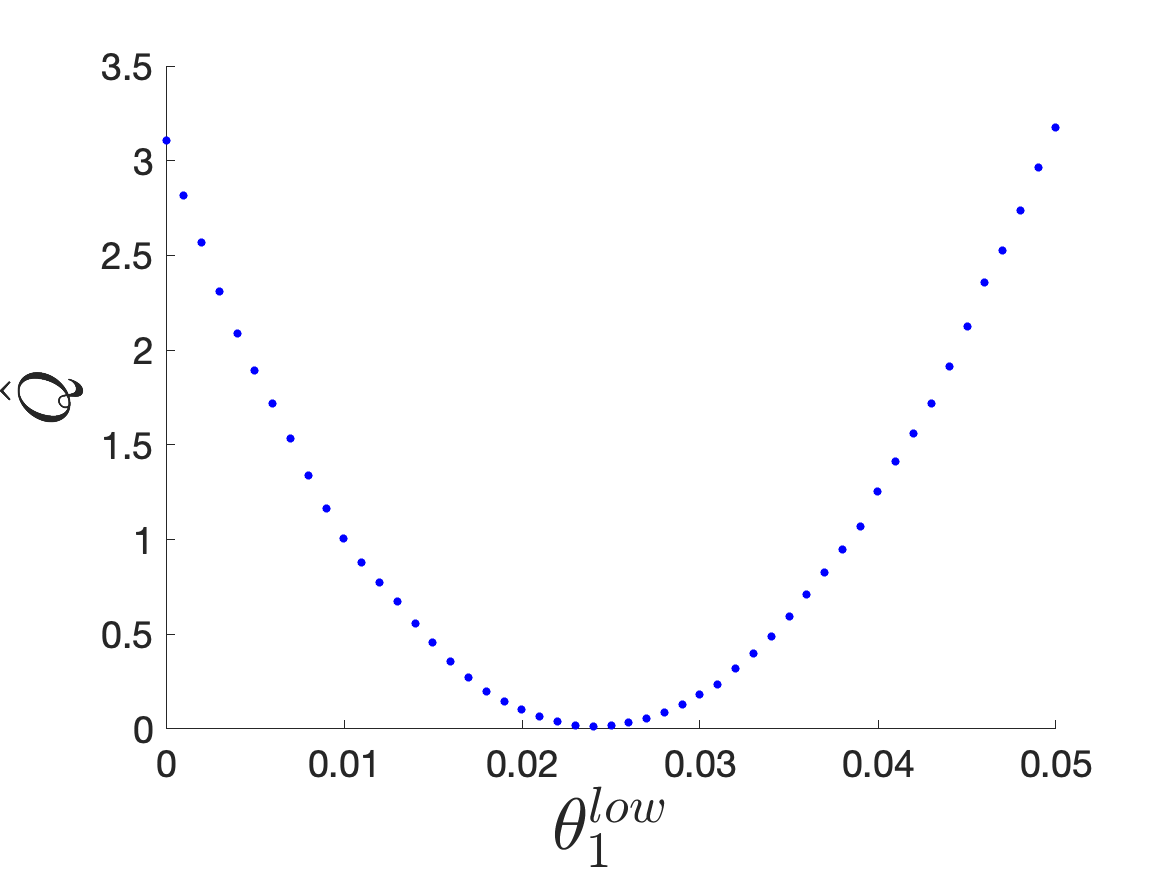}}
    \subfloat[$\alpha_1^{low}$]{\includegraphics[width=0.45\textwidth]{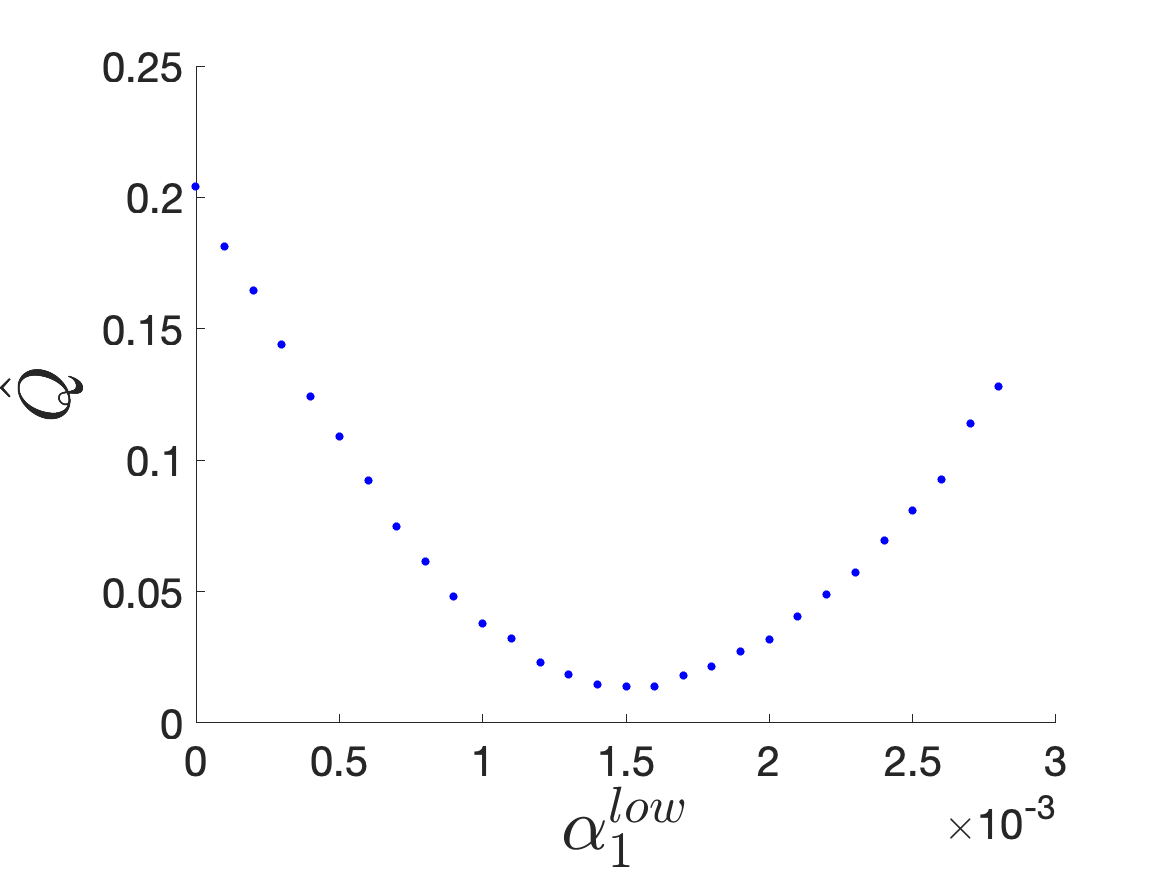}}

 \end{figure}

 \begin{figure}[H]
 \centering
 \begin{minipage}[b]{0.45\linewidth}
   \includegraphics [width=8cm]{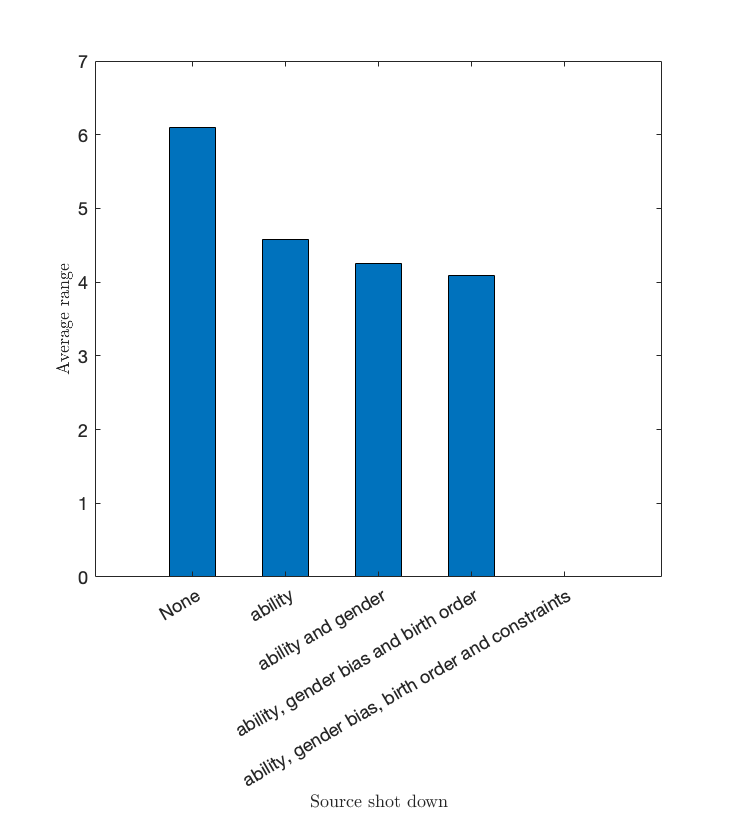}
   
 \end{minipage}
 \quad
 \begin{minipage}[b]{0.45\linewidth}
     \includegraphics [width=8cm]{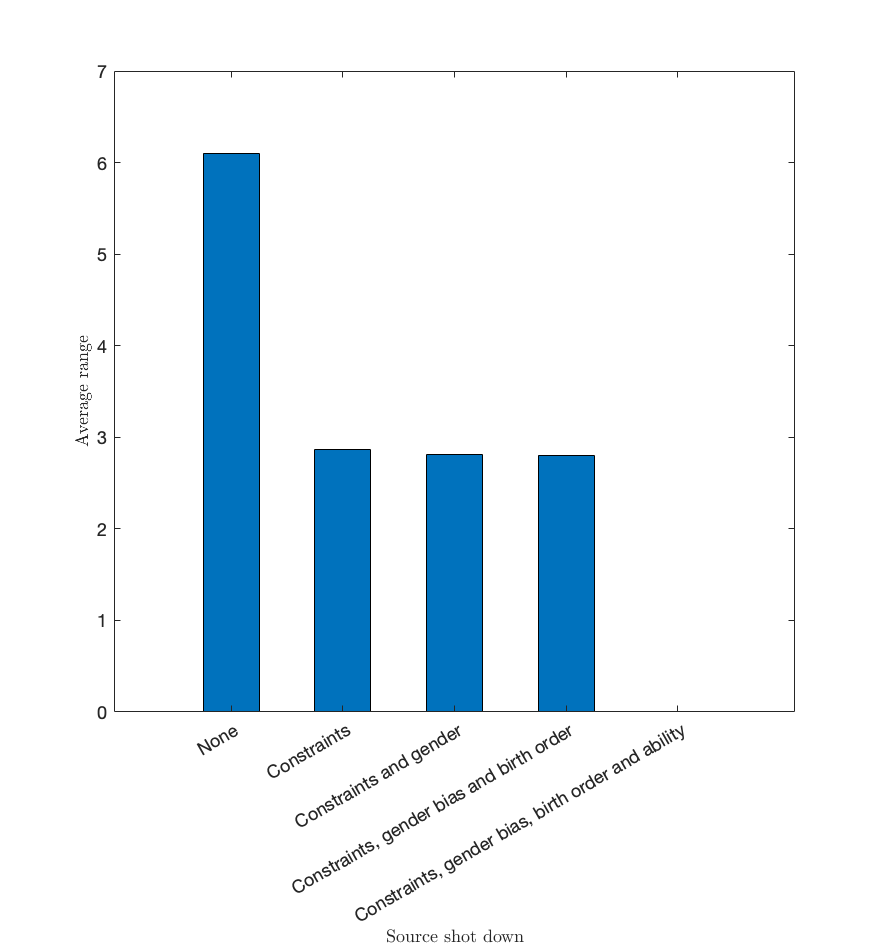}
     
\end{minipage}
\caption{Average Inequality after subsequent shot down of sources of inequality}
 \label{shot_down}
\end{figure}

\begin{figure}[H]
 \centering
 \begin{minipage}[b]{0.45\linewidth}
   \includegraphics [width=8cm]{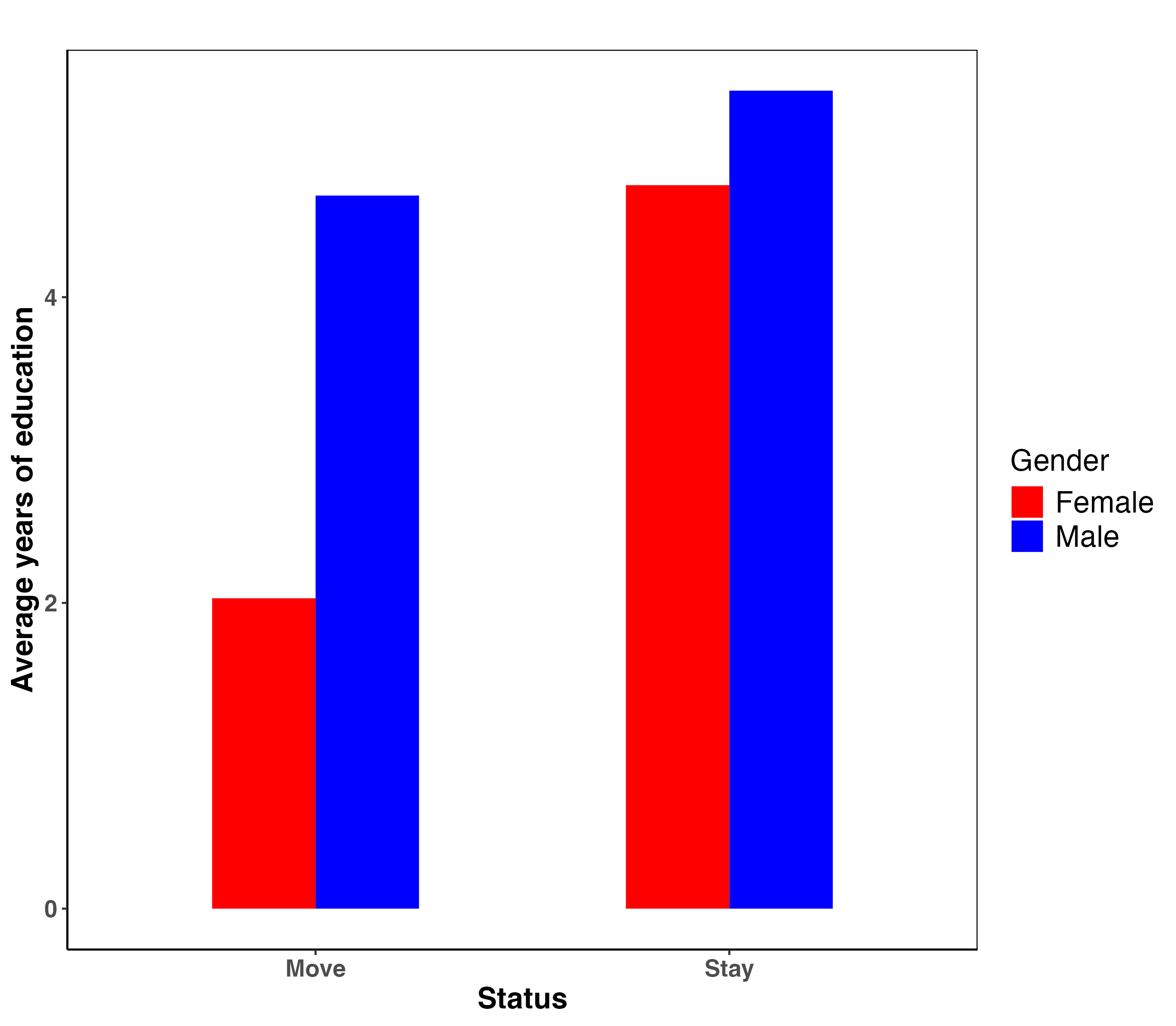}
   
   (a) Mean
 \end{minipage}
 \quad
 \begin{minipage}[b]{0.45\linewidth}
     \includegraphics [width=8cm]{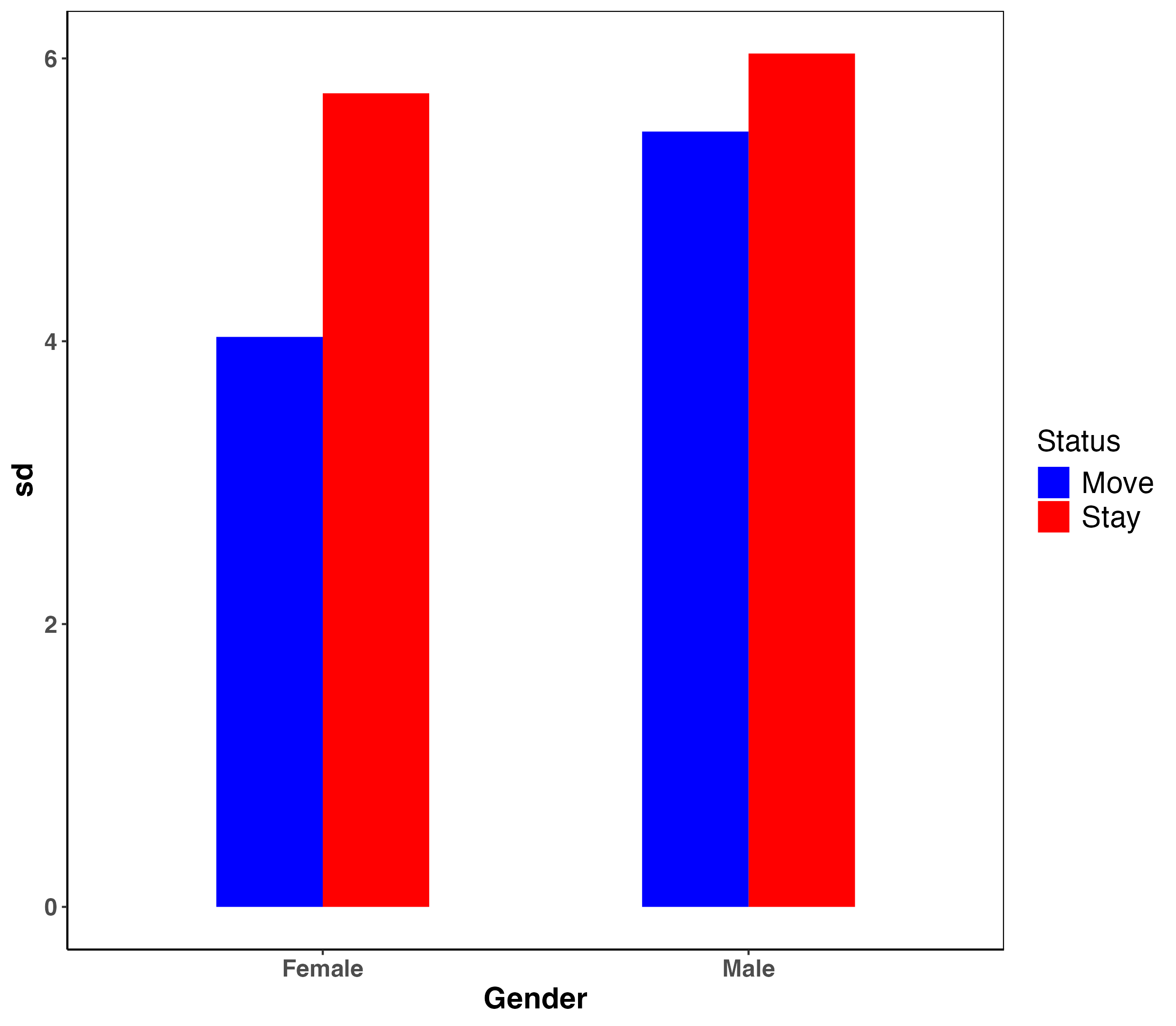}
     
     (b) Standard Deviation
\end{minipage}
\caption{Mean and standard deviation of adult between 25 and 40 years old}
 \label{move_saty_mean}
\end{figure}

 \begin{figure}[H]
 \centering
 \begin{minipage}[b]{0.45\linewidth}
   \includegraphics [width=8cm]{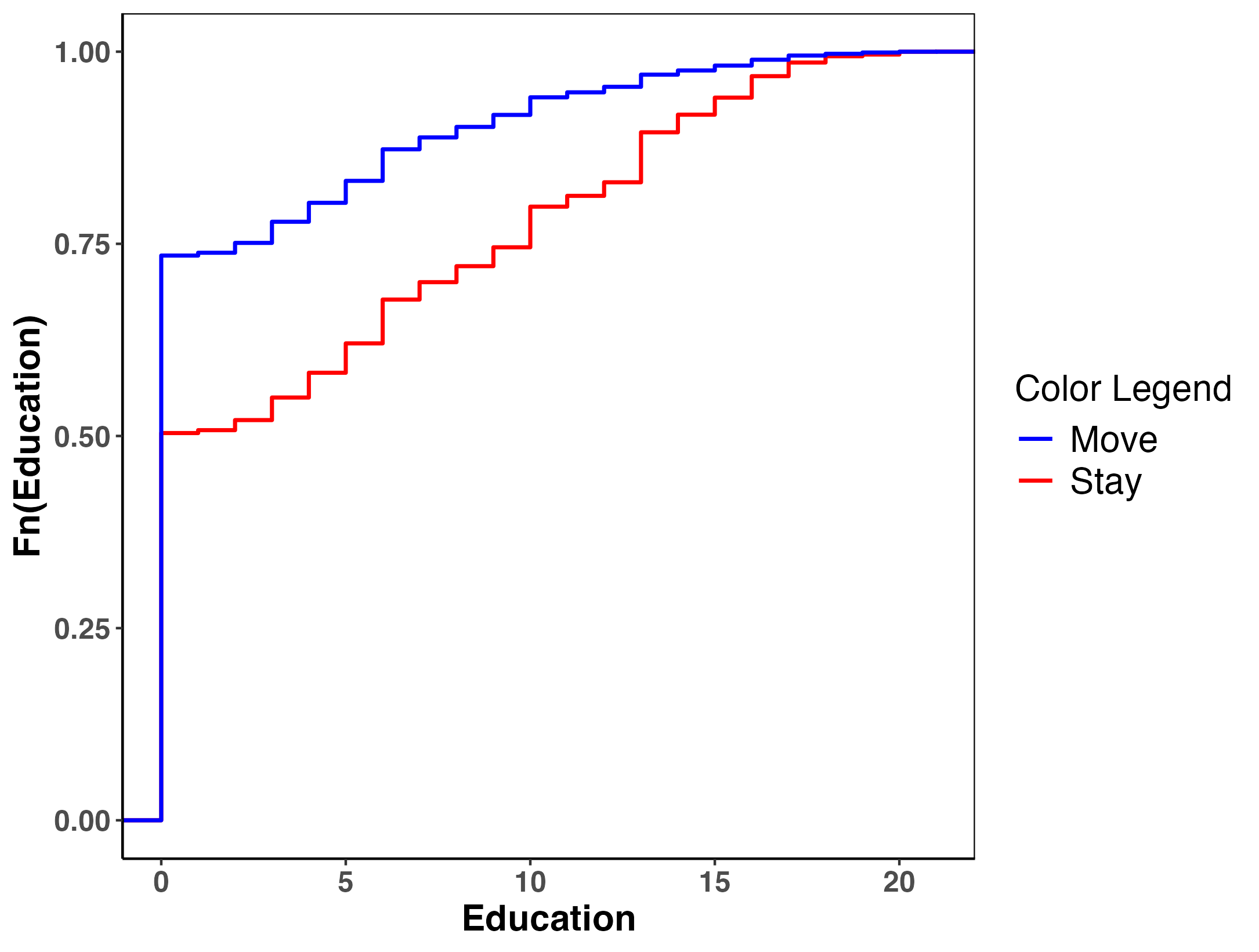}
   
   (a) Women
 \end{minipage}
 \quad
 \begin{minipage}[b]{0.45\linewidth}
     \includegraphics [width=8cm]{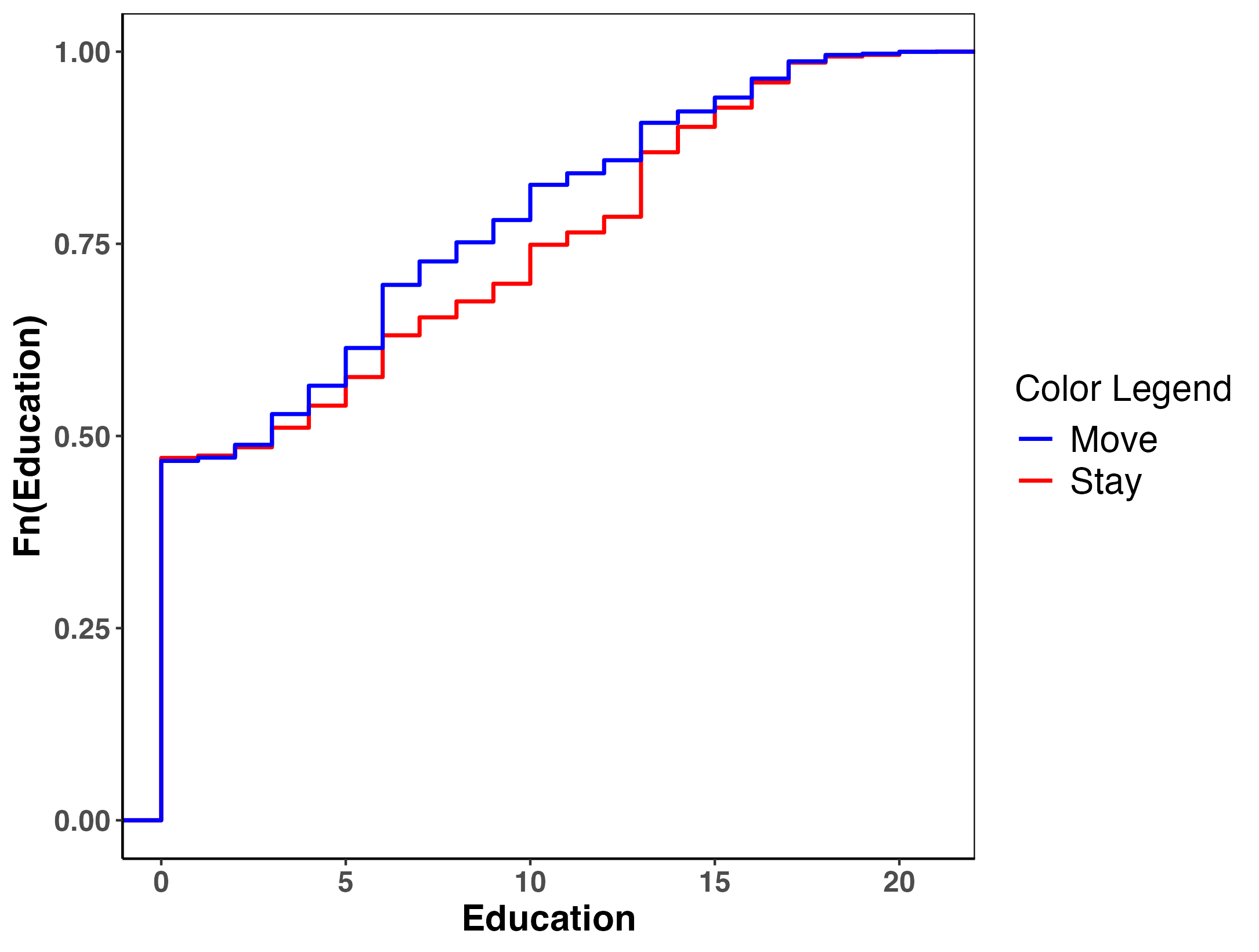}
     
     (b) Men
\end{minipage}
\caption{Empirical cdf of the years of education of adults between 25 and 40 years old}
 \label{move_saty}
\end{figure}



\newpage

\section*{Tables}

\begin{table}[H]
    \begin{center}
    \caption{Educational inequality on average education}
    {
\scalebox{0.83}{
\begin{tabular}{l c c c c c c}
\hline
 & (1) & (2) &  (3) &  (4) &  (5) &  (6) \\
 & \multicolumn{2}{c}{Full sample} & \multicolumn{2}{c}{Non-educated hh} &
  \multicolumn{2}{c}{ College educated hh} \\
\hline
Average education ($\Bar{q}$)     & $1.25^{*}$        & $1.04^{*}$        & $1.44^{*}$        & $1.26^{*}$        & $0.87^{*}$        & $0.64^{*}$        \\
                & $ [ 1.24;  1.27]$ & $ [ 1.02;  1.07]$ & $ [ 1.42;  1.46]$ & $ [ 1.23;  1.29]$ & $ [ 0.81;  0.94]$ & $ [ 0.54;  0.75]$ \\
$\Bar{q}^2$ & $-0.08^{*}$       & $-0.06^{*}$       & $-0.09^{*}$       & $-0.08^{*}$       & $-0.05^{*}$       & $-0.04^{*}$       \\
                & $ [-0.08; -0.08]$ & $ [-0.06; -0.06]$ & $ [-0.09; -0.09]$ & $ [-0.08; -0.08]$ & $ [-0.05; -0.04]$ & $ [-0.04; -0.03]$ \\
\hline
R$^2$           & $0.69$            & $0.71$            & $0.73$            & $0.74$            & $0.61$            & $0.62$            \\
Covariates     &         & \checkmark            &             & \checkmark            &            & \checkmark          \\
Num. obs.       & $32729$           & $32729$           & $19558$           & $19558$           & $1438$            & $1438$            \\
\hline
\multicolumn{7}{l}{\scriptsize{$^*$ Null hypothesis value outside the confidence interval. Covariates include parents' education, area of residence, religion, number of children,}}\\
\multicolumn{7}{l}{\scriptsize{and gender composition of the household.}}
\end{tabular}}
}

    \label{table:mean_variance_educ}
    \end{center}
\end{table}

\newpage

\section{Appendix A: Robustness of Estimates to Missing Siblings}
 
 The main sample used for this analysis comprises adult children who were living in the same household as their parents during the census period. This sample represents a specific subgroup within the larger population of adult children. Importantly, the decision for children to leave the parental home is often influenced by factors such as their occupation and educational accomplishments, making it an endogenous process. Moreover, the motives for leaving home frequently differ between daughters, commonly associated with marriage, and sons. Given these dynamics, there's a potential for bias in our estimates. This would be particularly concerning if, firstly, the children who remained at home are more similar to each other, and secondly, if they significantly differ from those who moved out.
The wide range in both educational attainment and gender among children residing in the same household as their parents suggests that the first concern may not be significant.


The second concern could lead to either overestimation--- if women who moved out are more educated compared to ones who stayed and men who moved out are less educated compared to the ones who stayed--- or underestimation--- if women who moved out are less educated and men who moved out are more educated, compared to those who remained at home.

In this section, I delve into the potential bias in estimating the effect of gender disadvantage on within-household inequality. To investigate this, I compare the educational attainment of adult women and men living in the same households as their parents to those who have moved out. The mean comparison between these two groups is presented in Figure  \ref{move_saty_mean}. This comparison suggests that the difference in average education between men and women is more pronounced in the sub-sample that is not included in my analysis. In addition we observe a clear first order stochastic dominance between the empirical distribution of the education of adult female living in the same households as their parents and those who do not (see Figure \ref{move_saty}). Such first order stochastic dominance is not as pronounced among men.   As a result, it implies that, if anything, I may be underestimating the effect of gender disadvantage. Consequently, my estimate of gender disadvantage can be interpreted as an estimate of the lower bound of the true parameter.

 
A similar argument to the one presented in the previous section also applies to the birth order disadvantage parameter. The decision for children to move out is closely linked to their age, with older children being more inclined to leave their parents' household. Consequently, we may have a selected sample of younger children in some households. In specific cases, children referred to as firstborns in certain households might actually be of a higher birth order. Additionally, more accomplished younger siblings may have already moved out. It's important to note that both of these situations would potentially bias our estimate of the birth order disadvantage parameter downward. In particular, if we maintain the assumption that firstborn children receive less education than other children, the older firstborn children with less education--- who already moved out of the family house--- are not included in our analysis. This leads to an underestimation of the birth order effect. In addition if high educated children are more likely to move out--- because they have better and stable socio-economic status--- we observe uniformly less educated children in our sample. In summary, we are likely to have in our sample, less educated children. On one hand, if the age effect dominates the education effect, we will have less firstborn children in our sample, which biases our estimate downward. On the other hand, if the education effect dominates the age effect, we have less second born children in our sample,  which also biases our estimate downward.

\section*{Appendix B}

\subsection{Appendix B1: Summary Statistics}

\begin{table}[H]
    \begin{center}
    \caption{Summary Statistics for all the individuals between 25 and 40 years old residing in the same household as their parents}
    {

\begin{tabular}{l l c c} 
&  & Mean &  Standard Deviation    \\
    [0.5ex] 
    \hline
& All & 5.11 & 5.94  \\
 [0.5ex] 
& Rural & 3.42 & 5.18 \\
 [0.5ex] 
 &Urban & 7.05 & 6.14  \\
 [0.5ex] 
 & Farmer& 2.91 & 4.83 \\
 [0.5ex] 
Education & Non-farmer & 6.75 & 6.15 \\
 [0.5ex] 
  & Non-Christian & 3.43 & 5.23 \\
 [0.5ex] 
 & Christian & 7.13 & 6.1 \\
 [0.5ex] 
& Non-educated parents & 3.45 & 5.16   \\
 [0.5ex] 
&College-educated parents  & 13.8 & 4.42  \\
 [0.5ex] 
\hline
 \multicolumn{2}{l}{No education} & 0.48 &  \\
\multicolumn{2}{l}{Female} & 0.38 &  \\
 [0.5ex]
 \multicolumn{2}{l}{Rural} &  0.53&  \\
 [0.5ex]
\multicolumn{2}{l}{Farmer} & 0.43 &  \\
 [0.5ex]
 \multicolumn{2}{l}{ Christian}& 0.45 &  \\
 [0.5ex] 
 \multicolumn{2}{l}{ Non-educated parents} & 0.72 &    \\
 [0.5ex] 
\multicolumn{2}{l}{ College-educated parents}  & 0.03 &   \\
 \multicolumn{2}{l}{ HWI} &0.03 & 0.98 \\
 [0.5ex] 
 \hline
 \multicolumn{2}{l}{ Number of children} & 5.35 & 4.1 \\
 [0.5ex]
 \multicolumn{2}{l}{ Number of children btw 25 and 40}& 2.1 & 1.51 \\
 [0.5ex] 
 \multicolumn{2}{l}{ Average educ in hh is 0} & 0.4 &  \\
 [0.5ex] 
 \multicolumn{2}{l}{ hh with only one child btw 25 and 40} & 0.45 &   \\
 [0.5ex] 
 \hline
\multicolumn{2}{l}{Number of observation} & \multicolumn{2}{c}{253,734} 
\end{tabular}
}
    \label{table:desc_stats}
    \end{center}
\end{table}

\subsection{Appendix B2: Model for households with Number of Children equal 3}

For $N_c =3$, let $ne \in \{0, 1, 2\}$ be number of children with no formal and education, and $\nu_h^L = 1\{ne_h = 0\}$, $\nu_h^M = 1\{ne_h = 1\}$, and $\nu_h^H = 1\{ne_h = 2\}$
\begin{equation}
    U(q_h, \theta) =  \nu_h^{L} \Big[ \sum_{i = 1}^{N_{c}} a_i. (q_i)^{\delta_{i, h}^{L}} - \alpha_{i}^{L} q_{i} \Big ] +\nu_h^{M} \Big \{  \sum_{i = 1}^{N_{c}} \Big [ e_{i}^{M}.  \big ( a_i. (q_i)^{\delta_{i, h}^{M}} - \alpha_{i}^{M} q_i \big ) \Big] \Big \} +  \nu_h^{H} \Big \{ \sum_{i = 1}^{N_{c}} \Big [ e_{i}^{H}.   \Big [ \big ( a_i. (q_i)^{\delta_{i, h}^{H}} - \alpha_{i}^{H} q_i \big )  \Big] \Big \}
\end{equation}

where,
\[\delta_{i, h}^{type} =  \gamma -  \theta_1^{type} Female_i \frac{1}{N_c - 1} \sum_{\{i, j \in h\}, j \neq i}(1 - Female_j), \text{ with } type \in \{low (L), medium (M), high (H)\}\]

\[ e_{i}^{M} = 1\{  \exists j:  \big ( a_i. (q_i)^{\delta_{i, h}^{M}} - \alpha_{i}^{M} q_i \big ) > \big ( a_j. (q_i)^{\delta_{j, h}^{M}} - \alpha_{j}^{M} q_j \big ) \},\] 
\[  e_{ i}^{H} = 1\{ \big ( a_i. (q_i)^{\delta_{i, h}^{H}} - \alpha_{i}^{H} q_i \big ) > \big ( a_j. (q_i)^{\delta_{j, h}^{H}} - \alpha_{j}^{H} q_j \big ), \forall j \neq i \},\] 

The vector of parameters of interest is \[\theta = \Big(\theta_1^{L}, \theta_1^{M}, \theta_1^{H}, \alpha_{(1)}^{L}, \alpha_{(2)}^{L}, \alpha_{(1)}^{M}, \alpha_{(2)}^{M}, \alpha_{(1)}^{H}, \alpha_{(2)}^{H} \Big)\]

$\alpha_{(3)}^{L}$, $\alpha_{(3)}^{M}$, $\alpha_{(3)}^{H}$ are normalized to 0.The dimension of $\theta$ is $1 \times 9$.

For each household $h$, [Parents' educ and number of children hold fixed], let

\begin{itemize}
\item $\nu_{h ne}^{\star}$ be the utility [due to unobserved financial, social  \& cultural constraints] from having $ne$ children with no formal education.
\item $ne \in \{0, 1, \dots, N_c - 1\}$ be the observed discrete outcome of number of children with no formal educ.
\end{itemize}
\[\nu_{hne}^{\star} = X_h' \beta_{ne} + \varepsilon_{hne}\]
[$X_h$ includes a cst, rural, farmer, HWI, religion, gender composition, and average education of children.]

The choice of children with no formal education is 
\[ne \text{ if } \nu_{hne}^{\star} > \nu_{hne'}^{\star} \forall ne' \neq ne \]
\[ne_h = argmax_{ne} X_h' \hat{\beta}_{ne} + \varepsilon_{hne}^{Sim} \text{, with } \varepsilon_{hne}^{Sim} \sim \mathcal{N}(0, \hat{\Sigma}) \]

    \begin{table}[H]
    \begin{center}
   \caption{Estimates of $\Hat{\theta}$,  ($N_c = 3$)}

\scalebox{0.83}{{
\begin{tabular}{lccccccccc} 
 \hline

   \addlinespace
 & $\hat{\theta}_1^{L}$ & $\hat{\alpha}_{1}^{L} - \hat{\alpha}_{2}^{L}$ & $\hat{\alpha}_{2}^{L} - \hat{\alpha}_{3}^{L}$ & $\hat{\theta}_1^{M}$ & $\hat{\alpha}_{1}^{M} - \hat{\alpha}_{2}^{M}$ & $\hat{\alpha}_{2}^{M}  - \hat{\alpha}_{3}^{M}$ & $\hat{\theta}_1^{H}$ & $\hat{\alpha}_{1}^{H} - \hat{\alpha}_{2}^{H}$ & $\hat{\alpha}_{2}^{H} - \hat{\alpha}_{3}^{H}$\\
 \addlinespace
Estimates &0.033$^{**}$ & 0.0026$^{**}$ & 0.0010$^{*}$& 0.0761$^{**}$& 0.0094$^{*}$& 0.0001$^{**}$& 0.1596$^{**}$& 0.0120$^{**}$& 0.0105$^{**}$  \\ [1ex]
\addlinespace
Standard errors & 0.0052  &  0.0006  &  0.0006   & 0.0058  &  0.0016 &   0.0019  &  0.0073  &  0.0019  &  0.0015  \\ [1ex]
\addlinespace
Number of observations &  \multicolumn{9}{c}{3644}  \\ [1ex]

 \hline
 \hline
\end{tabular}}
}


    \label{table:estimates_nc3}
    \end{center}
     \footnotesize{$**$ significant at 5\% level of significance, $*$ significant at 10\% level of significance.}
\end{table}

\subsection{Appendix B3: Generalized Households' utility function for any Number of Children}

 Let  $type \in \{ 0, 1, \dots N_c - 1\}$, where

\begin{itemize}
    \item $0$ corresponds to the least constrained household,

    \item  and $N_c - 1$ corresponds to the most constrained household.
\end{itemize}

\begin{equation}
    U(q_h, \theta) = \sum_{c = 2}^{N} \Big \{ 1\{ N_{c_h} = c \} . U^c(q_h, \theta) \Big \}  \text{ where,}
\end{equation}
\begin{itemize}
     \item $\delta_{i, h}^{type} =  \gamma -  \theta_1^{type} Female_i \frac{1}{N_{c_h} - 1} \sum_{\{i, j \in h\}, j \neq i}(1 - Female_j)$,
    \item $q_h = (q_{1, h}, \dots, q_{N_c, h})$
\end{itemize}

\newpage
\section{Appendix C: Additional Figures}
 \begin{figure}[H]
\begin{center}
 \begin{minipage}[b]{0.45\linewidth}
   \includegraphics [width=7cm]{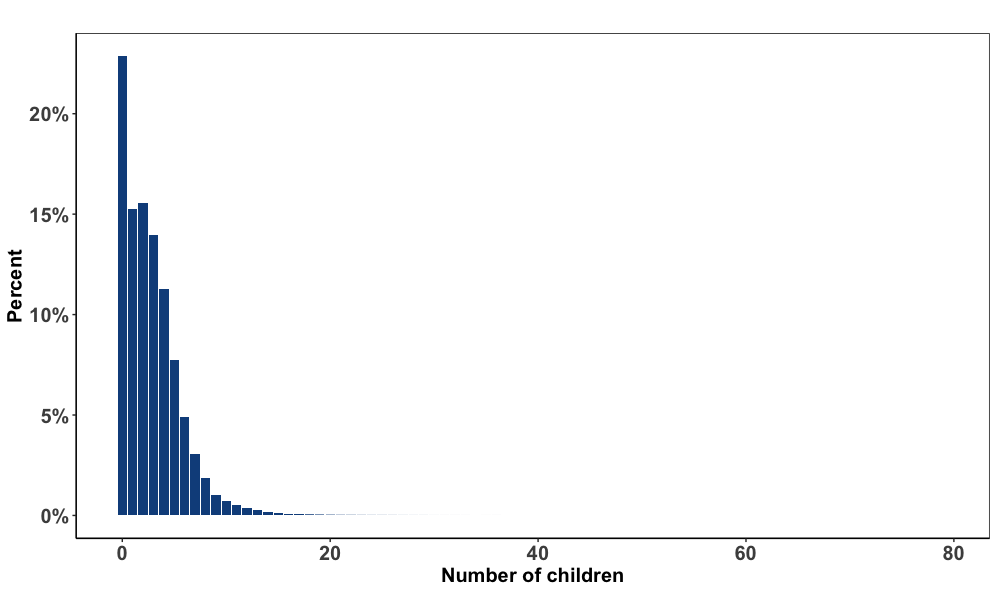}
   (a) All Children.
 \end{minipage}
 \quad
 \begin{minipage}[b]{0.45\linewidth}
    \includegraphics [width=7cm]{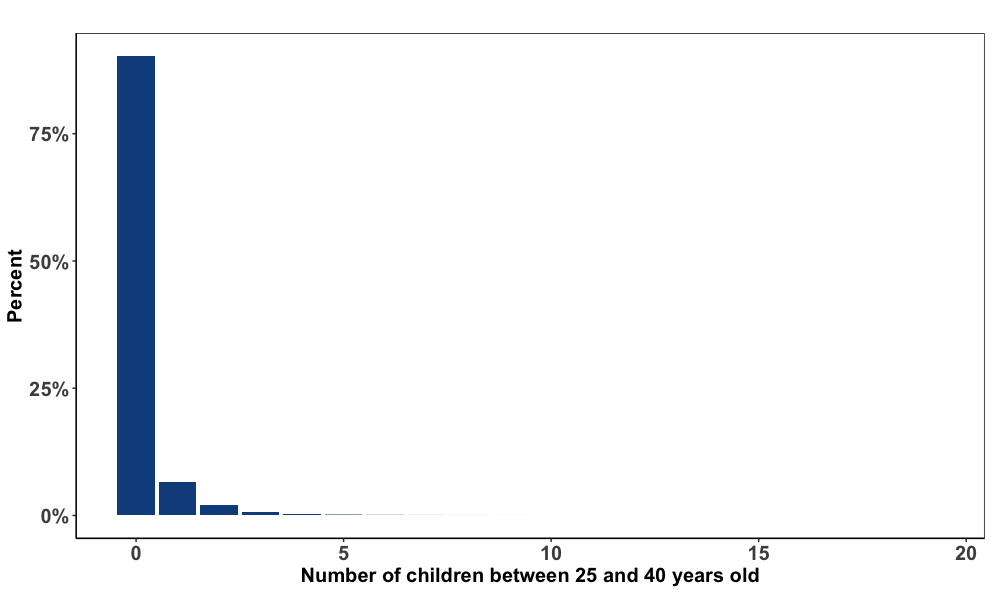}
     (b) Children between 25-40 years old.
\end{minipage}
  \caption{Histogram of number of children.}
 \label{NC_hist_appendix}
    \end{center}
\end{figure}

 \begin{figure}[H]
\begin{center}
 \begin{minipage}[b]{0.45\linewidth}
   \includegraphics [width=7cm]{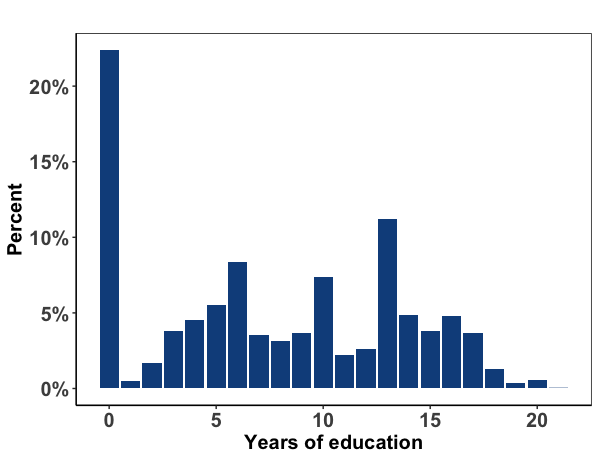}
   (a) Whole sample
 \end{minipage}
 \quad
 \begin{minipage}[b]{0.45\linewidth}
    \includegraphics [width=7cm]{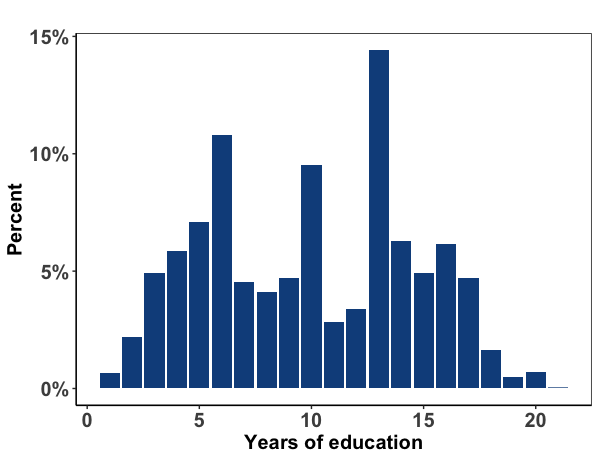}
     (b) Sample of educated individuals
\end{minipage}
  \caption{Histogram of years of education.}
 \label{hist_yrschool}
    \end{center}
\end{figure}

\begin{figure}[H]
 \begin{center}
 \begin{minipage}[b]{0.45\linewidth}
   \includegraphics [width=7cm]{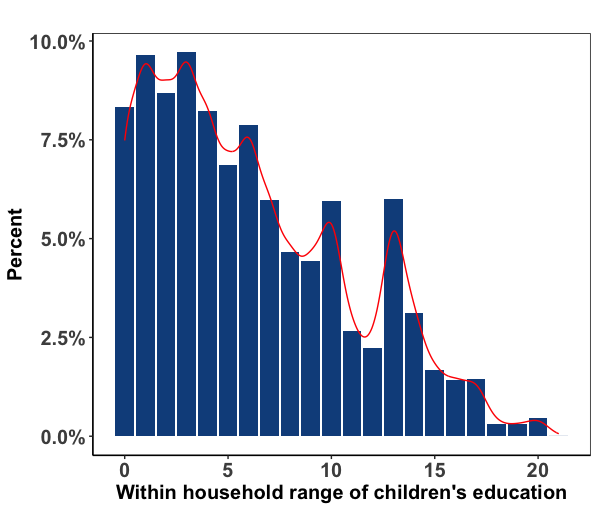}
 \end{minipage}
\caption{Distribution of within-household range of children's education.}
    \end{center}
\end{figure}

 \begin{figure}[H]
\begin{center}
   \includegraphics [width=13cm]{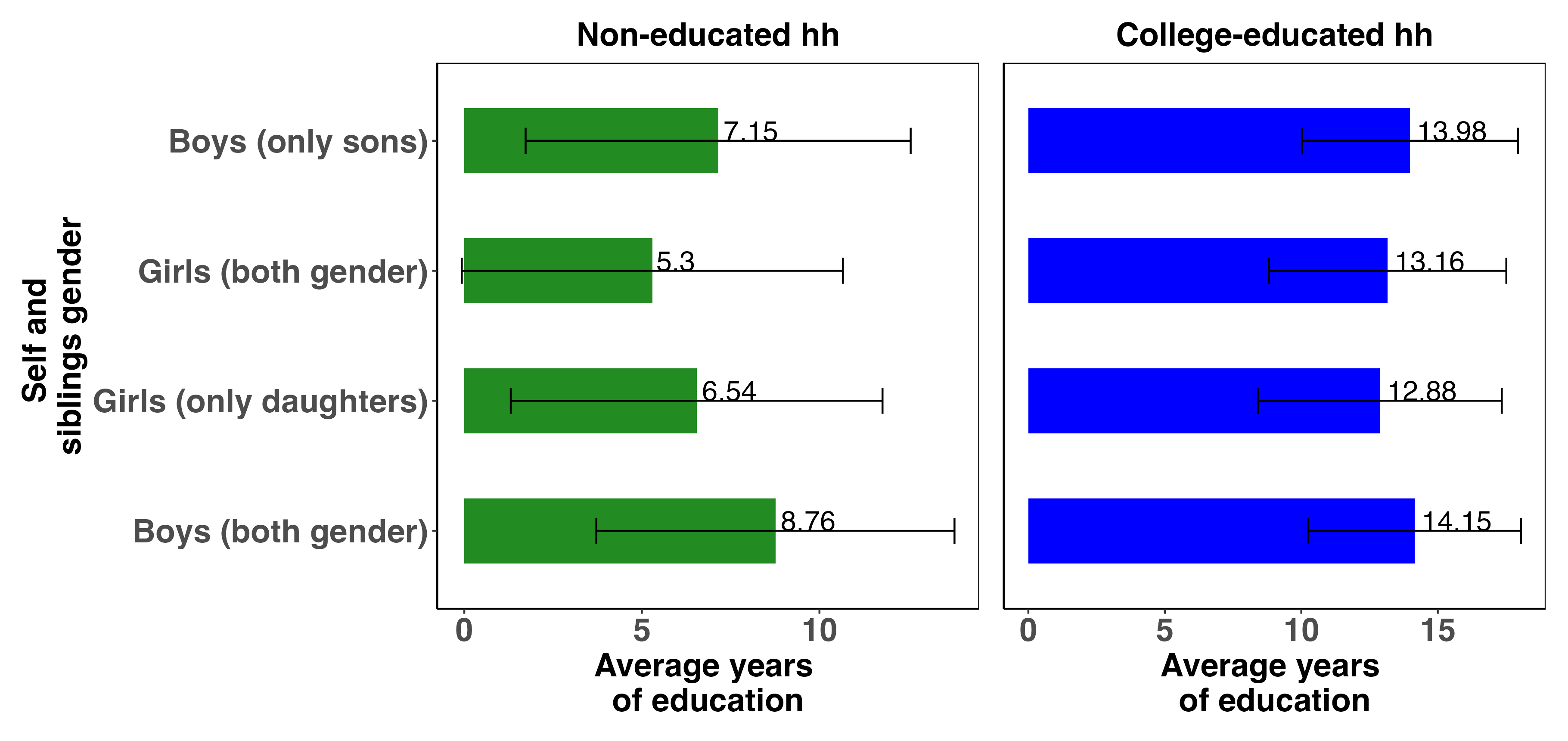}
\caption{Average years of education by gender and households gender composition and parents' education ($N_c = 2$)}
 \label{ineq_gender_appendix}
  \end{center}
\end{figure}

\begin{figure}[H]
\begin{center}
 \begin{minipage}[b]{0.45\linewidth}
   \includegraphics [width=9cm]{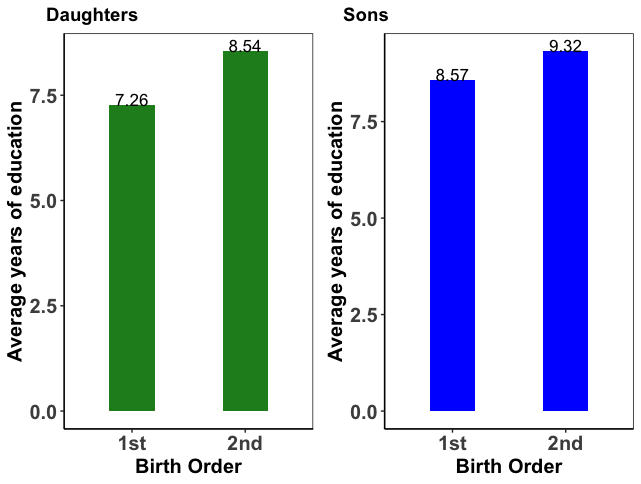}
 \end{minipage}
\caption{Average education by birth order for households with $N_c = 2$ (Benin, 2013)}
 \label{educ_diff_bo_nc2}
    \end{center}
\end{figure}

\begin{figure}[H]
\begin{center}

 \begin{minipage}[b]{0.45\linewidth}
   \includegraphics [width=9cm]{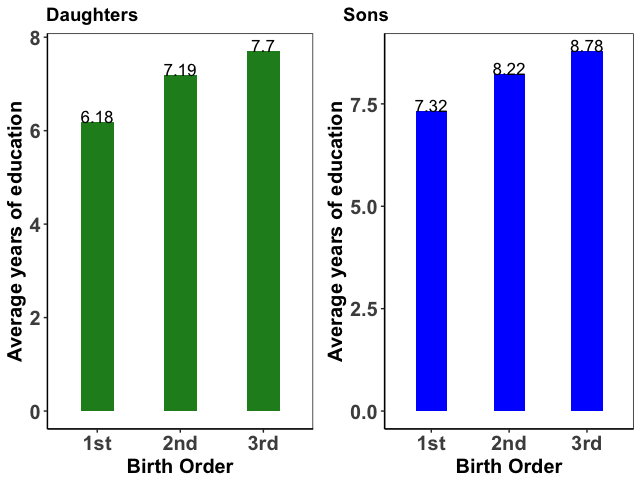}
 \end{minipage}
\caption{Average education by birth order for households with $N_c = 3$ (Benin, 2013)}
 \label{educ_diff_bo_nc3}
 \end{center}
\end{figure}

\begin{figure}[H]
 \centering
 \begin{minipage}[b]{0.45\linewidth}
   \includegraphics [width=9cm]{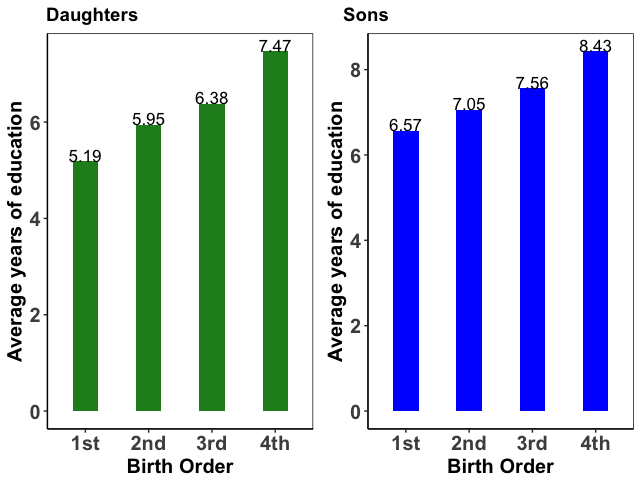}
 \end{minipage}
\caption{Average education by birth order for households with $N_c = 4$ (Benin, 2013)}
 \label{educ_diff_bo_nc4}
\end{figure}

\begin{figure}[H]
 \centering
 \begin{minipage}[b]{0.45\linewidth}
   \includegraphics [width=9cm]{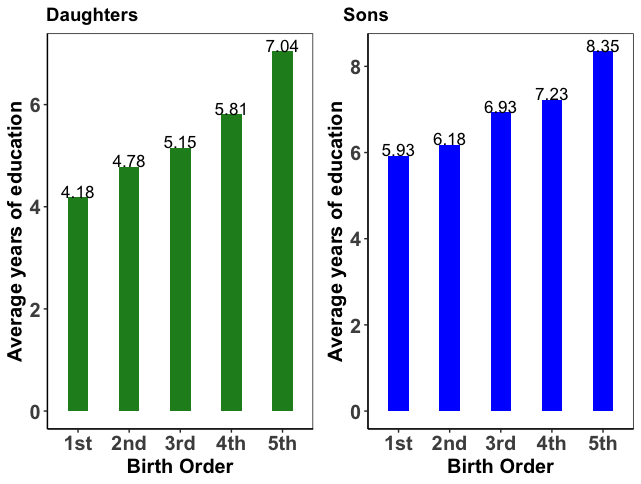}
 \end{minipage}
\caption{Average education by birth order for households with $N_c = 5$ (Benin, 2013)}
 \label{educ_diff_bo_nc5}
\end{figure}

 \begin{figure}[H]
\begin{center}
   \includegraphics [width=8cm]{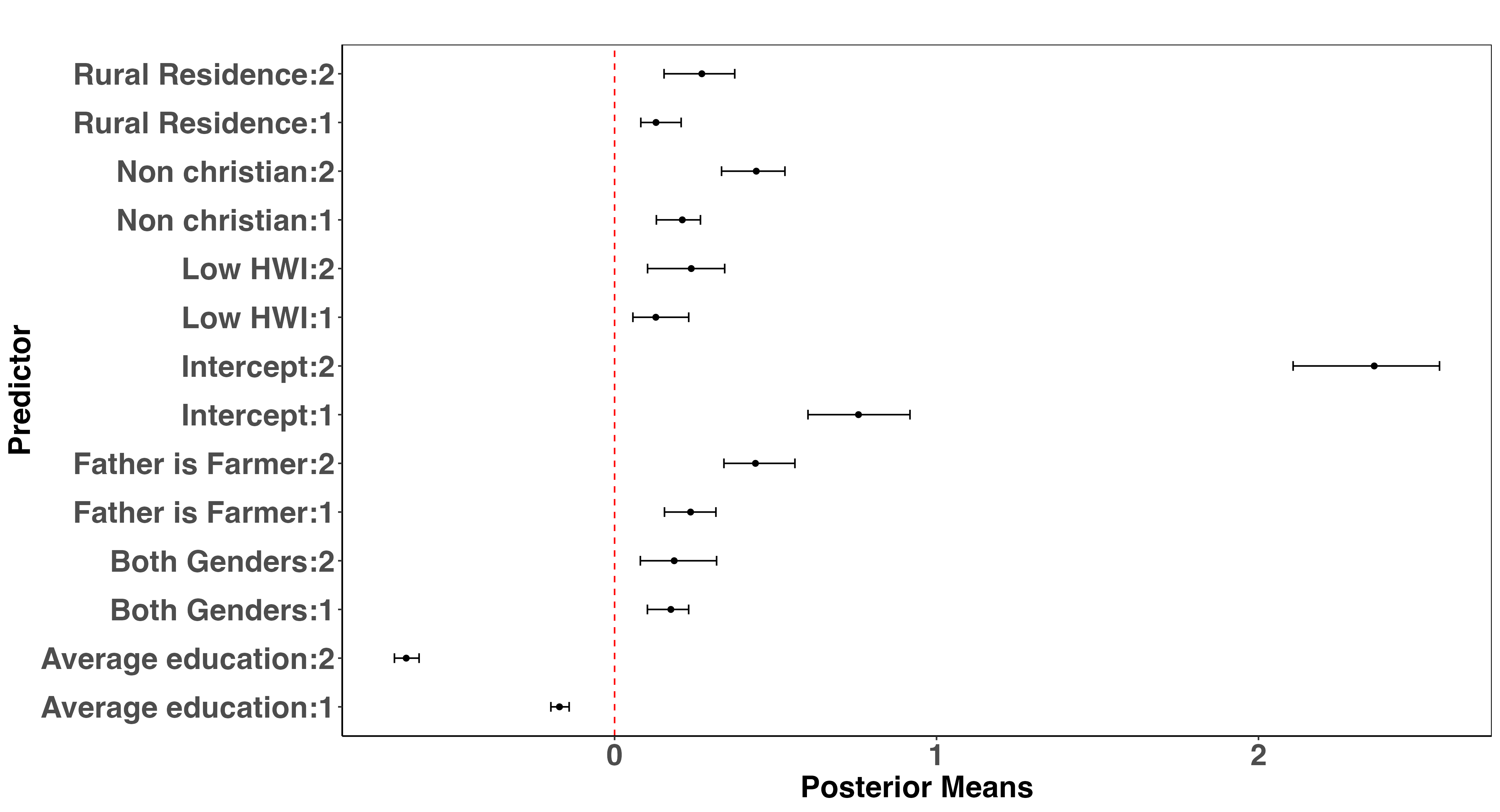}
   \caption{$\hat{\beta}$.}
    \end{center}
\end{figure}

 \begin{figure}[H]
\begin{center}
    \includegraphics [width=8cm]{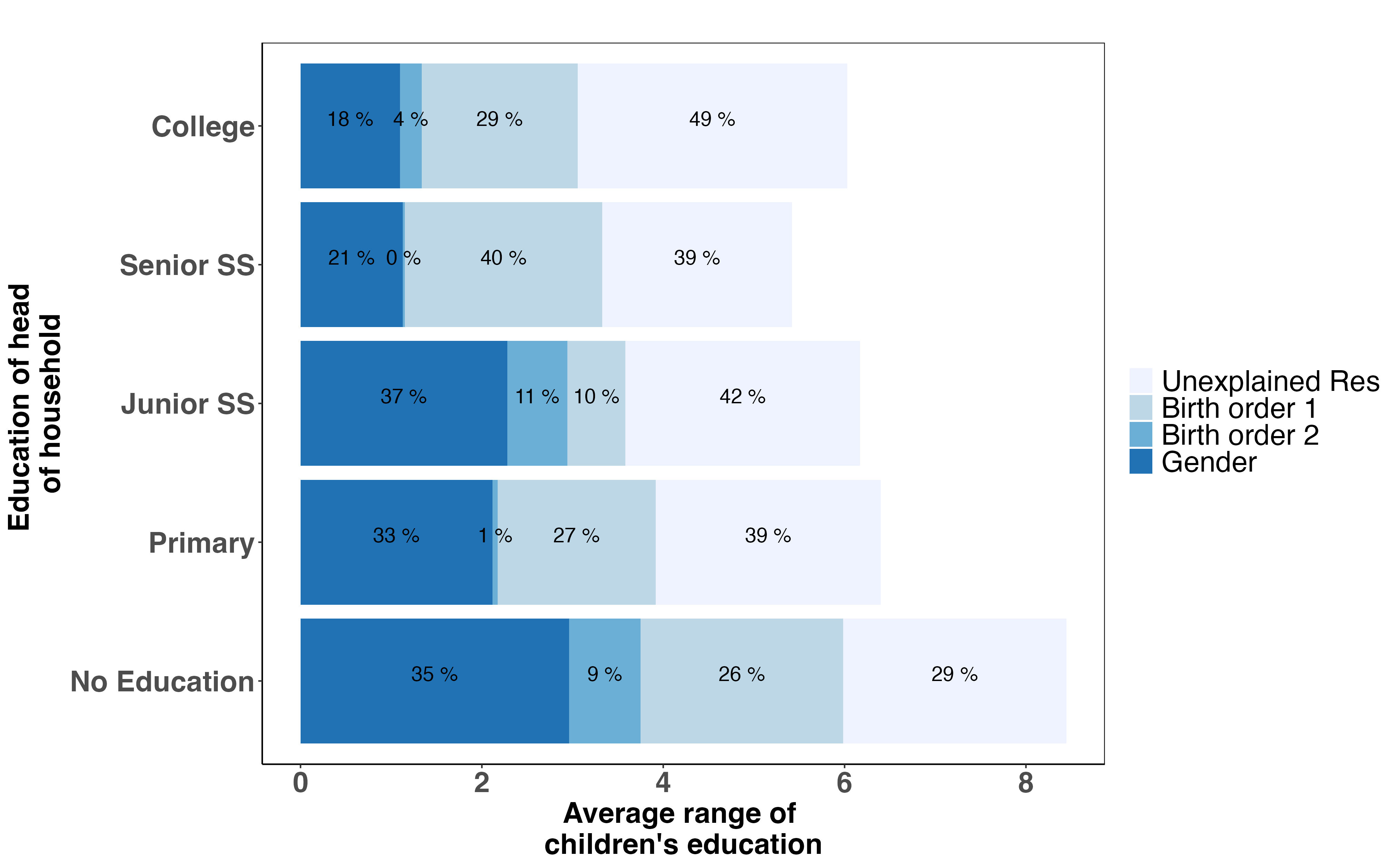}
  \caption{Effect of gender and birth order on within-household inequality in education and Inequality Decomposition ($N_c = 3$, Benin, 2013).}
 \label{NC_hist}
    \end{center}
\end{figure}

\newpage
\clearpage
\bibliography{references}

@article{becker1973interaction,
  title={On the Interaction between the Quantity and Quality of Children},
  author={Becker, Gary S and Lewis, H Gregg},
  journal={Journal of political Economy},
  volume={81},
  pages={S279--S288},
  year={1973},
  publisher={The University of Chicago Press}
}

@article{becker1976child,
  title={Child Endowments and the Quantity and Quality of Children},
  author={Becker, Gary and Tomes, Nigel},
  journal={Journal of political Economy},
  volume={84},
  pages={S143--S162},
  year={1976},
  publisher={The University of Chicago Press}
}

@article{ota2007within,
  title={The Within-household Schooling Decision: a Study of Children in Rural Andhra Pradesh},
  author={Ota, Masako and Moffatt, Peter G},
  journal={Journal of Population Economics},
  volume={20},
  pages={223--239},
  year={2007},
  publisher={Springer}
}

@article{nerlove1984investment,
  title={Investment in Human and Non-human Capital, Transfers among Siblings, and the Role of Government},
 author={Nerlove, Marc and Razin, Assaf and Sadka, Efraim},
  journal={Econometrica: Journal of the Econometric Society},
  pages={1191--1198},
  year={1984},
  publisher={JSTOR}
}

@article{li2008quantity,
  title={The Quantity-Quality Trade-off of Children in a Developing Country: Identification using Chinese Twins},
  author={Li, Hongbin and Zhang, Junsen and Zhu, Yi},
  journal={Demography},
  volume={45},
  pages={223--243},
  year={2008},
  publisher={Springer}
}

@book{montgomery1995tradeoff,
  title={The Tradeoff between Number of Children and Child Schooling: Evidence from Cote d'ivoire and Ghana},
  author={Montgomery, Mark},
  volume={112},
  year={1995},
  publisher={World Bank Publications}
}

@article{wantchekon2015education,
  title={Education and Human Capital Externalities: Evidence from Colonial Benin},
  author={Wantchekon, Leonard and Kla{\v{s}}nja, Marko and Novta, Natalija},
  journal={The Quarterly Journal of Economics},
  volume={130},
  pages={703--757},
  year={2015},
  publisher={MIT Press}
}

@article{maralani2008changing,
  title={The Changing Relationship between Family Size and Educational Attainment over the Course of Socioeconomic Development: Evidence from Indonesia},
  author={Maralani, Vida},
  journal={Demography},
  volume={45},
  pages={693--717},
  year={2008},
  publisher={Springer}
}

@article{conley2006parental,
  title={Parental Educational Investment and Children’s Academic Risk Estimates of the Impact of Sibship size and Birth order from Exogenous Variation in Fertility},
  author={Conley, Dalton and Glauber, Rebecca},
  journal={Journal of human resources},
  volume={41},
  pages={722--737},
  year={2006},
  publisher={University of Wisconsin Press}
}

@article{black2005more,
  title={The More the Merrier? The Effect of Family Size and Birth order on Children's Education},
  author={Black, Sandra E and Devereux, Paul J and Salvanes, Kjell G},
  journal={The Quarterly Journal of Economics},
  volume={120},
  pages={669--700},
  year={2005},
  publisher={MIT Press}
}

@article{biswas2000gender,
  title={Gender Disparities in Education in the Developing World},
  author={Biswas, Debashis},
  journal = {The Expression},
  year={2000}
}

@article{fergusson2006birth,
  title={Birth order and educational achievement in adolescence and young adulthood},
  author={Fergusson, David M and Horwood, L John and Boden, Joseph M},
  journal={Australian Journal of Education},
  volume={50},
  number={2},
  pages={122--139},
  year={2006},
  publisher={SAGE Publications Sage UK: London, England}
}

@article{de2010birth,
  title={Birth order, Family size and Educational Attainment},
  author={De Haan, Monique},
  journal={Economics of Education Review},
  volume={29},
  pages={576--588},
  year={2010},
  publisher={Elsevier}
}

@article{moshoeshoe2016birth,
  title={Birth Order Effects on Educational Attainment and Child Labour: Evidence from Lesotho},
  author={Moshoeshoe, Ramaele and others},
  journal={Economic Research Southern Africa (ERSA)},
  year={2016}
}

@article{weng2019family,
  title={Family Size, Birth Order and Educational Attainment: Evidence from China},
  author={Weng, Qian and Gao, Xia and He, Haoran and Li, Shi},
  journal={China Economic Review},
  volume={57},
  pages={101346},
  year={2019},
  publisher={Elsevier}
}

@article{esposito2020importance,
  title={The Importance of being Earliest: Birth Order and Educational Outcomes along the Socioeconomic Ladder in Mexico},
  author={Esposito, Lucio and Kumar, Sunil Mitra and Villase{\~n}or, Adri{\'a}n},
  journal={Journal of Population Economics},
  volume={33},
  pages={1069--1099},
  year={2020},
  publisher={Springer}
}

@article{ombati2012gender,
  title={Gender Inequality in Education in Sub-Saharan Africa},
  author={Ombati, Victor and Ombati, Mokua},
  journal={JWEE},
  number={3-4},
  pages={114--136},
  year={2012}
}

@article{osadan2014gender,
  title={Gender Equality in Primary Schools in Sub-Saharan Africa: Review and analysis},
  author={Osadan, Robert and Burrage, Irish A},
  journal={Wagadu: A Journal of Transnational Women's \& Gender Studies},
  volume={12},
  pages={10},
  year={2014}
}

@article{psaki2018measuring,
  title={Measuring Gender Equality in Education: Lessons from Trends in 43 Countries},
  author={Psaki, Stephanie R and McCarthy, Katharine J and Mensch, Barbara S},
  journal={Population \& Development Review},
  volume={44},
  number={1},
  year={2018}
}

@article{giannola2023parental,
  title={Parental Investments and Intra-household Inequality in Child Human Capital: Evidence from a Survey Experiment},
  author={Giannola, Michele},
  journal={The Economic Journal},
  volume={134},
  pages={671--727},
  year={2024},
  publisher={Oxford University Press}
}

@article{dizon2019parents,
  title={Parents’ Beliefs about their Children’s Academic Ability: Implications for Educational Investments},
  author={Dizon-Ross, Rebecca},
 author={Dizon-Ross, Rebecca},
  journal={American Economic Review},
  volume={109},
  pages={2728--2765},
  year={2019},
  publisher={American Economic Association 2014 Broadway, Suite 305, Nashville, TN 37203}
}

@article{thomas2003measuring,
  title={Measuring Education Inequality: Gini Coefficients of Education for 140 Countries, 1960-2000},
 author={Thomas, Vinod and Wang, Yan and Fan, Xibo},
  journal={Journal of Educational Planning and Administration},
  volume={17},
  number={1},
  year={2003}
}

@inproceedings{londono1990kuznetsian,
  title={Kuznetsian Tales with Attention to Human Capital},
  author={Londo{\~n}o, Juan Luis},
  booktitle={Third Inter-American Seminar in Economics, Rio de Janeiro, Brazil},
  year={1990}
}

@article{ram1990educational,
  title={Educational Expansion and Schooling Inequality: International Evidence and some Implications},
  author={Ram, Rati},
  journal={The Review of Economics and Statistics},
  pages={266--274},
  year={1990},
  publisher={JSTOR}
}

@article{unesco2021,
  title={Global Education Monitoring Report, 2021/2: Non-State Actors in Education: Who Chooses? Who Loses?},
  author={UNESCO},
  journal={United Nations Educational, Scientific and Cultural Organization.},
  year={2021},
  publisher={Paris: UNESCO, 2021}
}

@article{morrisson2013kuznets,
  title={The Kuznets Curve of Human Capital Inequality: 1870--2010},
 author={Morrisson, Christian and Murtin, Fabrice},
  journal={The Journal of Economic Inequality},
  volume={11},
  pages={283--301},
  year={2013},
  publisher={Springer}
}


\end{document}